 \documentclass [12pt,a4paper]{report}
\usepackage{isridef}

\begin{document}

\title{\bf\vspace{-3.5cm} The Heisenberg-Pauli Canonical Formalism\\
                               of Quantum Field Theory in the Rigorous\\
                          Setting of Nonlinear Generalized Functions\\
      ~\\
      ~\\
      {(Part I)}\\
      ~\\
      }
%                         =========================================

\author{ {\bf Jean-Fran\c{c}ois Colombeau}\\
             (jf.colombeau@wanadoo.fr)\\
                 {\bf Andre Gsponer}\\
         {\it Independent Scientific Research Institute}\\
              {\it Oxford, OX4 4YS, England}\\
       }

%\date{PROOF.23 ~~ \today}  % else on arXived version:

\date{ \normalsize \vspace{5cm} \emph{Category}: MP, mathematical physics.
     }

\maketitle

\newpage
~\\

\begin{center}
\medskip
{\Huge {\bf Abstract}}
%=====================

~\\

\end{center}

The unmodified Heisenberg-Pauli canonical formalism of quantum field theory applied to a self-interacting scalar boson field is shown to make sense mathematically in a framework of generalized functions adapted to nonlinear operations. 

 In this framework the operator-valued distributions defining the quantized fields are regularized \emph{ab initio} so that all operators and quantities subsequently derived from them make sense, and the functions and parameters characterizing the regularization are kept as general as possible until the end of the calculation so that that they do not constrain \emph{a priori} the physical predictions of the theory.

The free-field operators, their commutation relations, and the free-field Hamiltonian operator are calculated by a straightforward transcription of the usual formalism expressed in configuration space.  This leads to the usual results, which are essentially independent of the regularization, with the exception of the zero-point energy which may be set to zero if a particular regularization is chosen.

The calculations for the self-interacting field are more difficult, especially because of the well-known problems due to the unboundness of the operators and their time-dependent domains.  Nevertheless, a proper methodology is developed and a differentiation on time-dependent domains is defined.  The Heisenberg equations and the interacting-field equation are shown to be mathematically meaningful as operator-valued nonlinear generalized functions, which therefore provide an alternative to the usual Bogoliubov-Wightman interpretation of quantized fields as operator-valued distributions.

The equation for the time-evolution operator is proved using two different methods, but no attempt is made to calculate the scattering operator, and the applications to perturbation theory are left to a subsequent report.

The main conclusion is that the unmodified Heisenberg-Pauli calculations make sense mathematically, but since these calculations deal with generalized functions whose values are operators on a Hilbert space there remain serious needs of improvements, especially by mathematicians, due in particular to the unboundness of these operators and to the time dependence of their domains.

%%%Recalls: page (29)

%%%The interacting field operator $\bphi(t,\vec{x}\,)$ is for every $\{t,\vec{x}\}$ an equivalence class of unbounded operators $\bphi(\rho_\epsilon,t,\tau,\vec{x}\,)$ on the Fock space $\mathbb{F}$ mapping a dense subspace $\mathbb{D}(\rho_\epsilon,t,\tau,) \subset \mathbb{F}$ into itself.

\tableofcontents
%===============

%%\setcounter{chapter}{-1}
%%\chapter {Preface}
%%======================
%%\label{pref:0}

%%%%%%%%%%%%%%%%%% End of introduction %%%%%%%%%%%%%%%%%%%%%

%File: prer.30.tex          arXiv version 2       Date: 6 September 2008
%=====                
%           
%
\chapter{Introduction}
%======================
\label{prer:0}

This report contains the detailed proofs of the calculations presented in paper~\cite{C-G-P-2007}, which can be considered as a preface to the present report.

In this introductory chapter we summarize in Secs.\,\ref{fock:0} to \ref{summ:0} the basic concepts of quantum field theory and of the canonical formalism.

Then, in Sec.\,\ref{moll:0} and \ref{oper:0}, we define the operator-valued nonlinear generalized functions, as well as the suitable `mollifiers' and `dampers' used to regularize them.  

Throughout the report we will use the adjective `classical' to qualify objects defined in the ordinary setting of standard analysis and distribution theory, to distinguish them from those defined in the context of the theory of nonlinear generalized functions.  For brevity, we will refer to this context as the $\mathcal{G}$-context or setting, and we will use the prefix `$\mathcal{G}$-' to qualify objects and concepts arising in that setting: $\mathcal{G}$-function, $\mathcal{G}$-embedding, etc.

\section{Fock space and creation/annihilation operators}
%-------------------------------------------------------
\label{fock:0}
\setcounter{equation}{0}
\setcounter{definition}{0}
\setcounter{axiom}{0}
\setcounter{conjecture}{0}
\setcounter{lemma}{0}
\setcounter{theorem}{0}
\setcounter{corollary}{0}
\setcounter{proposition}{0}
\setcounter{example}{0}
\setcounter{remark}{0}
\setcounter{problem}{0}

The basic mathematical objects of quantum field theory (QFT) are \emph{functions} in a Hilbert space, which are interpreted as the states of elementary particles, and \emph{operators} acting on these functions, which correspond to the dynamical variables of the theory.  The main objective of QFT is to describe the creation/annihilation of these state-functions during physical interactions.

\subsection{Fock space}
%...................... 

The space of states called Fock space is the Hilbertian direct sum
\begin{equation}\label{fock:1}
        \mathbb{F} \DEF \bigoplus_{n=0}^\infty
                        \mathsf{L}_S^2\bigl((\mathbb{R}^3)^n,\mathbb{C} \bigr), 
\end{equation}
where, for $n>0$, ~$\mathsf{L}_S^2\bigl((\mathbb{R}^3)^n,\mathbb{C} \bigr)$ is the Hilbert space of complex valued symmetric square integrable functions on $(\mathbb{R}^3)^n$, with respect to the Lebesgue measure, and for $n=0$, $\mathsf{L}_S^2\bigl((\mathbb{R}^3)^n,\mathbb{C} \bigr)$ stands for the field of complex numbers. That is, an element of $\mathbb{F}$ is an infinite sequence $(f_n)_n, n=0, ..., \infty$, such that $|f_0|^2 + \sum_{n=1}^{\infty}(\|(f_n)\|_n)^2 < \infty$, where $\|~\|_n$ is the norm in $\mathsf{L}_S^2\bigl((\mathbb{R}^3)^n,\mathbb{C} \bigr)$, $f_0 \in \mathbb{C}$ is a constant, and $f_n$ stands for the symmetric function
\begin{equation}\label{fock:2}
  f_n ~:~ ({\vec \xi}_1,... , {\vec \xi}_n) \rightarrow
        ~({\rm Sym}~f_n)({\vec \xi}_1,... , {\vec \xi}_n), \quad {\vec \xi}_j \in \mathbb{R}^3,
\end{equation}
with
\begin{equation}\label{fock:3}
   ({\rm Sym}~f)({\vec \xi}_1,... , {\vec \xi}_n) = \frac{1}{n!}\sum_{\pi(~)}
        f({\vec \xi}_{\pi(1)},... , {\vec \xi}_{\pi(n)}),
\end{equation}
where $\pi(~)$ belongs to the set of the $n!$ permutations of $\{1,...,n\}$.

From the definition of a Hilbertian direct sum, we have 
\begin{align}
\label{fock:4}
    \|(f_n)\|_\mathbb{F} &= \Bigl( |f_0|^2 + \sum_{n=1}^{\infty}(\|(f_n)\|_n)^2 \Bigr)^{1/2},\\
\label{fock:5}
         \langle (f_n)|(g_n) \rangle_\mathbb{F} &= f_0^* \cdot {g_0}
                     + \sum_{n=1}^{\infty} \langle f_n|g_n \rangle_n,
\end{align}
where $^*$ is complex conjugation and $\langle ~|~ \rangle_n$ is the scalar product in $\mathsf{L}_S^2\bigl((\mathbb{R}^3)^n,\mathbb{C} \bigr)$, whose precise form will be defined in section \ref{fock:0}.3.

  In the following we shall use a dense subspace of $\mathbb{F}$, i.e., 
\begin{equation}\label{fock:6}
     \mathbb{D} \DEF \bigl\{~ (f_n)_n \in \mathbb{F} 
                     \text{~~such that $f_n=0$ for $n$ large enough}~\bigr\}. 
\end{equation}
$\mathbb{D}$ is said to be the family of the `states with a finite number of particles,' and the state $\mho \DEF (f_0,0,...)$ is called the `no-particle' or `vacuum' state.

\subsection{Creation and annihilation operators}
%............................................... 

If $\psi \in \mathsf{L}_S^2\bigl(\mathbb{R}^3\bigr)$ the creation operator $\mathbf{a}^+(\psi)$ is given by the formula
\begin{equation}\label{fock:7}
     \mathbf{a}^+(\psi)
\begin{pmatrix}
     f_0\\
     f_1\\
     f_2\\
     \vdots\\
     f_n\\
     \vdots
\end{pmatrix} 
     =
\begin{pmatrix}
     0 \\
     f_0 ~ \psi({\vec \xi}_1)\\
     \sqrt{2} ~{\rm Sym}( \psi({\vec \xi}_2) \otimes f_1)\\
     \vdots\\
     \sqrt{n} ~{\rm Sym}( \psi({\vec \xi}_n) \otimes f_{n-1})\\
     \vdots
\end{pmatrix} ,
\end{equation}
where ${\rm Sym}$ is the symmetrization operator \eqref{fock:3}. 

If $\psi \in \mathsf{L}_S^2\bigl(\mathbb{R}^3\bigr)$ the annihilation operator $\mathbf{a}^-(\psi)$ is given by the formula
\begin{equation}\label{fock:8}
     \mathbf{a}^-(\psi)
\begin{pmatrix}
     f_0\\
     f_1\\
     f_2\\
     \vdots\\
     f_n\\
     \vdots
\end{pmatrix} 
     =
\begin{pmatrix}
     \SCA \psi({\vec \xi}_1)| f_1({\vec \xi}_1) \LAR \\
     \sqrt{2} \SCA \psi({\vec \xi}_2)|
                        f_2({\vec \xi}_1,{\vec \xi}_2) \LAR\\
~~~~~\sqrt{3} \SCA \psi({\vec \xi}_3)|
                        f_3({\vec \xi}_1,{\vec \xi}_2,{\vec \xi}_3) \LAR\\
     \vdots\\
  ~~ \sqrt{n+1} \SCA \psi({\vec \xi}_{n+1})|
                        f_{n+1}({\vec \xi}_1,...,{\vec \xi}_{n+1}) \LAR\\
     \vdots
\end{pmatrix} ,
\end{equation}
so that, in particular, $\mathbf{a}^-(\psi)\mho =0$.  Due to the sesquilinearity of the scalar product the annihilation operator $\mathbf{a}^-(\psi)$ is antilinear while the creation operator $\mathbf{a}^+(\psi)$ is linear.  These operators are defined at least on the dense subspace $\mathbb{D}$ of $\mathbb{F}$, with values in $\mathbb{D}$. They are not bounded operators on $\mathbb{F}$ because of the coefficients $\sqrt{n}$ and $\sqrt{n+1}$.

%\subsection{Propositions}
%........................ 

The proofs of the following two propositions are straightforward verifications:
\begin{proposition}
%..................................
\label{fock:prop:1}
$\mathbf{a}^+(\psi)$ and $\mathbf{a}^-(\psi)$ are adjoint to each other, i.e., 
\begin{equation}\label{fock:9}
     \forall \Psi_1,\Psi_2 \in \mathbb{D} \qquad 
     \langle \mathbf{a}^+(\psi)\Psi_1|\Psi_2 \rangle_{\mathbb{F}} =
     \langle \Psi_1|\mathbf{a}^-(\psi)\Psi_2 \rangle_{\mathbb{F}}.
\end{equation}
\end{proposition}
\begin{proposition}
%..................................
\label{fock:prop:2}
$\mathbf{a}^+(\psi_1)$ and $\mathbf{a}^-(\psi_2)$ satisfy the commutation relations
\begin{align}
\label{fock:10}
[\mathbf{a}^+(\psi_1),\mathbf{a}^+(\psi_2)] &=0,\\
\label{fock:11}
[\mathbf{a}^-(\psi_1),\mathbf{a}^-(\psi_2)] &=0,\\
\label{fock:12}
[\mathbf{a}^-(\psi_1),\mathbf{a}^+(\psi_2)] &=
            \langle \psi_1 | \psi_2 \rangle~\mathbf{1}, 
\end{align}
where $[\mathbf{a},\mathbf{b}] \DEF \mathbf{a}\mathbf{b}-\mathbf{b}\mathbf{a}$.
\end{proposition}

\subsection{Scalar product}
%..........................
% References:  [B+D:2]p.27, [Costa]36-37,45-49, [Schweber]57-59, [Kastler]93

In the quantum theory of a spin-0 (or, `scalar' boson) field the argument of the creation and annihilation operators is a Lorentz-invariant scalar function $\psi(\xi) \in \mathbb{C}$ where the variable $\xi \DEF \{ t_\xi, \vec{\xi}\, \} =  \{ \xi_\mu \} \in \mathbb{R}^4$ and $\mu = 0,..,3$.\footnote{For mathematical clarity we systematically write $\{t,\vec{x}\,\} \in \mathbb{R}^4$ for a point in the Minkowski space, i.e., in this report, we do not take explicitly care of the difference in metric, nor of Lorentz invariance which is to a large extent ensured by the fact that we follow exactly the Heisenberg-Pauli calculations (it is well known that they lead a posteriori to a `formally' Lorentz invariant result).}  The scalar product of two $n$-particle states is then not given by the usual $\mathsf{L}_S^2\bigl((\mathbb{R}^3)^n,\mathbb{C} \bigr)$ formula 
\begin{align}
\label{fock:13}
 \langle f_n|g_n \rangle_n  = 
        \iiint \cdots \iiint
             f^*_n({\vec \xi}_1,... , {\vec \xi}_n) ~
             g_n  ({\vec \xi}_1,... , {\vec \xi}_n) ~ d^3\xi_1 \cdots d^3\xi_n,
\end{align}
because the 3-volume element $d^3\xi$ is not Lorentz invariant.  Instead, it is the relativistically invariant expression \cite[p.\,828]{WIGHT1955-}, \cite[p.\,93]{KASTL1961-},
\begin{align}
\nonumber
  \BRA f_n(\xi_1,..., \xi_n)\|g_n(\xi_1,..., \xi_n) \KET_n
  & =  (-i)^n\int \cdots \int d^3\xi_1 \cdots d^3\xi_n \\
\nonumber
  & \times f^*_n(\xi_1,..., \xi_n)
      \Bigl(\overleftarrow{\frac{\partial}{\partial t_{\xi_1}}}
    - \overrightarrow{\frac{\partial}{\partial t_{\xi_1}}} \Bigr) \cdots\\
\label{fock:14}
  & \times \Bigl(\overleftarrow{\frac{\partial}{\partial t_{\xi_n}}}
    - \overrightarrow{\frac{\partial}{\partial t_{\xi_n}}} \Bigr)
            g_n(\xi_1,..., \xi_n).
\end{align}
When $n=1$ this simplifies to
\begin{equation}\label{fock:15}
          \BRA  f_1(\xi) \| g_1(\xi) \KET
    \DEF  i \iiint_{t_\xi=\text{Cst.}} d^3\xi~\Bigl(
             f_1^* \frac{\partial g_1  }{\partial t_\xi} 
                 - \frac{\partial f_1^*}{\partial t_\xi} g_1 \Bigr),
\end{equation}
which is a special case of
\begin{equation}\label{foc:16}
          \BRA  f_1(\xi) \| g_1(\xi) \KET
     := i \iiint_\Sigma d^3\Sigma^\mu~\Bigl(
             f_1^* \frac{\partial g_1  }{\partial \xi_\mu} 
                 - \frac{\partial f_1^*}{\partial \xi_\mu} g_1 \Bigr),
\end{equation}
where $\Sigma$ is a space-like hypersurface, and $\mu$ a tensor index so that the contraction $\Sigma^\mu {\partial_{\xi_\mu}}$ is indeed invariant. Taking for $\Sigma$ the hyperplane  $t_\xi=\mathrm{Cst}$ leads to \eqref{fock:15}.

  The norm induced by these definitions is written $\|\cdot\|_{\mathsf{L}^2}$.

\section{Free-field states and operators}
%----------------------------------------
\label{stat:0}
\setcounter{equation}{0}
\setcounter{definition}{0}
\setcounter{axiom}{0}
\setcounter{conjecture}{0}
\setcounter{lemma}{0}
\setcounter{theorem}{0}
\setcounter{corollary}{0}
\setcounter{proposition}{0}
\setcounter{example}{0}
\setcounter{remark}{0}
\setcounter{problem}{0}

The complete specification of the Fock space depends on the representation chosen for the states.  From this representation derives the explicit form of the operators.

\subsection{Free-field states}
%.............................

In  spin-0 field theory the argument of the creation and annihilation operators is the relativistically invariant function
\begin{equation}\label{stat:1}
                \psi ~ : \qquad  \xi \mapsto \Delta_+(\xi-x),
\end{equation}
where $x = \{t,\vec{x}\} \in \mathbb{R}^4$ is a parameter.  The function
\begin{equation}\label{stat:2}
  \Delta_+(x) =
\frac{1}{(2\pi)^3}   \iiint \frac{d^3p}{2 E_p} \exp 
    i (\vec{p}\cdot \vec{x} - E_p t),
\end{equation}
where $E_p = \sqrt{|\vec{p}\,|^2+m^2}$ is often written as
\begin{equation}\label{stat:3}
  \Delta_{\pm}(x_1-x_2) =
   \iiint d^3p~\psi_p^{\pm}(x_1)\psi_p^{\pm}(-x_2),
\end{equation}
where
\begin{equation}\label{stat:4}
            \psi_p^{ }  (t,\vec{x}\,) \DEF \frac{1}{\sqrt{(2\pi)^3 2 E_p}}
            \exp   i (\vec{p}\cdot\vec{x} - E_pt),
\end{equation}
satisfies the orthogonality and normalization relations\footnote{This standard normalization ignores a factor $\exp{i(E_{p_1}-E_{p_2})t}$ on the right of \eqref{stat:6} because it disappears when evaluated on a test function.}
\begin{align}
\label{stat:5}
      \BRA \psi_{p_1}^{ }(x) \| \psi_{p_2}^{*}(x) \KET &= 0,\\
\label{stat:6}
      \BRA \psi_{p_1}^{ }(x) \| \psi_{p_2}^{ }(x) \KET &=
                                \delta^3(\vec{p_2}-\vec{p_1}),
\end{align}
because 
\begin{align}
\label{stat:7}
    \delta^3(\vec{x}\,) \DEF \frac{1}{(2\pi)^3} \iiint d^3p ~
            \exp ( \pm i \vec{p} \cdot \vec{x} ).
\end{align}
Using these relations it is easy to verify that
\begin{align}
\label{stat:8}
       \BRA \psi_p(\xi) \| \Delta_{+}^{*}(\xi-x)\KET 
      &= 0,\\
\label{stat:9}
       \BRA \psi_p(\xi) \| \Delta_{+}    (\xi-x)\KET 
      &=    \psi_p(-x),
\end{align}
and
\begin{align}
\label{stat:10}
       \BRA \Delta_{+}(\xi-x_1) \| \Delta_{+}^{*}(\xi-x_2)\KET 
      &= 0,\\
\label{stat:11}
       \BRA \Delta_{+}(\xi-x_1) \| \Delta_{+}    (\xi-x_2)\KET 
      &=    \Delta_{+}(x_1-x_2).
\end{align}
In particular, for $x_1=x_2=x$, we have
\begin{equation}\label{stat:12}
    \|\Delta_{+}(\xi-x)\|^2_{\mathsf{L}^2} = \Delta_{+}(0) =
      \frac{1}{(2\pi)^3}   \iiint \frac{d^3p}{2 E_p} =
      \lim_{p\rightarrow\infty} \frac{p^2}{\pi^2} = \infty,
\end{equation}
so that the state function \eqref{stat:1} is not in $\mathsf{L}_S^2\bigl(\mathbb{R}^3\bigr)$ as required by the definition of the Fock space: This lack of rigor is classically corrected by defining the operators depending on $\psi=\Delta_+(\xi-x)$ as distributions.

   Applying the creation operator $\mathbf{a}^+(\psi)$ defined by \eqref{fock:7} on the no-particle state $\mho = (f_0,0,...)$, the positive-energy free-field single-particle states is readily found to be $f_1 = f_0 \Delta_+(\xi-x)$.  Repetitive application of $\mathbf{a}^+(\psi)$ will then generate a basis of the Fock space $\mathbb{F}$, i.e., 
\begin{align}
\nonumber
b_0 &= f_0,\\
\nonumber
b_1 &= f_0 \Delta_+(\xi_1-x_1),\\
\nonumber
b_2 &= f_0  \sqrt{1/2}  \Bigl(
            \Delta_+(\xi_2-x_2)\Delta_+(\xi_1-x_1)
          + \Delta_+(\xi_1-x_2)\Delta_+(\xi_2-x_1)
                        \Bigr),\\
\nonumber
b_3 &= f_0   \sqrt{1/6}  \Bigl(
           \Delta_+(\xi_3-x_3) \Delta_+(\xi_2-x_2)\Delta_+(\xi_1-x_1)
          + (\text{5 terms})
                         \Bigr),\\
\label{stat:13}
b_4 &= ...,
\end{align}
which by linear superposition enables to densely populate the whole subspace $\mathbb{D} \subset \mathbb{F}$.

\subsection{Free-field operators}
%................................

The free-field operator is defined by
\begin{align}
\label{stat:14}
         {\bphi}_0(t,\vec{x}\,)
           = \mathbf{a}^+\bigl(\Delta_+(\xi-x) \bigr) 
           + \mathbf{a}^-\bigl(\Delta_+(\xi-x) \bigr), 
\end{align}
where $\Delta_+(\xi-x)$ is the function of the variable $\xi$ given by \eqref{stat:2}, and where the variable $x=\{t,\vec{x}\}$ corresponds to the space-time dependence of ${\bphi}_0$.  This function, however, is not in $\mathsf{L}_S^2\bigl(\mathbb{R}^3\bigr)$, so that the mathematical object that makes sense is
\begin{equation}\label{stat:15}
              {\bphi}_0(t,T) := \iiint_{\mathbb{R}^3}
              {\bphi}_0(t,\vec{x}\,) T(\vec{x}\,)~d^3x, 
\end{equation}
with $T(\vec{x}\,) \in \mathcal{D}$ a test function.  That is: ${\bphi}_0(t,\vec{x}\,)$ is a \emph{distribution} in the $\vec{x}$-variable whose values ${\bphi}_0(t,T)$ are densely defined linear unbounded operators on $\mathbb{F}$ (they map $\mathbb{D}$ into $\mathbb{D}$).  We set
\begin{equation}\label{stat:16}
  {\bpi}_0(t,\vec{x}\,) := \frac{\partial}{\partial t}{\bphi}_0(t,\vec{x}\,), 
\end{equation}
which is again a similar distribution ${\bpi}_0(t,T)$.

   Since by Proposition \ref{fock:prop:1} the operators $\mathbf{a}^+(\psi)$ and $\mathbf{a}^-(\psi)$ are adjoint to each other,  ${\bpi}_0(t,\vec{x}\,)$ is formally a symmetric operator.  However, as a distribution, we have:
\begin{proposition}
%..................
\label{stat:prop:1}
For real-valued $T \in \mathcal{D}$ the operator $\bphi_0(t,T)$ is a symmetric operator, i.e.,
\begin{equation}\label{stat:17}
     \forall \Psi_1,\Psi_2 \in \mathbb{D} \qquad 
     \langle \bphi_0(t,T)\Psi_1|\Psi_2 \rangle_{\mathbb{F}} =
     \langle \Psi_1|\bphi_0(t,T)\Psi_2 \rangle_{\mathbb{F}}.
\end{equation}
\end{proposition}
The proof follows at once from Proposition \ref{fock:prop:1}. \END
%
% NB: For symmetric vs self-adjoint, see A. Peres, and Riesz+Sz.Nagy
%

   In practice the field operator $\bphi_0$ is often written in terms of the function $\psi_p$ as
\begin{equation}\label{stat:18}
  \bphi_0(x) =
     \iiint d^3p ~\psi_p^*(x) \mathbf{a}^{+}\bigl(\psi_p(\xi)\bigr)
   + \iiint d^3p ~\psi_p  (x) \mathbf{a}^{-}\bigl(\psi_p(\xi)\bigr).
\end{equation}
One also introduces the abbreviations
\begin{align}
\label{stat:19}
         \mathbf{a}^+(x) \DEF \iiint d^3p ~\psi_p^*(x)
                              \mathbf{a}_p^+(\xi),
       \qquad \text{with} \qquad
         \mathbf{a}_p^+(\xi) \DEF \mathbf{a}^+\bigl(\psi_p(\xi)\bigr),
\end{align}
and
\begin{align}
\label{stat:20}
         \mathbf{a}^-(x) \DEF \iiint d^3p ~\psi_p  (x)
                              \mathbf{a}_p^-(\xi),
       \qquad \text{with} \qquad
         \mathbf{a}_p^-(\xi) \DEF \mathbf{a}^-\bigl(\psi_p(\xi)\bigr),
\end{align}
where $\psi_p(x)$ and $\psi_p(\xi)$ are given by \eqref{stat:4}.  Thus, \eqref{stat:14} can be expressed in the concise form
\begin{align}
\label{stat:21}
       \bphi_0(x) = \mathbf{a}^+(x) + \mathbf{a}^-(x), 
\end{align}
and similarly \eqref{stat:18} can be written
\begin{equation}\label{stat:22}
  \bphi_0(x) =
     \iiint d^3p ~\psi_p^*(x) \mathbf{a}_p^{+}
   + \iiint d^3p ~\psi_p  (x) \mathbf{a}_p^{-},
\end{equation}
which is the form that is frequently used in the introductory literature (where the $\xi$ dependence is generally left implicit) as the fundamental definition of $\bphi_0$.

\section{Summary of canonical formalism}
%---------------------------------------
\label{summ:0}
\setcounter{equation}{0}
\setcounter{definition}{0}
\setcounter{axiom}{0}
\setcounter{conjecture}{0}
\setcounter{lemma}{0}
\setcounter{theorem}{0}
\setcounter{corollary}{0}
\setcounter{proposition}{0}
\setcounter{example}{0}
\setcounter{remark}{0}
\setcounter{problem}{0}

In this section the customary Heisenberg-Pauli calculations (see for example \cite[p.\,21-22, 292-332]{WEINB1995-}) are summarized in the simplest non-trivial case, i.e., a self-interacting real scalar boson field. They are presented as purely formal calculations, i.e., they are not defined mathematically and are done by analogy with usual calculations on $\mathcal{C}^\infty$  functions.  The aim of the present report is to give them a rigorous mathematical sense. 

\subsection{Canonical commutation relations}
%...........................................

The fundamental postulates of the canonical formalism of QFT are (i) that the physical fields correspond to operators acting on the Fock space, and (ii) that these field operators obey the canonical equal-time commutation relations
\begin{align}
\label{summ:1}
[\bphi(t,\vec{x}_1),\bphi(t,\vec{x}_2)] &=0,\\
\label{summ:2}
[ \bpi(t,\vec{x}_1), \bpi(t,\vec{x}_2)] &=0,\\
\label{summ:3}
[\bphi(t,\vec{x}_1), \bpi(t,\vec{x}_2)] &=
            i \delta^3(\vec{x}_1-\vec{x}_2)~\mathbf{1}. 
\end{align}
The commutation relation \eqref{summ:3} implies that the fields $\bphi$ and $\bpi$ are necessarily distributions because the right-hand side is a Dirac $\delta$-function.   This difficulty is sometimes called `the initial singular problem of QFT.'  Considering $\bphi$ and $\bpi$ as distributions as in (\ref{stat:15}--\ref{stat:16}), and assuming $T_1(\vec{x}\,), T_2(\vec{x}\,) \in \mathcal{D}$, equation \eqref{summ:3} becomes
\begin{align}
\label{summ:4}
[{\bphi}(t,T_1), {\bpi}(t,T_2)] &=
       i \iiint_{\mathbb{R}^3}  T_1(\vec{x}\,) T_2(\vec{x}\,)~d^3x~\mathbf{1}, 
\end{align}
which is the  standard QFT interpretation of \eqref{summ:3}.

\subsection{Self-interacting-field equation}
%...........................................

Minimization of the action integral $\mathcal{A}( \phi, \partial_\mu\phi) = \int dt \iiint d^3x ~L\bigl( \phi(t,\vec{x}\,),\partial_\mu\phi(t,\vec{x}\,) \bigr)$, interpreted as a functional of the classical field $\phi$ and its first derivatives $\partial_\mu\phi$,  with
\begin{align}
\notag
     L \bigl( \phi(t,\vec{x}\,),\partial_\mu\phi(t,\vec{x}\,) \bigr)=
       &+ \frac{1}{2} \bigl(\partial_t\phi(t,\vec{x}\,)\bigr)^2 
       - \frac{1}{2} \sum_{1 \leq \mu \leq 3}
         \bigl(\partial_\mu \phi(t,\vec{x}\,)\bigr)^2 \\
\label{summ:5}  
       &- \frac{1}{2} m^2
         \bigl(\phi(t,\vec{x}\,)\bigr)^2 
       - \frac{g}{N+1} \bigl(\phi(t,\vec{x}\,)\bigr)^{N+1} , 
\end{align}
where $\partial_\mu = \partial/\partial x_\mu$ and $\partial_t = \partial/\partial t$, gives the {\it `self-interacting classical neutral spin-0 field equation:'} 
\begin{align}
\label{summ:6}
   \frac{\partial^2}{\partial t^2} \phi(t,\vec{x}\,) = 
   \sum_{1\leq\mu\leq 3} \frac{\partial^2}{\partial{x_\mu}^2} \phi(t,\vec{x}\,)
   - m^2 \phi(t,\vec{x}\,)
   - g \bigl(\phi(t,\vec{x}\,)\bigr)^N,
\end{align}
completed by the initial conditions at the time $t=\tau$
\begin{align}
\label{summ:7}
      \phi(\tau,\vec{x}\,) = \phi_{\text{ini}}(\tau,\vec{x}\,),
      \qquad \text{and} \qquad
      \frac{\partial}{\partial t} \phi             (\tau,\vec{x}\,) =
      \frac{\partial}{\partial t} \phi_{\text{ini}}(\tau,\vec{x}\,). 
\end{align}

   Heisenberg and Pauli's canonical quantization formalism consists of interpreting the field  $\phi$ and its canonical conjugate momentum $\pi\DEF \partial \phi(t,\vec{x}\,)/\partial t$ as operators $\bphi$, $\bpi$, operating upon states $\Psi$ in a Fock space $\mathbb{F}$, and satisfying the equal-time canonical commutation relations (\ref{summ:1}--\ref{summ:3}\,). Therefore:

\begin{definition}[Interacting-field equation]
%.............................................
\label{summ:defi:1}
The operator equation
\begin{align}
\label{summ:8}
   \frac{\partial}{\partial t} \bphi(t,\vec{x}\,) &= \bpi(t,\vec{x}\,),\\
\label{summ:9}
   \frac{\partial}{\partial t} \bpi (t,\vec{x}\,) &= 
   \sum_{1\leq\mu\leq 3} \frac{\partial^2}{\partial{x_\mu}^2} \bphi(t,\vec{x}\,)
   - m^2 \bphi(t,\vec{x}\,)
   - g \bigl(\bphi(t,\vec{x}\,)\bigr)^N,
\end{align}
completed by the initial conditions at the time $t=\tau$
\begin{align}
\label{summ:10}
      \bphi(\tau,\vec{x}\,) = \bphi_{\text{\rm ini}}(\tau,\vec{x}\,),
      \qquad \text{and} \qquad
      \bpi                 (\tau,\vec{x}\,) =
      \bpi_{\text{\rm ini}}(\tau,\vec{x}\,), 
\end{align}
as well as the equal-time commutations relations
\begin{align}
\label{summ:11}
[\bphi(t,\vec{x}_1),\bphi(t,\vec{x}_2)] &=0,\\
\label{summ:12}
[ \bpi(t,\vec{x}_1), \bpi(t,\vec{x}_2)] &=0,\\
\label{summ:13}
[\bphi(t,\vec{x}_1), \bpi(t,\vec{x}_2)] &=
            i \delta^3(\vec{x}_1-\vec{x}_2)~\mathbf{1}, 
\end{align}
is the field equation of a quantized self-interacting neutral spin-0 field.
\end{definition}
This is a wave equation with nonlinear second member. Since the initial condition is a pair of irregular distributions the solution is not expected to be more regular than a distribution for which the nonlinear term does not make sense in distribution theory (with further a `big' problem due to the fact one is confronted with unbounded operators).

\subsection{Hamiltonian function and Heisenberg equations}
%.........................................................

The interacting-field  equation is most easily solved in the Hamiltonian formalism.  One introduces the Hamiltonian operator
\begin{align}
\label{summ:14}  
    \mathbf{H}(\bphi,\bpi,t) &= \iiint_{\mathbb{R}^3}
                     \pmb{\mathcal{H}}
       \Bigl(  \bphi(t,\vec{x}\,),\bpi(t,\vec{x}\,)  \Bigr) ~d^3x, 
\end{align}
where $\pmb{\mathcal{H}}(t,\vec{x}\,)$ is the Hamiltonian density
\begin{align}
\notag
       \pmb{\mathcal{H}}(t,\vec{x}\,) &=
         \frac{1}{2} \bigl(\bpi(t,\vec{x}\,)\bigr)^2 
       + \frac{1}{2} \sum_{1 \leq \mu \leq 3}
         \bigl(\partial_\mu \bphi(t,\vec{x}\,)\bigr)^2 \\
\label{summ:15}  
       &+ \frac{1}{2} m^2
         \bigl(\bphi(t,\vec{x}\,)\bigr)^2 
       + \frac{g}{N+1} \bigl(\bphi(t,\vec{x}\,)\bigr)^{N+1}. 
\end{align}
Of course this definition involves classically unjustified products of distributions whose values are unbounded operators on a Hilbert space, and also an unjustified integration.  Ignoring these problems this formula leads to:
\begin{conjecture}[Heisenberg equations of motion]
%..................................................
\label{summ:conj:1}
Any pair of field operators $\bphi(t,\vec{x}\,)$ and $\bpi(t,\vec{x}\,)$ satisfying the commutation relations \emph{(\ref{summ:11}--\ref{summ:13})}, as well as the Heisenberg equations of motion
\begin{align}
\label{summ:16}
     \frac{\partial}{\partial t} \bphi(t,\vec{x}\,)   
     &= i[ \mathbf{H}, \bphi(t,\vec{x}\,) ] ,\\
\label{summ:17}
     \frac{\partial}{\partial t} \bpi (t,\vec{x}\,) 
     &= i[ \mathbf{H}, \bpi (t,\vec{x}\,) ] ,
\end{align}
with the initial conditions \eqref{summ:10}, where $\mathbf{H}$ is given by \emph{(\ref{summ:14}--\ref{summ:15})}, is a solution of the interacting-field equation \emph{(\ref{summ:8}--\ref{summ:13})}.

\end{conjecture}
Equations (\ref{summ:16}--\ref{summ:17}) can be written in the equivalent forms
\begin{align}
\label{summ:18}
     {\bphi}(t,\tau,\vec{x}\,)   =  \exp\bigl( i(t-\tau)\mathbf{H}(\tau)\bigr)
                              .{\bphi}_{\text{\rm ini}}(\tau,\vec{x}\,).
                               \exp\bigl(-i(t-\tau)\mathbf{H}(\tau)\bigr),\\
\label{summ:19}
     {\bpi }(t,\tau,\vec{x}\,)   =  \exp\bigl( i(t-\tau)\mathbf{H}(\tau)\bigr)
                              .{\bpi }_{\text{\rm ini}}(\tau,\vec{x}\,).
                               \exp\bigl(-i(t-\tau)\mathbf{H}(\tau)\bigr),
\end{align}
where $\tau$ is some initial time such that ${\bphi}(\tau,\tau,\vec{x}\,) = {\bphi}_{\text{\rm ini}}(\tau,\vec{x}\,)$ and ${\bpi}(\tau,\tau,\vec{x}\,) = {\bpi}_{\text{\rm ini}}(\tau,\vec{x}\,)$.  Again, the exponentials are not defined:  They should be defined as operators (expected to be unitary) on $\mathbb{F}$, and there is also a problem of composition of operators because the domains are not the same and, moreover, the domains of $\exp\bigl( \pm i(t-\tau)\mathbf{H}(\tau)\bigr)$ depend on time.

\begin{remark}[Time-independence of Hamiltonian]
%...............................................
\label{summ:rema:1}
For a Lagrangian density $L$ such as \eqref{summ:5}, which is invariant by time-translations, the Hamiltonian operator $\mathbf{H}(\bphi,\bpi,t)$, defined by \eqref{summ:14} as a function of time, is a constant of motion , i.e.,
\begin{align}
\label{summ:20}
     \frac{d}{d t} \mathbf{H} = 0.
\end{align}
This is the reason why $\mathbf{H}$ is customarily written as a time-independent quantity, as will be often done in the following unless we want to make explicit its dependence on $t$ or $\tau$ as parameters in individual terms or factors.
\end{remark}

\subsection{Interaction picture and scattering operator}
%.......................................................
%
In principle, knowing the operators $\bphi$ and $\bpi$ at some given reference time, the Heisenberg equations (\ref{summ:16}--\ref{summ:17}) enable to know the operators $\bphi(t,\vec{x}\,)$ and $\bpi(t,\vec{x}\,)$ at any later and earlier times, that is to solve the interacting field equations for all times.  Then, knowing the state vectors $\Psi_1, \Psi_2, ... \in \mathbb{D}$ of the system under consideration at that reference time, the quantum mechanical probability associated with an observable represented by an operator $\mathbf{A}$ is given by the formula
\begin{align}
\label{summ:21}
     \mathcal{P}_{\mathbf{A}}(1 \rightarrow 2) = | \SCA \Psi_2 |
              \mathbf{A}\bigl(\bphi,\bpi,t \bigr)
                   \Psi_1 \LAR |^2.
\end{align}
This procedure, in which the state vectors $\Psi_1$ and $\Psi_2$ are defined at a fixed time, while the time evolution is carried by the operators, is known as the \emph{Heisenberg picture}. (The opposite evolution pattern, in which the operators are constant while the state vectors depend on time, is called the \emph{Schr\"odinger picture}.)

   In practice, however, it is very difficult to solve the Heisenberg equations.  In particular, since the commutators (\ref{summ:11}--\ref{summ:13}) of the field operators are given by coupled equations which depend on the solutions of the  Heisenberg equations (\ref{summ:16}--\ref{summ:17} or \ref{summ:18}--\ref{summ:19}), which are also coupled equations because $\mathbf{H}$ depends on $\bphi(t,\vec{x}\,)$ and $\bpi(t,\vec{x}\,)$, and whose solutions cannot be known \emph{a priori}, even approximate solutions are difficult to obtain.  An elegant solution to this problem was found by introducing the \emph{interaction (or Dirac) picture}, that is by subjecting the operators and the states to a suitable unitary transformation.

\begin{remark}[Existence of interaction picture]
%...............................................
\label{summ:rema:2}
The existence of the unitary transformation to the interaction picture is not without problems, see, e.g., \emph{\cite[p.\,175]{BJORK1965-}, \cite[Sec.\,9.4]{BOGOL1990-}, \cite[Sec.\,4.5]{WIGHT1964-} }.  These difficulties are generally ignored in modern introductory textbooks, e.g., \emph{\cite[p.\,83]{PESKI1995-}, \cite[p.\,117]{MAGGI2005-}}, as will be done in this section which closely follows the expositions given in these texts.
\end{remark}

  One starts from an implicit solution of the Heisenberg equation of motion, which according to Conjecture \ref{summ:conj:1} is formally given by
\begin{equation}\label{summ:22}
    \bphi(t,\tau,\vec{x}\,) = \exp\bigl(i(t-\tau)\mathbf{H}\bigr)
                              .\bphi_{\text{\rm ini}}(\tau,\vec{x}\,).
                               \exp\bigl(-i(t-\tau)\mathbf{H}\bigr),
\end{equation}
where we have written $\mathbf{H}$ instead of $\mathbf{H}(\tau)$ since $\tau$ is just a parameter.  Then, to get an explicit solution one introduces the \emph{time evolution operator}
\begin{align}\label{summ:23}
    \mathbf{S}_\tau(t) := \exp\bigl( i(t-\tau)\mathbf{H}_0\bigr)
                 \exp\bigl(-i(t-\tau)\mathbf{H}\bigr), 
\end{align}
where $\mathbf{H}_0$ is the Hamiltonian of a free field, which, obviously, is identical to $\mathbf{H}$ when $g=0$. The operator $\mathbf{S}_\tau(t)$ is considered as a unitary operator on $\mathbb{F}$. Then one has (after a short calculation)
\begin{align}\label{summ:24}
        \bphi(t,\tau,\vec{x}\,)  = \bigl(\mathbf{S}_\tau(t)\bigr)^{-1}
                                  .\bphi_I(t,\tau,\vec{x}\,).
                                        \mathbf{S}_\tau(t),
\end{align}
where
\begin{equation}\label{summ:25}
    \bphi_I(t,\tau,\vec{x}\,) = \exp\bigl(i(t-\tau)\mathbf{H}_0\bigr)
                              .\bphi_{\text{\rm ini}}(\tau,\vec{x}\,).
                               \exp\bigl(-i(t-\tau)\mathbf{H}_0\bigr),
\end{equation}
and one finds that $\mathbf{S}_\tau(t)$ is solution of the differential equation 
\begin{align}
\label{summ:26}
        \frac{d}{dt} \mathbf{S}_\tau(t) &=
        - i \mathbf{H}_I(t,\tau) ~ \mathbf{S}_\tau(t),  \\
\label{summ:27}
        \mathbf{S}_\tau(\tau) &= \mathbf{1}, 
\end{align}
where $\mathbf{1}$ is the identity operator, and where
\begin{align}
\label{summ:28}
        \mathbf{H}_I(t,\tau) = \frac{g}{N+1} \iiint_{\mathbb{R}^3} 
        \bigl(\bphi_I(t,\tau,\vec{x}\,)\bigr)^{N+1}~d^3x, 
\end{align}
is the \emph{interaction} part of the self-interacting-field Hamiltonian (\ref{summ:14}--\ref{summ:15}) expressed in terms of the operator $\bphi_I$, i.e., Eq.~\eqref{summ:25}.  For this reason, the time-dependence induced by the unitary transformation \eqref{summ:24} is said to define the \emph{interaction picture}.

The differential equation \eqref{summ:26} is the starting point for an attempt (called \emph{`perturbation theory'}\,) to calculate $\mathbf{S}_\tau(t)$ by developing it in powers of the \emph{`coupling constant'} $g$, when $g$ is small. 

The numerical results of the theory are the limits  when $\tau \rightarrow -\infty, t \rightarrow +\infty$ of the scalar products
\begin{equation}
\label{summ:29}
\mathcal{P}(1 \rightarrow 2) =
  \lim_{\substack{
       \tau \rightarrow -\infty\\
          t \rightarrow +\infty}} ~
       |\SCA \Psi_2|\mathbf{S}_\tau(t)\Psi_1\LAR_\mathbb{F}|^2
      \DEF |\SCA \Psi_2|\mathbf{S}\Psi_1\LAR_\mathbb{F}|^2,
\end{equation}
where $\Psi_1, \Psi_2 \in \mathbb{D}$ are two normalized states.  The limit $\mathbf{S} \DEF \lim_{t \rightarrow \infty}\mathbf{S}_{-t}(t)$ is called \emph{scattering operator}, and  $\mathcal{P}(1 \rightarrow 2)$ is the transition probability from the state $\Psi_1$ to the state $\Psi_2$.

\begin{remark}[Fermi's golden rule]
%..................................
\label{summ:rema:3}
At the end of the calculation the transition amplitude $\SCA \Psi_2|\mathbf{S}\Psi_1\LAR_\mathbb{F}$ is generally proportional to an energy-momentum conservation $\delta$-function.  The transition probability \emph{\eqref{summ:29}} is therefore the square of a $\delta$-function, which is not defined in distribution theory.  The conventional way to deal with this problem is called Fermi's `golden rule,' which is explained in classic textbooks, e.g., \emph{\cite[p.\,101,  112]{BJORK1964-}}, while modern texts tend to ignore the difficulty by referring to `wave-packets' and smearing the $\delta$-functions.
\end{remark}

\subsection{Remark}
%..................

The above calculations form a rather intricate set of nonlinear calculations on distributions.  Schwartz's impossibility result,\footnote{This result concerns the impossibility of giving a natural definition of the product of two arbitrary generalized functions in an algebra containing the \emph{continuous} functions as a subalgebra.  It is however possible to define associative, commutative algebras containing the \emph{smooth} functions as a subalgebra.} see \cite[p.\,8]{COLOM1992-} and \cite[p.\,6]{GROSS2001-}, has been interpreted as the proof that these calculations cannot make sense even if we forget that they deal with unbounded operators on a Hilbert space (which makes them considerably more intricate than calculations in the case of bounded operators, close to scalar calculations).  But it appears now that Schwartz's impossibility result is circumvented  \cite[...]{COLOM1983-,COLOM1984-,COLOM1985-,COLOM1990-,COLOM1992-,GROSS2001-}, so that we may give a rigorous mathematical sense to these calculations, although with severe needs of improvements due the fact that they deal with unbounded operators.

\section{Mollifiers and dampers}
%-------------------------------
\label{moll:0}
\setcounter{equation}{0}
\setcounter{definition}{0}
\setcounter{axiom}{0}
\setcounter{conjecture}{0}
\setcounter{lemma}{0}
\setcounter{theorem}{0}
\setcounter{corollary}{0}
\setcounter{proposition}{0}
\setcounter{example}{0}
\setcounter{remark}{0}
\setcounter{problem}{0}

In this section we recall a few properties of Fourier transforms and define tools that will be extensively used in the application of nonlinear generalized functions to QFT.

\subsection{Fourier transformation}
%..................................

Let $\varphi \in \mathcal{S}(\mathbb{R}^3,\mathbb{C})$, i.e., Schwartz's space of smooth functions with steep descent.  We define its Fourier transform by\footnote{In the contemporary QFT literature the exponent has often the opposite sign, e.g., \cite[p.\,21]{PESKI1995-}.  We take the 3-dimensional case for definitiveness: The generalization to $\mathbb{R}^n$ is immediate.} 
\begin{equation}\label{moll:1}
  (\mathsf{F}\varphi)(\vec{p}\,) \DEF \FOU{\varphi}(\vec{p}\,)
 = \iiint_{\mathbb{R}^3} \rme^{i\vec{p} \cdot \vec{x}}
                         \varphi(\vec{x}\,) ~d^3x,
\end{equation}
and its inverse transformation by
\begin{equation}\label{moll:2}
  (\mathsf{F}^{-1}\varphi)(\vec{x}\,) \DEF 
  (2\pi)^{-3} \iiint_{\mathbb{R}^3} \rme^{-i\vec{p} \cdot \vec{x}}
                                     \varphi(\vec{p}\,) ~d^3p.
\end{equation}
Fourier's inversion theorem therefore reads
\begin{equation}\label{moll:3}
        \mathsf{F}^{-1}\varphi = (2\pi)^{-3} ~ \mathsf{F}\varphi^\vee,
\end{equation}
where
\begin{equation}\label{moll:4}
    \varphi^\vee(\vec{x}\,) =  \varphi(-\vec{x}\,).
\end{equation}
We also define the convolution of two functions $\varphi, \chi \in \mathcal{S}(\mathbb{R}^3,\mathbb{C})$ by the formula
\begin{equation}\label{moll:5}
    (\varphi \ast \chi)(\vec{x}\,) =  \iiint_{\mathbb{R}^3} 
     \varphi(\vec{x}-\vec{y}\,) \chi(\vec{y}\,) ~d^3y,
\end{equation}
and recall the theorem
\begin{equation}\label{moll:6}
    \mathsf{F} (\varphi \ast \chi) = (\mathsf{F} \varphi) (\mathsf{F}\chi).
\end{equation}
We finally remark that
\begin{equation}\label{moll:7}
   (\mathsf{F} \varphi) \in \mathbb{R}
   \qquad \Longleftrightarrow \qquad
   \varphi^\vee =  \varphi^*,
\end{equation}
as well as
\begin{align}
\label{moll:8} 
          (\mathsf{F} \varphi)(0)   =  \iiint_{\mathbb{R}^3} d^3x
                                      ~\varphi(\vec{x}\,),
%\end{align}
%
\quad \text{and} \quad
%
%\begin{align}
%\label{moll:9} 
          \varphi(0) =  (2\pi)^{-3}\iiint_{\mathbb{R}^3} d^3x
                       ~(\mathsf{F} \varphi)(\vec{x}\,).
\end{align}

\subsection{Suitable mollifiers and dampers}
%...........................................

An important procedure in the theory on nonlinear generalized functions is convolution by `mollifiers' (i.e., `smoothing kernels') which for technical reasons (discussed, for example, in \cite{GROSS2001-}) are Schwartz functions in $\mathcal{S}$ and have all moments vanishing:
\begin{definition}[Set of suitable mollifiers] 
%.............................................
\label{moll:defi:1}
A suitable mollifier $\rho$ is an element of the set $\mathcal{A}_\infty$ of \underline{real} functions such that $\rho(x) \in \mathcal{S}(\mathbb{R}^n,\mathbb{R})$ and $\FOU{\rho} = \mathsf{F}\rho \equiv 1$ in a neighborhood of 0 in $\mathbb{R}^n$.  For the applications to QFT considered in this report we further require that $\mathsf{F}\rho$ has compact support.  Thus $\FOU{\rho} \in \mathcal{D}(\mathbb{R}^n,\mathbb{C}) \subset \mathcal{S}(\mathbb{R}^n,\mathbb{C})$, whereas $\rho \in \mathcal{S}(\mathbb{R}^n,\mathbb{R})$. 
\end{definition}
More elaborate versions of mollifiers are used to insure that nonlinear generalized functions have properties such as diffeomorphism invariance, extensions to manifolds, etc., which are not needed here. (See, e.g., \cite{GROSS2001-}.) 
\begin{proposition}[Convolution of suitable mollifiers] 
%......................................................
\label{moll:prop:1}
The convolution product of any number of suitable mollifiers is again a suitable mollifier, i.e.,
\begin{align}
\label{moll:9}
     \forall \rho, \chi, ... \in \mathcal{A}_\infty 
     \quad \Rightarrow \quad  \rho \ast \chi \ast \cdots \in \mathcal{A}_\infty.
\end{align}
\end{proposition}
Proof: By Eq.~\eqref{moll:6}, $\mathsf{F}(\rho \ast \chi) = (\mathsf{F} \rho) (\mathsf{F}\chi)$.  Thus, if $Z$ is a neighborhood of zero in $\mathbb{R}^n$ such that $\zeta \in Z \Rightarrow (\mathsf{F} \rho)(\zeta) = (\mathsf{F}\chi)(\zeta) = 1$, then $\mathsf{F}(\rho \ast \chi)(\zeta) = 1$ so that $\mathsf{F}(\rho \ast \chi)$ is a suitable mollifier.  By induction, since the convolution operation is commutative and associative, the proposition is true for the convolution product of any number of suitable mollifiers. \END
\begin{definition}[Set of suitable dampers] 
%..........................................
\label{moll:defi:2}
A suitable damper $\FOU{\chi}$ is an element of the set $\mathcal{B}_\infty$ of $\mathcal{C}^\infty$ \underline{real} functions $\FOU{\chi} : \mathbb{R}^n \rightarrow \mathbb{R}$ with compact support  and such that $\FOU{\chi} = \mathsf{F}\chi \equiv 1$ in a neighborhood of 0 in $\mathbb{R}^n$.  Thus $\FOU{\chi} \in \mathcal{D}(\mathbb{R}^n,\mathbb{R}) \subset \mathcal{S}(\mathbb{R}^n,\mathbb{R})$, whereas $\chi \in \mathcal{S}(\mathbb{R}^n,\mathbb{C})$.  Evidently, a product of dampers is again a damper
\begin{align}
\label{moll:10}
     \forall \FOU{\chi}, \FOU{\eta}, ... \in \mathcal{B}_\infty 
     \quad \Rightarrow \quad  \FOU{\chi} \,\FOU{\eta} \cdots \in \mathcal{B}_\infty.
\end{align}
\end{definition}
The concepts of suitable mollifier and damper are closely related.  But, because of \eqref{moll:7}, $\rho \in \mathcal{A}_\infty \Rightarrow \FOU{\rho} \in \mathcal{B}_\infty$ only if $\rho^\vee = \rho$.  Conversely, $\FOU{\chi} \in \mathcal{B}_\infty \Rightarrow  \chi \in \mathcal{A}_\infty$ only if $\FOU{\chi}^\vee = \FOU{\chi}$.  Since it is important in QFT to keep the mollifiers and dampers in a form as general as possible one cannot a priori impose symmetry conditions such as $\rho^\vee = \rho$ or $\FOU{\chi}^\vee = \FOU{\chi}$ on them.  The two concepts have to remain independent.

The main features of suitable mollifiers derive from the theorem:
\begin{theorem}[Moments of suitable mollifiers] 
%..................................................
\label{moll:theo:1}
Let $\rho \in \mathcal{A}_\infty$ and $\FOU{\chi} \in \mathcal{B}_\infty$.  Then,  $\forall \alpha_j \in \mathbb{N}, j=1,...,n$ such that $|\alpha| \DEF \alpha_1 + \alpha_2 + \cdots +\alpha_n \neq 0$,
\begin{equation}
\label{moll:11}
   \int_{\mathbb{R}^n} \rho(x) ~d^nx = 1,
       \qquad \text{and} \qquad
   \int_{\mathbb{R}^n} x^\alpha\rho(x) ~d^nx = 0,
\end{equation}
as well as
\begin{equation}
\label{moll:12}
   \int_{\mathbb{R}^n} \chi(x) ~d^nx = 1,
       \qquad \text{and} \qquad
   \int_{\mathbb{R}^n} x^\alpha\chi(x) ~d^nx = 0,
\end{equation}
where $\chi = (\mathsf{F}^{-1}\FOU{\chi})$ and $x^\alpha \DEF x_1^{\alpha_1}  x_2^{\alpha_2} \cdots x_n^{\alpha_n}$.
\end{theorem}
Proof: We consider only the one dimensional case, i.e., $x\in\mathbb{R}$, since the generalization to $n>1$ is obvious.  We start with $\rho \in \mathcal{A}_\infty$:  As $\rho \in \mathcal{S}$, we can write $\rho$ as the inverse Fourier transformation $\rho = (\mathsf{F}^{-1}\FOU{\rho})$.  We therefore calculate the moment
\begin{equation}\label{moll:13}
  \int_{x\in\mathbb{R}}dx~ x^\alpha(\mathsf{F}^{-1}\FOU{\rho})(x)
               = \int_{p\in\mathbb{R}}dp~ \FOU{\rho}(p)  \frac{1}{2\pi}
                 \int_{x\in\mathbb{R}}dx~ x^\alpha\rme^{-ipx},
\end{equation}
where we have interchanged the order of the integrations, which is allowed because $\FOU{\rho} \in \mathcal{S}$ since $\rho \in \mathcal{S}$.  For $\alpha=0$ the right-most integral is just the Dirac distribution $\delta(p)$, so that in that case \eqref{moll:13} is equal to $\FOU{\rho}(0)$, i.e., $1$ by definition of $\rho$.  This proves the first equation of \eqref{moll:11}.  When $\alpha\geq 1$ the right-most integral is the inverse Fourier transform of $x^\alpha$, which is related to the derivatives of the $\delta$-function by the distributional identity
\begin{equation}\label{moll:14}
  (\mathsf{F}^{-1}x^\alpha)(p) = i^\alpha \delta^{(\alpha)}(p).
\end{equation}
Therefore, replacing and integrating by parts, 
\begin{align}
\nonumber
  \int_{x\in\mathbb{R}}dx~ x^\alpha \rho(x)
       & = i^\alpha\int_{p\in\mathbb{R}}dp~ \FOU{\rho}(p) \delta^{(\alpha)}(p)\\
\label{moll:15}
       & = (-i)^\alpha\int_{p\in\mathbb{R}}dp~ \FOU{\rho}^{(\alpha)}(p) \delta(p)
         = (-i)^\alpha \FOU{\rho}^{(\alpha)}(0) = 0,
\end{align}
provided $\alpha\geq 1$, because by definition $\FOU{\rho} \equiv 1$ in the neighborhood of $0$. This proves \eqref{moll:11}.  For the case $\FOU{\chi} \in \mathcal{B}_\infty$ we define $\rho$ as $\rho = (\mathsf{F}^{-1}\FOU{\chi})$.  The only difference is that now $\rho \in \mathbb{C}$.  But this does not change anything in the proof so that \eqref{moll:12} is true. (More rigorous proofs to be found in the mathematical literature.) \END

   Let us justify the terminology `mollifier' and `damper' by taking the three-dimensional case as an example.  Then, for $\epsilon > 0$ the function 
\begin{align}
\label{moll:16}
       \rho_\epsilon (\vec{x}\,)
  \DEF \frac{1}{\epsilon^3} \rho\Bigl( \frac{\vec{x}}{\epsilon}\Bigr),
       \qquad \text{such that} \qquad
       \iiint d^3x~\rho_\epsilon(\vec{x}\,)=1,
\end{align}
is a representation of the three-dimensional Dirac $\delta$-function in the limit $\epsilon \rightarrow 0$.  Thus the \emph{scaled mollifier} $\rho_\epsilon (\vec{x}\,)$ can be seen as a $\delta$-function that has been smeared, or `mollified,' about $|\vec{x}\,| =0$.  It has area~$1$, approximate height $1/2\epsilon$, and approximate width $2\epsilon$.  On the other hand, the Fourier transform of that function, i.e., the \emph{scaled damper} $\FOU{\rho}_\epsilon (\vec{p}\,)$
\begin{align}
\label{moll:17}
       \FOU{\rho}_\epsilon (\vec{p}\,)
  \DEF \FOU{\rho}(\epsilon\vec{p}\,),
       \qquad \text{such that} \qquad
       \FOU{\rho}_\epsilon(0) \equiv 1,
\end{align}
can be seen as a cut-off, or `damping factor,' that smoothly vanishes when $|\vec{p}\,|~\rightarrow~\infty$. It has height~$1$ in a neighborhood of the origin, approximate area $2/\epsilon$, and approximate width $2/\epsilon$.

  However, because of the constraints on the moments of $\rho$ implied by Definition~\ref{moll:defi:1}, the details of the functions $\rho_\epsilon$ and $\FOU{\rho}_\epsilon$ can be much more complicated then those of the customary representations $\delta_\epsilon$ of the classical Dirac $\delta$-function, and of their Fourier transforms, which satisfy $\FOU{\delta}_\epsilon(\vec{x}\,)=1$ only at the \emph{point} $\vec{x}=0$, rather than in a whole neighborhood of $\vec{x}=0$.

   Moreover, $\rho_\epsilon$ can be interpreted as an approximation (or, better, as a generalization) of order $\epsilon^{q+1}$, $\forall q \in \mathbb{N}$, of the $\delta$-function with regards to the sifting property $\int dx~f(x) \delta(x-a) = f(a)$.  This enables to integrate expressions containing a mollifier as follows:
\begin{proposition}[Mollifier integration] 
%.........................................
\label{moll:prop:2}
Let $\rho \in \mathcal{A}_\infty$ and $\FOU{\chi} \in \mathcal{B}_\infty$ (or else $\FOU{\rho} \in \mathcal{B}_\infty$ or $\chi \in \mathcal{A}_\infty$) and let $f(\vec{x}\,) \in \mathcal{O}_{\text{\emph{M}}}(\mathbb{R}^3,\mathbb{C})$, i.e., $f(\vec{x}\,) \in \mathcal{C}^\infty$ and each of its derivatives do not grow faster than a power of $|\vec{x}\,|$ at infinity in $\mathbb{R}^3$.  Then, $\forall q \in \mathbb{N}$,
\begin{align}
\label{moll:18}
   \iiint d^3x~ \FOU{\chi}(\epsilon\vec{x}\,)~ f(\vec{x}\,)
                \frac{1}{\epsilon^3} \rho(\frac{\vec{x}-\vec{a}\,}{\epsilon})
   = \FOU{\chi}(\epsilon\vec{a}\,)~ f(\vec{a}\,) + \OOO(\epsilon^{q+1}).
\end{align}
\end{proposition}
Proof: Let us make the change of variable $\vec{x} = \vec{a} + \epsilon \vec{z}$, i.e.,
\begin{equation}
\label{moll:19}
   J = \iiint d^3x~ \FOU{\chi}(\epsilon\vec{x}\,)~ f(\vec{x}\,)
       \frac{1}{\epsilon^3} \rho(\frac{\vec{x}-\vec{a}\,}{\epsilon})
     = \iiint d^3z~ \FOU{\chi}(\epsilon\vec{a} + \epsilon^2\vec{z}\,)
                   ~f(\vec{a} + \epsilon \vec{z}\,)
       \rho(\vec{z}\,),
\end{equation}
and, since $f, \FOU{\chi} \in \mathcal{C}^\infty$, let us make the Taylor expansions with remainders
\begin{align}
\nonumber
 f(\vec{a} + \epsilon \vec{z}\,)
    &= f(\vec{a}\,)
     + \epsilon|\vec{z}\,|\bigl(\rmD^{1}f\bigr)(\vec{a}\,)
     + \frac{\epsilon^2|\vec{z}\,|^2}
            {2!}    \bigl(\rmD^{2}f\bigr)(\vec{a}\,) \\
\label{moll:20}
    &+ ... + \frac{(\epsilon|\vec{z}\,|)^{q+1}}
            {(q+1)!} \bigl(\rmD^{q+1}f\bigr)(\vec{a}
           + \theta\epsilon\vec{z}\,),
\end{align}
and
\begin{align}
\nonumber
 \FOU{\chi}(\epsilon\vec{a} + \epsilon^2\vec{z}\,)
    &= \FOU{\chi}(\epsilon\vec{a}\,)
     + \epsilon^2|\vec{z}\,|\bigl(\rmD^{1}\FOU{\chi}\bigr)(\epsilon\vec{a}\,)
     + \frac{\epsilon^4|\vec{z}\,|^2}
            {2!}    \bigl(\rmD^{2}\FOU{\chi}\bigr)(\epsilon\vec{a}\,) \\
\label{moll:21}
    &+ ... + \frac{(\epsilon^{2}|\vec{z}\,|)^{q+1}}
            {(q+1)!} \bigl(\rmD^{q+1}\FOU{\chi}\bigr)(\epsilon\vec{a}
           + \theta\epsilon^2\vec{z}\,),
\end{align}
where the abusive one-dimensional notation $\bigl(\rmD^{n}f\bigr)$ is used for the derivatives.  As $\rho \in \mathcal{A}_\infty$ we have
\begin{equation}
\label{moll:22}
   \iiint d^3z~  \rho(\vec{z}\,) = 1,
   \quad \text{and} \quad
   \iiint d^3z~ |\vec{z}\,|^n \rho(\vec{z}\,) = 0,
   \quad \forall n \in  \mathbb{N}.
\end{equation}
Therefore, when the expansions \eqref{moll:20} and \eqref{moll:21} are inserted in \eqref{moll:19}, all terms in $|\vec{z}\,|^n$ with $1 < n < q+1$ multiplied by a quantity independent of $z$ are zero.  Consequently, it remains
\begin{align}
\nonumber
   J &= \FOU{\chi}(\epsilon\vec{a}\,)~f(\vec{a}\,)\\
\nonumber
     &+ \epsilon^{q+1} \iiint d^3z~  \rho(\vec{z}\,)
       \sum_{n=0}^{q} A_n\epsilon^{2n}|\vec{z}\,|^{q+1+n}
                     \bigl(\rmD^{q+1}f\bigr)
                        (\vec{a} + \theta\epsilon\vec{z}\,)\\
\nonumber
     &+ \epsilon^{2q+2} \iiint d^3z~  \rho(\vec{z}\,)
       \sum_{n=0}^{q} B_n\epsilon^{n}|\vec{z}\,|^{{q+1+n}}
                       \bigl(\rmD^{q+1}\FOU{\chi}\bigr)
                        (\epsilon\vec{a} + \theta\epsilon^2\vec{z}\,)\\
\label{moll:23}
     &+ \epsilon^{3q+3} \iiint d^3z~  \rho(\vec{z}\,) 
                    C |\vec{z}\,|^{2q+2} \bigl(\rmD^{q+1}f\bigr)
                        (\vec{a} + \theta\epsilon\vec{z}\,)
                \bigl(\rmD^{q+1}\FOU{\chi}\bigr)
                       ~(\epsilon\vec{a} + \theta\epsilon^2\vec{z}\,),
\end{align}
where $A_n,B_n,$ and $C$ are independent of $\vec{z}$.   On the other hand, the derivatives depend on $\vec{z}, \epsilon$, and $\theta \in ]0,1[$.  In the case of $\rmD^{q+1}f$, since $f(\vec{z}\,) \in \mathcal{O}_{\text{M}}(\mathbb{R}^3,\mathbb{C})$, it is bounded by a power of $|\vec{z}\,|$.  But in the case of $\rmD^{q+1}\FOU{\chi}$ we have to consider the possibility that $\theta$ can tend to $0$ as $\epsilon \rightarrow 0$ while $|\vec{z}\,| \rightarrow \infty$ when $\vec{z}$ ranges in $\mathbb{R}^3$.  This means that the bound on the corresponding derivative is of the type $|\bigl(\rmD^{q+1} \FOU{\chi}\bigr)(\epsilon,\theta,\vec{z}\,)| \leq \Cst$, where the constant is independent of $\epsilon$ and $\vec{z}$.  However, since $\rho \in \mathcal{A}_\infty$ (or else $\FOU{\rho} \in \mathcal{B}_\infty$) implies that $\rho \in \mathcal{S}$, the product of $\rho$ with $|\vec{z}\,|^{n}$ is still in $\mathcal{S}$ for whatever $n$.  Consequently, the integrals in \eqref{moll:23} make sense and are thus bounded, which proves the proposition.  \END

   Finally, we remark that a damper $\FOU{\chi}$ appearing in an integral such as
\begin{equation}
\label{moll:24}
   \iiint d^3x ~\FOU{\chi}(\epsilon\vec{x}\,)~f(\vec{x}\,),
\end{equation}
is acting as a regularization for a large class of functions $f(\vec{x}\,)$ which otherwise would lead to a diverging integral.  But since $\FOU{\rho} \in \mathcal{B}_\infty$ this regularization has additional properties in the $\mathcal{G}$-context, which is why we refer to it as a suitable damper.  For instance, the following proposition specifies under which conditions an integrand containing a damper can be simplified:

\begin{proposition}[Damper simplification]
%.........................................
\label{moll:prop:3}
Let $f(\vec{x}\,) \in \mathcal{S}(\mathbb{R}^3,\mathbb{C})$ and $\FOU{\chi} \in \mathcal{B}_\infty$ (or else $\chi \in \mathcal{A}_\infty$).  Then, $\forall q \in \mathbb{N}$,
\begin{align}
\label{moll:25}
   \iiint d^3x~ \FOU{\chi}(\epsilon\vec{x}\,)~f(\vec{x}\,)
   = \iiint f(\vec{x}\,) ~d^3x + \OOO(\epsilon^{q+1})
\end{align}
\end{proposition}
Proof: Let us make the Taylor expansion with remainder
\begin{align}
\nonumber
 \FOU{\chi}(\epsilon \vec{x}\,)
    &= \FOU{\chi}(0)
     + \epsilon|\vec{x}\,|\bigl(\rmD^{1}\FOU{\chi}\bigr)(0)
     + \frac{\epsilon^2|\vec{x}\,|^2}
            {2!}    \bigl(\rmD^{2}\FOU{\chi}\bigr)(0) \\
\label{moll:26}
    &+ ... + \frac{(\epsilon|\vec{x}\,|)^{q+1}}
            {(q+1)!} \bigl(\rmD^{q+1}\FOU{\chi}\bigr)(\theta\epsilon\vec{x}\,),
\end{align}
where the abusive one-dimensional notation $\bigl(\rmD^{n}\FOU{\chi}\bigr)$ is used for the derivatives.  As $\FOU{\chi} \in \mathcal{B}_\infty$ (or else $\chi \in \mathcal{A}_\infty$) we have
\begin{equation}
\label{moll:27}
   \FOU{\chi}(0) = 1,
   \quad \text{and} \quad
   \bigl(\rmD^{n+1}\FOU{\chi}\bigr)(0) = 0,
   \quad \forall n \in  \mathbb{N}.
\end{equation}
Therefore, all terms in $|\vec{x}\,|^n$  with $1 < n < q+1$ are zero in \eqref{moll:25}.  Consequently
\begin{align}
\nonumber
    \iiint d^3x~ f(\vec{x}\,)
   \FOU{\chi}(\epsilon\vec{x}\,)
   &=    \iiint d^3x~ f(\vec{x}\,)\\
\label{moll:28}
   &+ \epsilon^{q+1} \iiint d^3x~ f(\vec{x}\,)
       \frac{|\vec{x}\,|^{q+1}}{(q+1)!}
       \bigl(\rmD^{q+1}\FOU{\chi}\bigr)(\theta\epsilon\vec{x}\,).
\end{align}
In this expression the remainder depends on $\vec{x}, \epsilon$, and $\theta \in ]0,1[$ which depends also on $\vec{x}$, and there is no control on the product $\theta\epsilon\vec{x}$ when $\vec{x}$ ranges in $\mathbb{R}^3$.  Thus $\theta$ can tend to $0$ as $\epsilon \rightarrow 0$ while $|\vec{x}\,| \rightarrow \infty$.   This means that we only have a bound of the type $|\bigl(\rmD^{q+1} \FOU{\chi}\bigr)(\theta\epsilon\vec{x}\,)| \leq \Cst$, where the constant is independent of $\epsilon$ and $\vec{x}$.\footnote{In any case, as $\FOU{\chi} \in \mathcal{D}$, there is always a bound  $|\bigl(\rmD^{q+1} \FOU{\chi}\bigr)(\vec{z}\,)| \leq \beta$ for some real number $\beta$ independent of $\vec{z}$.} But since $f(\vec{x}\,)$ is in $\mathcal{S}$ and $\bigl(\rmD^{q+1}\FOU{\chi}\bigr)(\theta\epsilon\vec{x}\,)$ is bounded their product with $|\vec{x}\,|^{q+1}$ is still in $\mathcal{S}$.  The integral in \eqref{moll:28} which defines the factor $C^{q+1}(\theta\epsilon)$ is therefore bounded, which proves the proposition.  \END

\subsection{Sifting property of the powers of the $\delta$-function}
%...................................................................

   The following proposition, which is a straightforward generalization of Prop.\,\ref{moll:prop:2}, specifies how the sifting property $\int dx~f(x) \delta(x-a) = f(a)$ of the $\delta$-function generalizes to its $m$-th power $\delta^m(x)$ interpreted as the $\mathcal{G}$-function $\rho_\epsilon^m(x)$: 

\begin{proposition}[Sifting property of $\rho_\epsilon^m$] 
%.........................................................
\label{moll:prop:4} 
Let $m$ and $N$ be two fixed integers, and let the moments of $\rho \in \mathcal{S}(\mathbb{C}^3,\mathbb{C})$ defined as\footnote{To be fully general and consistent we use the three-dimensional multi-index notation, i.e., we define  $n=(n_1,n_2,n_3)$, $0=(0,0,0)$, and $|n|=n_1+n_2+n_3$, and write $\vec{z}\,^n = z_1^{n_1} z_2^{n_2} z_3^{n_3}$. Then  $\rmD^{|n|}f(\vec{a}\,)$ is a $|n|$-linear form acting on $\vec{z}\,^n$, e.g., as in  $\rmD^{|n|}f(\vec{a}\,).\vec{z}\,^n$. }
\begin{equation}
\label{moll:29}
  M[^m_n] \DEF   \iiint d^3z~ \vec{z}\,^n \rho^m(\vec{z}\,) ,
\end{equation}
be such that
\begin{equation}
\label{moll:30}
    M[^m_0] \in \mathbb{R},
   \qquad \text{and} \quad
    M[^m_n] = 0,
   \qquad \forall n \text{  with  } |n| \in [1,N].
\end{equation}
Let further $\FOU{\chi} \in \mathcal{B}_\infty$ (or else $\chi \in \mathcal{A}_\infty$) and let $f(\vec{x}\,) \in \mathcal{O}_{\text{\emph{M}}}(\mathbb{R}^3,\mathbb{C})$, i.e., $f(\vec{x}\,) \in \mathcal{C}^\infty$ and each of its derivatives do not grow faster than a power of $|\vec{x}\,|$ at infinity in $\mathbb{R}^3$.  Then,
\begin{align}
\nonumber
 \iiint d^3x~ \FOU{\chi}(\epsilon\vec{x}\,)~ f(\vec{x}\,)
              \frac{1}{\epsilon^{3m}} \rho^m(\frac{\vec{x}-\vec{a}\,}{\epsilon})
      &= \frac{M[^m_0]}{\epsilon^{3(m-1)}}
         \FOU{\chi}(\epsilon\vec{a}\,)~ f(\vec{a}\,) \\
\label{moll:31}
      &+ \OOO(\frac{\epsilon^{N+1}}{\epsilon^{3(m-1)}}).
\end{align}
\end{proposition}
Proof: Let us make the change of variable $\vec{x} = \vec{a} + \epsilon \vec{z}$, i.e.,
\begin{align}
\nonumber
   J &= \iiint d^3x~ \FOU{\chi}(\epsilon\vec{x}\,)~ f(\vec{x}\,)
        \frac{1}{\epsilon^{3m}} \rho^m(\frac{\vec{x}-\vec{a}\,}{\epsilon})\\
\label{moll:32}
     &= \iiint d^3z~ \FOU{\chi}(\epsilon\vec{a} + \epsilon^2\vec{z}\,)
                   ~f(\vec{a} + \epsilon\vec{z}\,)
        \frac{\rho^m(\vec{z}\,)}{\epsilon^{3(m-1)}},
\end{align}
and, since $f, \FOU{\chi} \in \mathcal{C}^\infty$, let us make the Taylor expansions with remainders
\begin{align}
\nonumber
 f(\vec{a} + \epsilon \vec{z}\,)
    &= f(\vec{a}\,)
     + \epsilon\bigl(\rmD^{1}f\bigr)(\vec{a}\,).\vec{z}
     + \frac{\epsilon^2}{2!}    
            \bigl(\rmD^{2}f\bigr)(\vec{a}\,) .\vec{z}\,^2\\
\label{moll:33}
    &+ ... + \frac{\epsilon^{N+1}}{(N+1)!} 
            \bigl(\rmD^{N+1}f\bigr)(\vec{a}
           + \theta\epsilon\vec{z}\,).\vec{z}\,^{N+1},
\end{align}
and
\begin{align}
\nonumber
 \FOU{\chi}(\epsilon\vec{a} + \epsilon^2\vec{z}\,)
    &= \FOU{\chi}(\epsilon\vec{a}\,)
     + \epsilon^2\bigl(\rmD^{1}\FOU{\chi}\bigr)(\epsilon\vec{a}\,).\vec{z}\,
     + \frac{\epsilon^4}{2!}    
            \bigl(\rmD^{2}\FOU{\chi}\bigr)(\epsilon\vec{a}\,).\vec{z}\,^2 \\
\label{moll:34}
    &+ ... + \frac{(\epsilon^{2})^{N+1}}
            {(N+1)!} \bigl(\rmD^{N+1}\FOU{\chi}\bigr)(\epsilon\vec{a}
           + \theta\epsilon^2\vec{z}\,).\vec{z}\,^{N+1},
\end{align}
where the standard one-dimensional notation $\bigl(\rmD^{|n|}f\bigr)$ is used for the derivatives.  As $\rho$ satisfies the condition \eqref{moll:30} we have
\begin{equation}
\label{moll:35}
   \iiint d^3z~  \rho^m(\vec{z}\,) = M[^m_0],
   \quad \text{and} \quad
   \iiint d^3z~ \vec{z}\,^n \rho^m(\vec{z}\,) = 0,
   \quad \forall n \text{  with  } |n| \in  [1,N].
\end{equation}
Therefore, when the expansions \eqref{moll:33} and \eqref{moll:34} are inserted in \eqref{moll:32}, all terms in $\vec{z}\,^n$ with $1 < |n| < N+1$ multiplied by a quantity independent of $z$ are zero.  Consequently, it remains
\begin{align}
\nonumber
   J &= \frac{M[^m_0]}{\epsilon^{3(m-1)}}
        \FOU{\chi}(\epsilon\vec{a}\,)~f(\vec{a}\,) \\
\nonumber
     &+ \frac{\epsilon^{N+1}}{\epsilon^{3(m-1)}} \iiint d^3z~  \rho^m(\vec{z}\,)
       \sum_{n=0}^{N} \epsilon^{2n}A_n.\vec{z}\,^n
                      \bigl(\rmD^{N+1}f\bigr)
                     (\vec{a} + \theta\epsilon\vec{z}\,).\vec{z}\,^{N+1}\\
\nonumber
     &+ \frac{\epsilon^{2N+2}}{\epsilon^{3(m-1)}} \iiint d^3z~ \rho^m(\vec{z}\,)
       \sum_{n=0}^{N} \epsilon^{n}B_n.\vec{z}\,^n
                      \bigl(\rmD^{N+1}\FOU{\chi}\bigr)
           (\epsilon\vec{a} + \theta\epsilon^2\vec{z}\,).\vec{z}\,^{{N+1}}\\
\label{moll:36}
     &+ \frac{\epsilon^{3N+3}}{\epsilon^{3(m-1)}}\iiint d^3z~  \rho^m(\vec{z}\,)
                    C \bigl(\rmD^{N+1}f\bigr)
                        (\vec{a} + \theta\epsilon\vec{z}\,).\vec{z}\,^{N+1} 
                \bigl(\rmD^{N+1}\FOU{\chi}\bigr)
               ~(\epsilon\vec{a} + \theta\epsilon^2\vec{z}\,).\vec{z}\,^{N+1},
\end{align}
where the $|n|$-linear forms $A_n,B_n,$ and the constant $C$ are bounded independently of $\vec{z}$.   On the other hand, the derivatives depend on $\vec{z}, \epsilon$, and $\theta \in ]0,1[$.  In the case of $\rmD^{N+1}f$, since $f(\vec{z}\,) \in \mathcal{O}_{\text{M}}(\mathbb{R}^3,\mathbb{C})$, it is bounded by a power of $|\vec{z}\,|$.  But in the case of $\rmD^{N+1}\FOU{\chi}$ we have to consider the possibility that $\theta$ can tend to $0$ as $\epsilon \rightarrow 0$ while $|\vec{z}\,| \rightarrow \infty$ when $\vec{z}$ ranges in $\mathbb{R}^3$.  This means that the bound on the corresponding derivative is of the type $|\bigl(\rmD^{N+1} \FOU{\chi}\bigr)(\epsilon,\theta,\vec{z}\,)| \leq \Cst$, where the constant is independent of $\epsilon$ and $\vec{z}$.  However, since $\rho \in \mathcal{S}$, the product of $\rho^m$ with $|\vec{z}\,|^{n}$ is still in $\mathcal{S}$ for whatever $n$.  Consequently, the integrals in \eqref{moll:36} make sense and are thus bounded, which proves the proposition.  \END

  When $m=1$ this proposition is equivalent to Prop.\,\ref{moll:prop:2} because the conditions \eqref{moll:30} can be satisfied with $\rho \in \mathbb{R}$ for $\forall N \in \mathbb{N}$ so that $\rho$ can actually be a suitable mollifier.

  When $m\geq 1$, however, the conditions \eqref{moll:30} on the moments $M[^m_n]$ are impossible unless $\rho$ is complex valued.  In that case, for at least $m=2$, it is shown in Ref.\,\cite{COLOM2008X} that the conditions \eqref{moll:30} can be satisfied for all $n$ up to some fixed $N$, and that it is even possible to have $M[^2_0]=0$.

  Finally, this proposition is also applicable when $\rho^m$ is replaced by $|\rho|^m$.  But in that case it will not be possible to satisfy the conditions $M[^m_n]=0$ when $n_1,n_2$, and $n_3$ are even unless $\rho$ is identically zero, so that Eq.\eqref{moll:31} will apply only with $N=0$ or $1$.

\section{Operator-valued nonlinear generalized functions}
%--------------------------------------------------------
\label{oper:0}
\setcounter{equation}{0}
\setcounter{definition}{0}
\setcounter{axiom}{0}
\setcounter{conjecture}{0}
\setcounter{lemma}{0}
\setcounter{theorem}{0}
\setcounter{corollary}{0}
\setcounter{proposition}{0}
\setcounter{example}{0}
\setcounter{remark}{0}
\setcounter{problem}{0}

In this section we define operator-valued nonlinear generalized functions and give some general information on their use in QFT.  As much as possible the notations will be those of reference \cite{GROSS2001-}.

\subsection{Bounded sets structures on $\mathbb{D}$ and $\mathsf{L}(\mathbb{D})$}
%......................................................

In Section \ref{fock:0} we have defined the Fock space $\mathbb{F}$ and the dense vector subspace $\mathbb{D}$ of states with a finite number of particles.   We now define appropriate structures of `bounded sets' on $\mathbb{D}$, as well as on the algebra $\mathsf{L}(\mathbb{D})$ of all linear `continuous' maps from $\mathbb{D}$ into $\mathbb{D}$.  This concept will replace the classical concept of continuity for linear maps between normed spaces, and will enable us to define the concept of operator valued generalized functions on $\mathbb{D}$.

\subsubsection{1. \underline{Bounded sets structure on $\mathbb{D}$}}
% ...................................................................

%
\begin{definition}[Bounded subset of $\mathbb{D}$] 
%.................................................
\label{oper:defi:1}
A bounded subset of $\mathbb{D}$ is a family $B \DEF \{\Phi_i\}_{i\in I}$, where $I$ is any set of indices, such that 
\begin{align}
\label{oper:1}
\left\{ \quad
\begin{array}{l}
           \exists N \in \mathbb{N} \text{ such that }
    (\Phi_i)_n =0 \text{ ~if~ } n\geq N, \text{ and},\\
           \exists M > 0 \text{ such that }
   \| (\Phi_i)_n \|_{\mathsf{L}_S^2\bigl((\mathbb{R}^3)^n,\mathbb{C} \bigr)}
   \leq M, \forall n \leq N,
\end{array}
\quad \right.
\end{align}
for $\Phi_i \in B, \forall i \in I$ (i.e., $N,M$ independent of $i$, they depend only on $B$), where $(\Phi_i)_n$ is the $n$th component of $\Phi_i$, i.e.,
\begin{align}
\label{oper:2}
     \Phi_i =
\begin{pmatrix}
      (\Phi_i)_0\\
      (\Phi_i)_1\\
       \vdots\\
      (\Phi_i)_n\\
       \vdots
\end{pmatrix} ,
\quad\text{with}\quad (\Phi_i)_0 \in \mathbb{C}, \quad\text{and}\quad
(\Phi_i)_n \in \mathsf{L}_S^2\bigl((\mathbb{R}^3)^n,\mathbb{C} \bigr).
\end{align}
\end{definition}
We have the obvious properties:
\begin{itemize}
{\it
\item[$\cdot$] All elements of\, $\mathbb{D}$ are bounded and no straight line $\mathbb{R}.\Phi, \Phi\neq 0,$ is bounded;

\item[$\cdot$] Every subset contained in a bounded set is bounded;

\item[$\cdot$] The sum of two bounded sets is bounded;

\item[$\cdot$] Any homothetic of a bounded set is bounded;

\item[$\cdot$] Any bounded set is contained in a bounded disk.
}
\end{itemize}
A disk $d$ in a vector space over $\mathbb{R}$ or $\mathbb{C}$ is so defined that it has the two properties of being a `convex' and `balanced' set:
\begin{align}
\label{oper:3}
\left\{ 
\begin{array}{l}
           d \text{~convex~~~~} \Longleftrightarrow
             \{ f_1,f_2 \in d; \alpha, \beta \in \mathbb{R}^+;
             \alpha+\beta = 1 \} \Rightarrow \alpha f_1 + \beta f_2 \in d,\\
           d \text{~balanced~} \Longleftrightarrow \{ f \in d, \lambda \in \mathbb{C}, |\lambda| \leq 1 \} \Rightarrow \lambda f \in d.
\end{array}
\quad \right.
\end{align}
Note that if $B$ is a bounded disk then $\bigcup_{n \in \mathbb{N}} nB$ is a vector space.  If $f \in \bigcup_{n} nB$ then
\begin{equation}
\label{oper:4}
   \| f \|_B \DEF \inf \{ \lambda > 0 \text{  such that  } f \in \lambda B \},
\end{equation}
is a norm on $\bigcup_{n} nB$ which then becomes a normed space.   We shall always use this norm on the vector space $\bigcup_{n} nB$.

With this structure of bounded disks $\mathbb{D}$ appears thus as an `union directed by inclusions' of normed spaces, and a subset of $\mathbb{D}$ is bounded iff it is contained and bounded in one of these normed spaces (in the usual sense of bounded sets in a normed space).

\subsubsection{2. \underline{Bounded sets structure on
 % ...................................................
$\mathsf{L}(\mathbb{D})$}}
%.........................

Let $\mathsf{L}(\mathbb{D})$ denote the vector space of all linear maps $\mathbb{D} \rightarrow \mathbb{D}$ that map any bounded set of $\mathbb{D}$ into another bounded set of $\mathbb{D}$.  Such linear maps are said to be `bounded.'\footnote{However, these maps can be unbounded operators in the Hilbert space sense:  For instance, the free-field operators are `bounded maps' from $\mathbb{D}$ into $\mathbb{D}$ while they are unbounded operators on the Fock space $\mathbb{F}$ in the usual sense.} Then, if $u,v \in \mathsf{L}(\mathbb{D})$, the composition products $u \circ v$ and $v \circ u$ are also in $\mathsf{L}(\mathbb{D})$ (but $u \circ v \neq v \circ u$ in general).  For example, if $\psi \in \mathsf{L}_S^2(\mathbb{R}^3)$, then the mapping $\mathbf{a}^{\pm}(\psi)$ is in $\mathsf{L}(\mathbb{D})$, as will be seen in Section \ref{embo:0}.

We now equip $\mathsf{L}(\mathbb{D})$ with a structure of `bounded sets' as follows:
\begin{definition}[Bounded subset in $\mathsf{L}(\mathbb{D})$] 
%.............................................................
\label{oper:defi:2}
A subset $\mathfrak{B} \DEF \{u_i\}_{i \in I} \subset \mathsf{L}(\mathbb{D})$ is said to be bounded if\, $\forall B$ bounded set in $\mathbb{D}$, $\{u_i(f)\}_{i \in I, f \in B}$ is another bounded set in $\mathbb{D}$, i.e., the maps $u_i$ are uniformly bounded on any bounded set of $\mathbb{D}$.
\end{definition}
In other words, setting $u_i(B) = \{u_i(f)\}_{f \in B}$, this definition implies that $\forall B$ bounded set in $\mathbb{D}$, $\bigcup_{i \in I}  u_i(B)$ is another bounded set in $\mathbb{D}$.

   These bounded sets have all the natural properties of bounded sets that were stated above.  Note in particular that if $\{u_i\}_{i \in I}$ and $\{v_j\}_{j \in J}$ are two bounded sets in $\mathsf{L}(\mathbb{D})$, then from the definition the sets $\{u_i \circ v_j\}_{i \in I, j \in J}$ and  $\{v_j \circ u_i\}_{j \in J, i \in I}$ are again two bounded sets in $\mathsf{L}(\mathbb{D})$.

\begin{remark}[Integration of functions
%....................................................
$\mathbb{R}^n \rightarrow \mathsf{L}(\mathbb{D})$. ]
%..................................................
\label{oper:rema:1}
We briefly note that one can prove that one can choose the bounded disks $\mathfrak{B} \subset \mathsf{L}(\mathbb{D})$ so that the normed spaces $\bigcup_n n\mathfrak{B}$ are Banach spaces, which permits here to use the theory of integration of functions valued in Banach spaces.
\end{remark}

\subsection{Definition of the algebra
% ........................................
$\mathcal{G}\bigl(\mathbb{R}^n,\mathsf{L}(\mathbb{D})\bigr)$}

The definition of the operator valued generalized function algebra $\mathcal{G}\bigl(\mathbb{R}^n,\mathsf{L}(\mathbb{D})\bigr)$, with $n=3$ or $4$ in practice, is a straight forward extension of the definition of the scalar valued algebra $\mathcal{G}(\mathbb{R}^n,\mathbb{C})$, e.g., \cite[Definition 3.1]{STEIN2006-}. 

\begin{definition}[$\mathcal{C}^\infty$ functions]
%.................................................
\label{oper:defi:3}
A function $f: \mathbb{R}^n \rightarrow \mathsf{L}(\mathbb{D})$ with $n \in \mathbb{N}$ is of class $\mathcal{C}^\infty$ iff\, $\forall R > 0$, $\forall N \in \mathbb{N}$, $\exists \mathfrak{B}$ bounded disk in $\mathsf{L}(\mathbb{D})$ such that $f$ maps $\{ |x| < R \} \subset \mathbb{R}^n$ into the normed space $\bigcup_n  n\mathfrak{B}$ and
\begin{align}
\label{oper:5}
f\Bigr|_{\{ |x| < R \}} : \{ |x| < R \} \rightarrow \bigcup_n  n\mathfrak{B},
\end{align}
is of class $\mathcal{C}^N$.
\end{definition}
The reader should not bother about $\mathcal{C}^\infty$ properties that cause indeed no problem here.\footnote{In reference \cite{COLOM1982-} the `structure of bounded sets' and the concept of `$\mathcal{C}^\infty$ functions between these structures' are developed in detail.}

The functions $f_\epsilon$ considered in the following are assumed to be $\mathcal{C}^\infty$ from $\mathbb{R}^n$ into $\mathsf{L}(\mathbb{D})$.

Let us denote $\mathbb{N} = \{ 0,1,2,...\}$ and recall the standard multi-index notation
\begin{align}
\label{oper:6}
 \rmD^\alpha = \frac{\partial^{|\alpha|}}
                   {(\partial x_1)^{\alpha_1} \cdots (\partial x_n)^{\alpha_n}},
\end{align}
where $\alpha \in \mathbb{N}^n$ and $|\alpha| = \alpha_1 + \alpha_2 + \cdots \alpha_n$.
\begin{definition}[Moderate functions] 
%.....................................
\label{oper:defi:4}
The space $\mathcal{E}_{\text{\emph{M}}}$, where the letter $\emph{M}$ stands for `moderate growth' functions, is
\begin{align}
\nonumber
    \mathcal{E}_{\text{\emph{M}}}\bigl(\mathbb{R}^n,\mathsf{L}(\mathbb{D})\bigr) \DEF
 \Bigl\{ &\{f_\epsilon\} : \forall K \text{~compact in~} \mathbb{R}^n,
                        \forall \alpha \in \mathbb{N}^n,\\
\nonumber
         &\exists \mathfrak{B} \text{~bounded set in~}  \mathsf{L}(\mathbb{D}),
          \exists N \in \mathbb{N}\\
\nonumber
         & \text{~such that~}, \forall x \in K, \\
\label{oper:7}
         & \rmD^\alpha f_\epsilon(x) \in \frac{1}{\epsilon^N}\mathfrak{B}
           \text{~~~as~} \epsilon \rightarrow 0^+
  \Bigr\}.
\end{align}
\end{definition}
\begin{definition}[Negligible functions]
%.......................................
\label{oper:defi:5}
The space $\mathcal{N}$ of `negligible' functions is
\begin{align}
\nonumber
    \mathcal{N}\bigl(\mathbb{R}^n,\mathsf{L}(\mathbb{D})\bigr) \DEF
 \Bigl\{ &\{f_\epsilon\} : \forall K \text{~compact in~} \mathbb{R}^n,
                        \forall \alpha \in \mathbb{N}^n,\\
\nonumber
         &\exists \mathfrak{B} \text{~bounded set in~}
               \mathsf{L}(\mathbb{D})\\
\nonumber
         &\text{~such that~},    
           \forall x \in K, \forall q \in \mathbb{N}, \exists C_q > 0, \\
\label{oper:8}
         & \rmD^\alpha f_\epsilon(x) \in C_q\epsilon^q\mathfrak{B}
           \text{~~~as~} \epsilon \rightarrow 0^+
  \Bigr\}.
\end{align}
\end{definition}
\begin{definition}[Special algebra] 
%..................................
\label{oper:defi:6}
The special algebra of operator valued nonlinear functions is the quotient space
\begin{align}
\label{oper:9}
       \mathcal{G}  \bigl(\mathbb{R}^n,\mathsf{L}(\mathbb{D})\bigr) \DEF 
 \frac {\mathcal{E}_{\text{\emph{M}}}\bigl(\mathbb{R}^n,\mathsf{L}(\mathbb{D})\bigr)}
       {\mathcal{N}  \bigl(\mathbb{R}^n,\mathsf{L}(\mathbb{D})\bigr)}.
\end{align}
\end{definition}
Thus a nonlinear generalized function $f \in \mathcal{G}  \bigl(\mathbb{R}^n,\mathsf{L}(\mathbb{D})\bigr)$ denoted by `$f$'$=$`$\{ f_\epsilon \}$' is an equivalence class of moderate sequences of smooth functions modulo negligible ones; it is represented by a moderate sequence of smooth functions $\{ f_\epsilon \}$. The space $\mathcal{E}_{\text{M}}\bigl(\mathbb{R}^n,\mathsf{L}(\mathbb{D})\bigr)$ is a differential algebra with pointwise operations and, since the space of negligible functions is a differential ideal, $\mathcal{G}  \bigl(\mathbb{R}^n,\mathsf{L}(\mathbb{D})\bigr)$ is also a differential algebra --- which however is not commutative since it is an operator valued algebra.

   The vector space of distributions whose values are bounded linear maps from $\mathbb{D}$ into $\mathbb{D}$ is now embedded into the algebra $\mathcal{G} \bigl(\mathbb{R}^n,\mathsf{L}(\mathbb{D})\bigr)$ following the convention of \cite[p.\,59]{GROSS2001-}, i.e., via convolution with a mollifier $\rho^\vee$ according to the formulas \ref{moll:5}.  More precisely, we choose a suitable mollifier $\rho$ belonging to the set $\mathcal{A}_\infty$ defined in Definition \ref{moll:defi:1}.  Then:
\begin{definition}[Embedding of distributions] 
%.............................................
\label{oper:defi:7}
Let $\rho \in \mathcal{A}_\infty$. The embedding $\iota(\gamma) \in \mathcal{G}  \bigl(\mathbb{R}^n,\mathsf{L}(\mathbb{D})\bigr)$ of a distribution $\gamma \in \mathcal{D}'(\mathbb{R}^n,\mathsf{L}(\mathbb{D})\bigr)$ is the convolution
\begin{align}
\label{oper:10}
       \iota(\gamma)_\epsilon \DEF \gamma \ast \rho^\vee_\epsilon.
 \end{align}
\end{definition}
Distributions which are not compactly supported are embedded via a localized version of \eqref{oper:10} using a standard sheaf theoretic construction, as is done for instance in \cite{GROSS2001-}.  Continuous functions are embedded as a consequence of the inclusion $\mathcal{C} \subset \mathcal{D}'$.  This embedding by convolution can also be applied to $\mathcal{C}^\infty$ functions which then coincides with the trivial embedding $f \rightarrow (f_\epsilon = f, \forall \epsilon)$ because $\rho \in \mathcal{A}_\infty$ and because of the quotient \eqref{oper:9}.

    In the $\mathcal{G}$-theory one may frequently interpret the results in terms of distributions using the concept of `association:'
\begin{definition}[Association] 
%..............................
\label{oper:defi:8}
A generalized function $f \in \mathcal{G}  \bigl(\mathbb{R}^n,\mathsf{L}(\mathbb{D})\bigr)$ is said to be associated with zero iff for one (hence any) representative $\{ f_\epsilon \}$, we have $\forall \varphi \in \mathcal{C}^\infty(\mathbb{R}^n)$ with compact support
\begin{align}
\label{oper:11}
       \int f_\epsilon(x) \varphi(x) ~d^nx 
       \rightarrow  0  \quad \text{in} \quad \mathsf{L}(\mathbb{D})
       \quad \text{as} \quad \epsilon \rightarrow 0^+,
 \end{align}
and we then write $f \ASS 0$.  
\end{definition}
In a natural sense, $f_\epsilon \varphi$ is a continuous function from $\mathbb{R}^n$ into a normed space $\bigcup_n n\mathfrak{B} \subset \mathsf{L}(\mathbb{D})$; one shows further that this normed space can be chosen as a Banach space so that the integral makes sense.  Then convergence to $0$ in $\mathsf{L}(\mathbb{D})$ means:  $\exists \mathfrak{B}$ bounded in $\mathsf{L}(\mathbb{D})$ such that $\forall \chi > 0, \exists \eta > 0$ such that
\begin{align}
\label{oper:12}
       0 < \epsilon < \eta \Rightarrow 
       \int f_\epsilon(x) \varphi(x) ~d^nx \in \chi \mathfrak{B}.
 \end{align}

We say that two generalized functions $f$ and $g$ are associated, or that $f$ is associated with $g$, and we write $f \ASS g$, iff $f-g$ is associated with zero.  Associated to zero objects (numbers, functions) are also called `infinitesimals.'

   Association is an equivalence relation which respects addition and differentiation.  But by the Schwartz impossibility results it cannot respect multiplication in general. Indeed, it is impossible to define an intrinsic multiplication on the space of distributions whose restriction to continuous functions would be the pointwise multiplication of continuous functions. However, $\mathcal{G}  \bigl(\mathbb{R}^n,\mathsf{L}(\mathbb{D})\bigr)$ contains the smooth functions as a faithful subalgebra (their multiplication is the usual pointwise product of smooth functions), as well as the distributions as a linear subspace.  Multiplication of distributions can be handled in this framework by making proper use of the concepts of equality (`$=$') and of association (`$\ASS$').  This will be further discussed in Section \ref{oper:0}.5:  Indeed, the new product in $\mathcal{G}$ of continuous functions, which is not equal in  $\mathcal{G}$ to the pointwise product, will be only associated to the pointwise product.

\subsection{Use of nonlinear generalized functions}
%..................................................

There are basically two lines of application of nonlinear generalized functions in physics:

\begin{itemize}

\item The first one is best suited for presentations in short form and consists in using explicitly the $\mathcal{G}$-context by \emph{working with abstract generalized functions} (i.e., objects depending only on $x$ and $t$), and introducing representatives (i.e., objects depending on $\rho, \epsilon ,x,$ and $t$) only when needed (in definition, some proofs, etc.). One is then as close as possible to the formal calculations.

\item The second line consists in \emph{working with representatives of generalized functions} throughout the calculations.  It is heavier: $\rho,\epsilon,x,$ and $t$ everywhere, and it may give the false impression that the $\mathcal{G}$-context is only a regularization.  But this method has the advantage to manipulate only classical objects, which might be clearer and permit immediate discussion with anybody unaware of the $\mathcal{G}$-theory for improvement.

\end{itemize}
These two lines are similar to the two main approaches to applying classical distribution theory, the former one referring to the original Sobolev-Schwartz definition of distributions as functionals, and the latter to the Mikusinski sequential approach which is often preferred by physicists.

   In this report, which we address to readers that are new to the $\mathcal{G}$-theory, all QFT calculations are done on representatives.  In simple terms, we therefore consider `moderate' function, i.e., objects $f(\rho_\epsilon,...)$ --- where the suspension points are usual variables such as $t$ and $\vec{x}$ --- that are bounded by
\begin{align}
\label{oper:13}
       f(\rho_\epsilon,...) \leq \frac{\text{constant}}{\epsilon^N},
 \end{align}
for some $N \in \mathbb{N}$ and $\epsilon$ small enough (with uniform bounds on compact sets in the variables $t,\vec{x},...$ as well as all $t,\vec{x},...$ partial derivatives). 

   Two such objects $f_1(\rho_\epsilon,...)$ and $f_2(\rho_\epsilon,...)$ are then identified iff the difference $f_1-f_2$ is `negligible,' meaning that it can be `identified to zero' in the sense that $\forall q \in \mathbb{N}, \exists C_q > 0, \eta_q >0$ such that $|(f_1-f_2)(\rho_\epsilon,...)| \leq C_q \epsilon^q$ if $0 < \epsilon < \eta_q$ (with uniform bounds on compact sets for the functions and their partial derivatives as above) $C_q$ being a constant depending on $q$.  The concept of generalized functions permits therefore to replace (by definition) bounds of the type `$|\text{object}| \leq C_q \epsilon^q, \forall q$' by `$\text{object} = 0$' in the $\mathcal{G}$-context. That is: In order to calculate we are forced to consider `regularized objects,' but then when the presence of the regularizations gives rise to `very small quantities' $\leq C_q \epsilon^q, \forall q$ as $\epsilon \rightarrow 0^+$, which may be nonzero in the classical setting, they are declared equal to zero in the $\mathcal{G}$-setting (by definition).

   In summary, it is the proper consideration given to functions associated to zero which enables to overcome Schwartz's multiplication-impossibility theorem.  It all derives from $\mathcal{G}$ canonically containing  $\mathcal{D}'$ as a subspace, and the smooth (i.e., $\mathcal{C}^\infty$) functions as a faithful subalgebra, while (consistent with Schwartz's impossibility theorem) the simply continuous functions are not a subalgebra.  Thus, if $f$ and $g$ are two continuous functions on $\Omega$, their new product $f \odot g$ in $\mathcal{G}(\Omega)$ differs from their classical product $f \cdot g$ in $\mathcal{C}(\Omega)$ by an `infinitesimal function' $\imath \ASS 0$, i.e., a function associated to zero (which is our definition of `infinitesimal') so that
\begin{align}
\label{oper:14}
       f \odot g = f \cdot g + \imath,
 \end{align}
where $\imath =0$ when $f,g \in \mathcal{C}^\infty(\Omega)$.  Thus, the $\mathcal{G}$-embedding of continuous functions, as well as of distributions, are only \emph{`subalgebras modulo infinitesimals.'}  For any two elements $f,g \in \mathcal{G}(\Omega)$ one has therefore
\begin{align}
\label{oper:15}
       (f + \imath_1) \odot (g + \imath_2) = f \cdot g + \imath_3,
 \end{align}
where $\imath_1, \imath_2,$ and $\imath_3$ are infinitesimal functions.  Then, if one drops all infinitesimals one obtains the equality of the two product: one says that they are `associated' and on writes
\begin{align}
\label{oper:16}
       f \odot g \ASS f \cdot g.
 \end{align}
As illustrated by this example, one observes that, in all calculations making sense within distribution theory, dropping the infinitesimals gives nothing other than the classical calculations with distributions. (The infinitesimals can be dropped freely at any stage, or only at the end.)  In this way, the theory of nonlinear generalized functions is perfectly coherent with classical mathematics and distribution theory.  But in $\mathcal{G}$-calculations which do not make sense within distribution theory, dropping the infinitesimals usually leads to nonsense.

   To give a concrete example let us consider the commutations relations (\ref{summ:1}--\ref{summ:3}), which only make sense as distributions.  Thus, embedding these equations in $\mathcal{G}$, we can \emph{a priori} conclude that
\begin{align}
\label{oper:17}
\iota\Bigl( [\bphi(t,\vec{x}_1),\bphi(t,\vec{x}_2)] \Bigr)_\epsilon & \ASS 0,\\
\label{oper:18}
\iota\Bigl( [ \bpi(t,\vec{x}_1), \bpi(t,\vec{x}_2)] \Bigr)_\epsilon & \ASS 0,\\
\label{oper:19}
\iota\Bigl( [\bphi(t,\vec{x}_1), \bpi(t,\vec{x}_2)] \Bigr)_\epsilon & \ASS 
       \iota\Bigl( i \delta^3(\vec{x}_1-\vec{x}_2) \Bigr)_\epsilon~\mathbf{1}. 
\end{align}
However, doing the embeddings explicitly and calculating the commutators, it will be found that
\begin{align}
\label{oper:20}
\iota\Bigl( [\bphi(t,\vec{x}_1),\bphi(t,\vec{x}_2)] \Bigr)_\epsilon & = 0,\\
\label{oper:21}
\iota\Bigl( [ \bpi(t,\vec{x}_1), \bpi(t,\vec{x}_2)] \Bigr)_\epsilon & = 0,\\
\label{oper:22}
\iota\Bigl( [\bphi(t,\vec{x}_1), \bpi(t,\vec{x}_2)] \Bigr)_\epsilon & \ASS 
       \iota\Bigl( i \delta^3(\vec{x}_1-\vec{x}_2) \Bigr)_\epsilon~\mathbf{1}, 
\end{align}
where $\delta^3$ is the Dirac distribution at $0$ in $\mathbb{R}^3$.  Therefore, the first two commutation relations are actually satisfied as equalities  (an essential requirement in a local field theory), whereas association in the third commutation relation implies that infinitesimal contributions may have to be taken into account in further calculations.  For instance, to calculate in $\mathcal{G}$ with these commutation relations, we shall need an equality in $\mathcal{G}$ in equation \eqref{oper:22}.  Therefore, we shall write 
\begin{align}
\label{oper:23}
     [\bphi_\mathcal{G}(t,\vec{x}_1), \bpi_\mathcal{G}(t,\vec{x}_2)] = 
        i \delta^3_\mathcal{G}(\vec{x}_1-\vec{x}_2)~\mathbf{1}, 
\end{align}
where $\delta^3_\mathcal{G}$ is now an element of $\mathcal{G}$ precisely defined by the commutator $[\bphi_\mathcal{G}(t,\vec{x}_1),$ $\bpi_\mathcal{G}(t,\vec{x}_2)]$, that is not necessarily a distribution, but a $\mathcal{G}$-function that will have the properties needed in the sequel, which include in particular all those classically attributed to a `Dirac delta function.'  (In the $\mathcal{G}$ context there are several different Dirac delta functions.)  To a large extent we can therefore set aside the concept of distributions and replace it by the more general one of $\mathcal{G}$-functions.

%File: free.39.tex        arXiv version 2           Date: 6 September 2008
%=====              
%
%     "cos" in (2.5.4) - (2.5.19) replaced by "exp"
%     sign restored to (t-t_\xi) in footnote page 54 
%     (2.7.33) corrected 
%     All signs starting in (2.6.10) corrected!
%     (2.6.13) --> (2.6.16)  \xi  
%     Proposition 2.2.1  OK with "1" 
%     "zero-point energy" part updated on 5 Sept.

\chapter{The free field}
%=======================
\label{free:0}

In this chapter we define the $\mathcal{G}$-embeddings of the state functions, creation/annihila\-tion operators, and Hamiltonian operator of a free spin-$0$ field.  %%%% As we go along, and verify the basic properties of these objects, we determine the minimal characteristics of the mollifier and dampers compatible with those properties, so that they may belong to classes as large as possible. 

\section{$\mathcal{G}$-embedding of free-field state functions}
%--------------------------------------------------------------
\label{embs:0}
\setcounter{equation}{0}
\setcounter{definition}{0}
\setcounter{axiom}{0}
\setcounter{conjecture}{0}
\setcounter{lemma}{0}
\setcounter{theorem}{0}
\setcounter{corollary}{0}
\setcounter{proposition}{0}
\setcounter{example}{0}
\setcounter{remark}{0}
\setcounter{problem}{0}

To set QFT in the $\mathcal{G}$-context we begin by embedding the classical single-particle state function $\psi=\Delta_+$ used as argument of the creation/annihilation operators $\mathbf{a}^\pm(\psi)$.  That is, instead of working with $\psi$ defined by (\ref{stat:1} -- \ref{stat:2}) as
\begin{align}
\nonumber
     \vec{\xi} \mapsto
     \psi(\xi,x)
    &= \Delta_+(\xi - x)\\
\label{embs:1}
    &=\frac{1}{(2\pi)^3}   \iiint \frac{d^3p}{2 E_p}
     \exp i\Bigl(\vec{p}\cdot (\vec{\xi} - \vec{x}\,) - E_p (t_\xi - t)\Bigr),
\end{align}
where  $\xi = \{t_\xi, \vec{\xi}\,\}$ and $x = \{t, \vec{x}\,\} \in \mathbb{R}^4$, we shall work with the nonlinear generalized function  $\iota(\psi) \in \mathcal{G}$ which is obtained by embedding $\psi$ as a distribution in the variable $\vec{\xi}$ according to Definition \ref{oper:defi:7}, i.e., as the convolution
\begin{equation}\label{embs:2}
   \iota(\psi)_\epsilon  \DEF \psi \ast \rho_\epsilon^\vee,
\end{equation}
where
\begin{equation}\label{embs:3}
    \rho_\epsilon(\vec{\xi}\,) \DEF \frac{1}{\epsilon^3}
                               \rho\bigl( \frac{\vec{\xi}}{\epsilon} \bigr).
\end{equation}
Here $\epsilon$ is a strictly positive real parameter that is arbitrarily close to 0, and $\rho \in \mathcal{A}_\infty$ is a suitable mollifier function, i.e., by Definition \ref{moll:defi:1}, such that its Fourier transform $\FOU{\rho}$ is in $\mathcal{S}(\mathbb{R}^3,\mathbb{C})$ with compact support and identical to 1 on a 0-neighborhood in $\mathbb{R}^3$.  Therefore, with \eqref{moll:4} and \eqref{moll:5},
\begin{align}
\nonumber
        \iota(\psi)_\epsilon
      &= \iiint d^3y~ 
             \frac{1}{\epsilon^3} 
             \rho\bigl( \frac{\vec{y}-\vec{\xi}}{\epsilon} \bigr)
             \psi(t_\xi,\vec{y},x)\\
\label{embs:4}
      &= \iiint d^3z~ 
             \rho(\vec{z})
             \psi(t_\xi,\vec{\xi}+\epsilon\vec{z},x),
\end{align}
where we made the change of variable $\vec{y}-\vec{\xi}=\epsilon\vec{z}$, $d^3y=\epsilon^3~d^3z$.  Substituting $\psi$ from \eqref{embs:1} we get
\begin{align}
\nonumber
        \iota(\psi)_\epsilon
       &=\frac{1}{(2\pi)^3}   \iiint \frac{d^3p}{2 E_p}
     \exp i\Bigl(\vec{p}\cdot (\vec{\xi} - \vec{x}\,) - E_p (t_\xi - t)\Bigr),\\
\label{embs:5}
       &\times \iiint d^3z~ 
             \rho(\vec{z}) \exp(i\epsilon\vec{p}\cdot\vec{z}\,) ,
\end{align}
which by the definition \eqref{moll:1} of the Fourier transform gives
\begin{equation}\label{embs:6}
         \iota(\psi)_\epsilon(\xi,x)
     =\frac{1}{(2\pi)^3}   \iiint \frac{d^3p}{2 E_p}
      \FOU{\rho}(\epsilon\vec{p}\,)
     \exp i\Bigl(\vec{p}\cdot (\vec{\xi} - \vec{x}\,) - E_p (t_\xi - t)\Bigr).
\end{equation}
Comparing with \eqref{embs:1} this suggests defining
\begin{align}
\label{embs:7}
     \Delta_\epsilon(\FOU{\rho}, x) \DEF
     \frac{1}{(2\pi)^3}   \iiint \frac{d^3p}{2 E_p}
     \FOU{\rho}(\epsilon\vec{p}\,)
     \exp i (\vec{p}\cdot \vec{x} - E_p t),
\end{align}
as the embedding of the function $\Delta_+$.  The embedded form of equation \eqref{embs:1} is therefore
\begin{align}
\label{embs:8}
     \iota \Bigl( \vec{\xi} \rightarrow \psi(\xi,x) \Bigr)_\epsilon
           = \Delta_\epsilon(\FOU{\rho}, \xi - x).
\end{align}

   Of course, from the definition \eqref{fock:7} of the creation operator, the embedding of $\psi$ directly corresponds to the embedding of the single-particle substate $f_1=f_0 \psi$ of $\mathbf{a}^+(\psi)\mho \in \mathbb{F}$.  The generalization to multiparticle states $(f_n)(\vec{\xi}_1, ...,\vec{\xi}_n) \in \mathbb{D}$, i.e., to Fock-space states generated by linear superpositions of the basis elements \eqref{stat:13}, is straightforward:  All classical functions $\Delta_+(\xi_j-x_j)$ are interpreted as distributions in the variables $\vec{\xi}_j$ and thus simply replaced by the generalized functions $\Delta_\epsilon(\FOU{\rho}, \xi_j-x_j)$, i.e., 
\begin{align}
\label{embs:9}
                 \iota \Bigl( (f_n)_n \Bigr)_\epsilon 
         = f_0   \sqrt{1/n!}  \Bigl(
                 \Delta_\epsilon(\FOU{\rho},\xi_1-x_1)
                 \Delta_\epsilon(\FOU{\rho},\xi_2-x_2) \cdots
                     + \cdots \Bigr),
\end{align}
whereas the vacuum state is left unchanged
\begin{align}
\label{embs:10}
                 \iota \Bigl( (f_0) \Bigr)_\epsilon 
         =  (f_0).
\end{align}
\begin{proposition}
%...................
\label{embs:prop:1}
The embedded multiparticle states $\psi_{n,\epsilon} = \iota \bigl( (f_n)({\xi}_j-{x}_j) \bigr)_\epsilon$, with $n\neq 0$ and $\epsilon \neq 0$, are elements of the space $\mathcal{S}\bigl( (\mathbb{R}^3)^n,\mathbb{C}\bigr)$ in the variables $\vec{\xi}_j-\vec{x}_j$.
\end{proposition}
Proof: Since the embedded states are polynomial in the functions $\Delta_\epsilon(\FOU{\rho},\xi_j-x_j)$, and $\mathcal{S}$ is an algebra with respect to pointwise multiplication, it is enough to prove that $\Delta_\epsilon(\FOU{\rho},x) \in \mathcal{S}(\mathbb{R}^3,\mathbb{C})$.  In view of this we rewrite \eqref{embs:7} as
\begin{align}
\label{embs:11}
     \Delta_\epsilon(\FOU{\rho}, x) =
     \iiint d^3p ~
     \exp i (\vec{p}\cdot \vec{x})
     \Bigl(
     \frac{1}{(2\pi)^3 2 E_p}\exp i (- E_p t)
     \FOU{\rho}(\epsilon\vec{p}\,)
     \Bigr),
\end{align}
which shows that $\Delta_\epsilon(\FOU{\rho},x)$ is the Fourier transform of the expression in the big parenthesis.  However, since $\FOU{\rho}(\epsilon\vec{p}\,) \in \mathcal{S}$ when $\epsilon \neq 0$, and $|E_p^{-1}\exp i (- E_p t)| = 1/\sqrt{m^2+|\vec{p}\,|^2}$ this expression also belongs to $\mathcal{S}$.  Consequently, as the Fourier transform is an isomorphism of $\mathcal{S}$, it follows that $\Delta_\epsilon(\FOU{\rho},x) \in \mathcal{S}$. \END

%%%{\bf Remark on notation.  For the moment we will write $\Delta_\epsilon(\FOU{\rho}, x)$ and later use the abbreviated form $\Delta_\epsilon(x)$ when it will be sure that the embedding is in fact `canonical,' i.e., independent of the choice of $\FOU{\rho}$, provided $\FOU{\rho}$ belongs to a restricted class to be defined at the end of this chapter.}

%
\begin{proposition}[Equivalence of embeddings]
%.............................................
\label{embs:prop:2}
Let $\rho$ and $\eta \in \mathcal{A}_\infty$.  Then,
\begin{align}
\label{embs:12}
     \Delta_\epsilon(\FOU{\rho}\,\FOU{\eta}, x)
   = \Delta_\epsilon(\FOU{\rho \ast \eta}, x)
   = \Delta_\epsilon(\FOU{\rho}, x)
   + \OOO(\epsilon^{q+1}), \qquad \forall q \in \mathbb{N}.
\end{align}
That is, all embeddings \eqref{embs:7} with any $\rho \in \mathcal{A}_\infty$ are equivalent modulo an infinitesimal quantity $\OOO(\epsilon^{q+1}) \in \mathbb{C}$.
\end{proposition}
Proof: Let us write the left-hand side of \eqref{embs:12} as
\begin{align}
\label{embs:13}
     \Delta_\epsilon(\FOU{\rho}\,\FOU{\eta}, x) =
    \iiint d^3p~\FOU{\rho}(\epsilon\vec{p}\,)
               ~\FOU{\eta}(\epsilon\vec{p}\,)~f(\vec{p}\,),
\end{align}
where the function
\begin{align}
\label{embs:14}
     f(\vec{p}\,) =
     \frac{1}{(2\pi)^32 E_p} \exp i (\vec{p}\cdot \vec{x} - E_p t),
\end{align}
evidently belongs to $\mathcal{O}_{\text{M}}(\mathbb{R}^3,\mathbb{C})$.  We now make the Taylor expansion with remainder
\begin{align}
\nonumber
 \FOU{\eta}(\epsilon \vec{p}\,)
    &= \FOU{\eta}(0)
     + \epsilon|\vec{p}\,|\bigl(\rmD^{1}\FOU{\eta}\bigr)(0)
     + \frac{\epsilon^2|\vec{p}\,|^2}
            {2!}    \bigl(\rmD^{2}\FOU{\eta}\bigr)(0) \\
\label{embs:15}
    &+ ... + \frac{(\epsilon|\vec{p}\,|)^{q+1}}
            {(q+1)!} \bigl(\rmD^{q+1}\FOU{\eta}\bigr)(\theta\epsilon\vec{p}\,),
\end{align}
where the abusive one-dimensional notation $\bigl(\rmD^{n}\FOU{\eta}\bigr)$ is used for the derivatives.  As $\eta \in \mathcal{A}_\infty$ we have by definition
\begin{equation}
\label{embs:16}
   \FOU{\eta}(0) = 1,
   \quad \text{and} \quad
   \bigl(\rmD^{n+1}\FOU{\eta}\bigr)(0) = 0,
   \quad \forall n \in  \mathbb{N}.
\end{equation}
Therefore, when this development is inserted in \eqref{embs:13} all terms in $|\vec{p}\,|^n$  with $1 < n < q+1$ are zero.  Consequently
\begin{align}
\nonumber
   \iiint d^3p~\FOU{\rho}(\epsilon\vec{p}\,)
              ~\FOU{\eta}(\epsilon\vec{p}\,)~f(\vec{p}\,)
   =    \iiint &d^3p~\FOU{\rho}(\epsilon\vec{p}\,)~f(\vec{p}\,)\\
\label{embs:17}
   + \epsilon^{q+1} \iiint &d^3p~\FOU{\rho}(\epsilon\vec{p}\,)~ f(\vec{p}\,)
       \frac{|\vec{p}\,|^{q+1}}{(q+1)!}
       \bigl(\rmD^{q+1}\FOU{\eta}\bigr)(\theta\epsilon\vec{p}\,).
\end{align}
In this expression the remainder depends on $\vec{p}, \epsilon, \text{and~} \theta \in ]0,1[$, and there is no control on the product $\theta\epsilon\vec{p}$ when $\vec{p}$ ranges in $\mathbb{R}^3$.  Thus $\theta$ can tend to $0$ as $\epsilon \rightarrow 0$ while $|\vec{p}\,| \rightarrow \infty$.   This means that we only have a bound of the type $|\bigl(\rmD^{q+1} \FOU{\eta}\bigr)(\theta\epsilon\vec{p}\,)| \leq \Cst$, where the constant is independent of $\epsilon$ and $\vec{p}$.  However, when $\epsilon\neq 0$, the damper $\FOU{\rho}(\epsilon\vec{p}\,) \in \mathcal{S}$.  Then, as $f(\vec{p}\,) \in \mathcal{O}_{\text{M}}(\mathbb{R}^3,\mathbb{C})$ and as  $\bigl(\rmD^{q+1}\FOU{\eta}\bigr)$ is bounded, their product with $|\vec{p}\,|^{q+1}$ is still in $\mathcal{S}$.  The integral on the second line of \eqref{embs:17} which defines the coefficient of $\OOO(\epsilon^{q+1})$ in \eqref{embs:12} is therefore bounded in that case.  As for the case $\epsilon=0$, Eq.~\eqref{embs:12} is an identity, so that the proposition is proved.  \END

\section{Scalar product, norm, and normalization in $\mathcal{G}$}
%------------------------------------------------------------------
\label{norm:0}
\setcounter{equation}{0}
\setcounter{definition}{0}
\setcounter{axiom}{0}
\setcounter{conjecture}{0}
\setcounter{lemma}{0}
\setcounter{theorem}{0}
\setcounter{corollary}{0}
\setcounter{proposition}{0}
\setcounter{example}{0}
\setcounter{remark}{0}
\setcounter{problem}{0}

   Having embedded the state functions, the next step is to verify that they can be normalized, which was not possible in the classical setting where $\|\Delta_+(\xi-x)\|_{\mathsf{L}^2} = \infty$ as shown by \eqref{stat:12}.  The reason is of course that $\Delta_+$ is a distribution and therefore should be integrated only after being multiplied by a test function.  To deal with this in a systematic way we redefine the formal scalar product \eqref{fock:14} as a distribution and take a suitable damper for the test function:
\begin{definition}[Scalar product and norm in $\mathcal{G}$]
%...........................................................
\label{norm:defi:1}
Let $\FOU{\chi} \in \mathcal{B}_\infty$ and let $\epsilon >0$.  The $\mathcal{G}$-embedded form of the 1-particle scalar product \eqref{fock:15} is then
\begin{equation}\label{norm:1}
          \BRA  f_1(\xi) \| g_1(\xi) \KET_\mathcal{G}
    \DEF  i \iiint_{t_\xi=\text{Cst.}}  d^3\xi~ \Bigl(
             f_1^* \frac{\partial g_1  }{\partial t} 
                 - \frac{\partial f_1^*}{\partial t} g_1 \Bigr)
            \FOU{\chi}(\epsilon\vec{\xi}\,),
\end{equation}
and its generalization to $n$-particle states consists of introducing $n$ factors $\FOU{\chi}(\epsilon\vec{\xi}_j)$ in the formal definition  \eqref{fock:14}.  The norm induced by \eqref{norm:1} and its $n$-particle generalization are written $\|\cdot\|_{\mathsf{L}^2_\mathcal{G}}$.
\end{definition}
With this definition we can now derive the embedded versions of formulas (\ref{stat:5}--\ref{stat:6}) and (\ref{stat:8}--\ref{stat:11}), which we shall frequently need in the following.  For the first pair we take the wave-function $\psi_p(\xi)$ defined by \eqref{stat:4} and  easily get
\begin{align}
\label{norm:2}
      \BRA \psi_{p_1}(\xi\,)
        \| \psi_{p_2}(\xi\,) \KET_\mathcal{G}
        =  \frac{E_{p_1} + E_{p_2}}{2\sqrt{E_{p_1} E_{p_2}}}
            \rme^{i(E_{p_1}-E_{p_2})t_\xi}
                 \iiint \frac{d^3\xi}{(2\pi)^3}
                 \rme^{i(\vec{p}_2-\vec{p}_1)\cdot\vec{\xi}}
                 \FOU{\chi}(\epsilon \vec{\xi}\,),
\end{align}
where by inverse Fourier transform
\begin{align}
\label{norm:3}
     \iiint \frac{d^3\xi}{(2\pi)^3}
                 \rme^{i(\vec{p}_2-\vec{p}_1)\cdot\vec{\xi}} ~
                 \FOU{\chi}(\epsilon \vec{\xi}\,) 
     =\frac{1}{\epsilon^3} \chi(\frac{\vec{p}_1-\vec{p}_2}{\epsilon}),
\end{align}
so that the embedded form of \eqref{stat:6} is
\begin{align}
\label{norm:4}
      \BRA \psi_{p_1}(\xi\,)
        \| \psi_{p_2}(\xi\,) \KET_\mathcal{G}
        =  \frac{E_{p_1} + E_{p_2}}{2\sqrt{E_{p_1} E_{p_2}}}
            \rme^{i(E_{p_1}-E_{p_2})t_\xi}
         \frac{1}{\epsilon^3} \chi(\frac{\vec{p}_1-\vec{p}_2}{\epsilon}),
\end{align}
which has the expected form since it tends to the $\delta$-function in the limit $\epsilon \rightarrow 0$.  Similarly, the embedded form of \eqref{stat:5} is 
\begin{align}
\label{norm:5}
      \BRA \psi_{p_1}(\xi\,)
        \| \psi_{p_2}^*(\xi\,) \KET_\mathcal{G}
        =  \frac{E_{p_1} - E_{p_2}}{2\sqrt{E_{p_1} E_{p_2}}}
            \rme^{i(E_{p_1}+E_{p_2})t_\xi}
         \frac{1}{\epsilon^3} \chi(\frac{\vec{p}_1+\vec{p}_2}{\epsilon}),
\end{align}
which gives zero in the limit $\epsilon \rightarrow 0$.

   To embed formula \eqref{stat:9} we first use \eqref{stat:4} to rewrite \eqref{embs:7} as
\begin{align}
\label{norm:6}
     \Delta_\epsilon(\FOU{\rho},\xi-x) 
      = \iiint \frac{d^3p}{\sqrt{(2\pi)^3 2 E_p}} \FOU{\rho}(\epsilon\vec{p}\,)
        \psi_p(\xi-x),
\end{align}
where $\psi_p(\xi-x)=\psi_p(\xi)~\psi_p(-x)$  and then use \eqref{norm:4} to get
\begin{align}
\nonumber
   \BRA \psi_p(\xi) \| \Delta_\epsilon(\FOU{\rho},\xi-x) \KET_\mathcal{G}
 = \iiint d^3p_2 ~\FOU{\rho}(\epsilon\vec{p}_2) \psi_{p_2}(-x) 
   \BRA \psi_p(\xi) \| \psi_{p_2}(\xi) \KET_\mathcal{G}\\
\label{norm:7}
 = \iiint d^3p_2 ~\FOU{\rho}(\epsilon\vec{p}_2) \psi_{p_2}(-x)
     \frac{E_p+E_{p_2}}
          {2\sqrt{E_p  E_{p_2}}} \exp i(E_p - E_{p_2} )t_\xi ~
     \frac{1}{\epsilon^3} \chi(\frac{\vec{p}-\vec{p}_2}{\epsilon}).
\end{align}
This expression is significantly more complicated than \eqref{stat:9}. But since $\FOU{\chi} \in \mathcal{B}_\infty$ and $\rho \in \mathcal{A}_\infty$, Proposition \ref{moll:prop:2} enables to reduce it to a much simpler form.  Indeed, we can write
\begin{align}
\label{norm:8}
   \BRA \psi_p(\xi) \| \Delta_\epsilon(\FOU{\rho},\xi-x) \KET_\mathcal{G}
 =      \iiint d^3p_2~ \FOU{\rho}(\epsilon\vec{p}_2) ~h(\vec{p},\vec{p}_2)
         \frac{1}{\epsilon^3} \chi(\frac{\vec{p}-\vec{p}_2}{\epsilon}),
\end{align}
where the function
\begin{align}
\label{norm:9}
  h(\vec{p},\vec{p}_2) = \psi_{p_2}(-x)
         \frac{E_p+E_{p_2}}{2\sqrt{E_p  E_{p_2}}} \exp i(E_p - E_{p_2} )t_\xi,
\end{align}
clearly belongs to $\mathcal{O}_{\text{M}}(\mathbb{R}^3,\mathbb{C})$ in the variable $\vec{p}_2 \in \mathbb{R}^3$ because $E_p=\sqrt{m^2 + {\vec{p}\,}^2}$.  Thus, applying Proposition \ref{moll:prop:2} we get
\begin{align}
\label{norm:10}
   \BRA \psi_p(\xi) \| \Delta_\epsilon(\FOU{\rho},\xi-x) \KET_\mathcal{G}
 = \FOU{\rho}(\epsilon\vec{p}\,)\psi_p(-x)
 + \OOO(\epsilon^{q+1}), \qquad \forall q \in \mathbb{N},
\end{align}
which is identical to \eqref{stat:9} in the limit $\epsilon \rightarrow 0$.  As for  \eqref{stat:8}, complex-conjugating $\Delta_\epsilon$ in \eqref{norm:7} leads to the substitution $E_{p_2} \rightarrow - E_{p_2}$ in \eqref{norm:9} so that
\begin{align}
\label{norm:11}
   \BRA \psi_p(\xi) \| \Delta_\epsilon^*(\FOU{\rho},\xi-x) \KET_\mathcal{G}
 = \OOO(\epsilon^{q+1}), \qquad \forall q \in \mathbb{N}.
\end{align}

    We now apply the same method to embed and reduce formulas (\ref{stat:10}--\ref{stat:11}).  With the help of \eqref{stat:4} and \eqref{norm:4} we obtain
\begin{align}
\nonumber
   \BRA \Delta_\epsilon(\FOU{\rho},\xi-x_1) \|
 &      \Delta_\epsilon(\FOU{\rho},\xi-x_2) \KET_\mathcal{G} 
 =       \iiint d^3p_1~\FOU{\rho}^*(\epsilon \vec{p}_1)
         \iiint d^3p_2~\FOU{\rho}  (\epsilon \vec{p}_2)\\
\nonumber
 &\times  \frac{1}{(2\pi)^3} \frac{(E_{p_1}+E_{p_2})}{4E_{p_1}E_{p_2}}
   ~ \exp \Bigl( i(E_{p_1}-E_{p_2})t_\xi  -i(E_{p_1}t_1-E_{p_2}t_2) \Bigr)\\
\label{norm:12}
 &\times  \exp i(\vec{p}_1\cdot \vec{x}_1 - \vec{p}_2\cdot \vec{x}_2 \,)
        ~ \frac{1}{\epsilon^3}
          \chi \bigl(\frac{\vec{p}_1-\vec{p}_2}{\epsilon}\bigr).
\end{align}
Then, as $\FOU{\chi} \in \mathcal{B}_\infty$ and $\rho \in \mathcal{A}_\infty$, we rewrite the $p_2$-integral as
\begin{align}
\label{norm:13}
      \iiint d^3p_2~ \FOU{\rho}(\epsilon\vec{p}_2)~ h(\vec{p}_1,\vec{p}_2)
         \frac{1}{\epsilon^3} \chi(\frac{\vec{p}_1-\vec{p}_2}{\epsilon}),
\end{align}
where the function
\begin{align}
\nonumber
  h(\vec{p}_1,\vec{p}_2)
  &= \frac{1}{(2\pi)^3} \frac{(E_{p_1}+E_{p_2})}{4E_{p_1}E_{p_2}} 
   ~ \exp \Bigl( i(E_{p_1}-E_{p_2})t_\xi  -i(E_{p_1}t_1-E_{p_2}t_2) \Bigr)\\
\label{norm:14}
 &\times  \exp i(\vec{p}_1\cdot \vec{x}_1 - \vec{p}_2\cdot \vec{x}_2 \,),
\end{align}
has the requested properties for applying Proposition \ref{moll:prop:2}.  Thus, Eq.~\eqref{norm:12} reduces to
\begin{align}
\nonumber
   \BRA \Delta_\epsilon(&\FOU{\rho},\xi-x_1) \|
       \Delta_\epsilon(\FOU{\rho},\xi-x_2) \KET_\mathcal{G} 
  = \frac{1}{(2\pi)^3} \iiint \frac{d^3p}{2 E_p} \FOU{\rho}^*(\epsilon\vec{p})\\
\nonumber
  &\times   \Bigl( \FOU{\rho}(\epsilon\vec{p})
          ~ \exp \bigl(  -iE_{p}(t_1-t_2) \bigr)
            \exp \bigl( i\vec{p}\cdot (\vec{x}_1 - \vec{x}_2\,) \bigr)
            + \OOO(\epsilon^{q+1}) \Bigr),\\
\label{norm:15}
  &~\qquad \qquad \qquad \qquad \qquad \qquad \qquad \qquad
      \qquad \qquad \qquad\forall q \in \mathbb{N}.
\end{align}
Since $\FOU{\rho}^*\FOU{\rho}=\FOU{\rho^\vee\ast\rho}$ and the integral over the remainder is bounded, this can be written in the compact form\footnote{The order $\OOO(\epsilon^{q-1})$ of the remainder is derived in Proposition \ref{norm:prop:1}.}
\begin{align}
\label{norm:16}
   \BRA \Delta_\epsilon(\FOU{\rho},\xi-x_1) \|
        \Delta_\epsilon(\FOU{\rho},\xi-x_2) \KET_\mathcal{G} 
   =  \Delta_\epsilon(\FOU{\rho^\vee\ast\rho}, x_1-x_2)
   +  \OOO(\epsilon^{q-1}),
\end{align}
where $\Delta_\epsilon(\FOU{\rho^\vee\ast\rho}, x_1-x_2)$ is the embedded $\Delta_+$ function \eqref{embs:7}.  Finally, complex-conjugating $\Delta_\epsilon(\FOU{\rho},\xi-x_2)$ in \eqref{norm:12} leads to the substitution $E_{p_2} \rightarrow - E_{p_2}$ in \eqref{norm:14} so that the embedded from of \eqref{stat:10} is
\begin{align}
\label{norm:17}
   \BRA \Delta_\epsilon(\FOU{\rho},\xi-x_1) \|
        \Delta_\epsilon^*(\FOU{\rho},\xi-x_2) \KET_\mathcal{G} 
   =   \OOO(\epsilon^{q-1}).
\end{align}

  Having derived these formulas, we prove the proposition:

\begin{proposition}[Norm and normalizability of $\Delta_\epsilon(\FOU{\rho},\xi-x)$]
%........................................................................
\label{norm:prop:1}
Let $\rho \in \mathcal{A}_\infty$ and $\FOU{\chi} \in \mathcal{B}_\infty$.  Then,  for $\epsilon > 0$ small enough,
\begin{align}
\label{norm:18}
     \| \Delta_\epsilon(\FOU{\rho},\xi-x) \|^2_\mathcal{G}
     = {\rm N}^2(\FOU{\rho},\epsilon) +  \OOO(\epsilon^{q-2}),
\end{align}
where
\begin{align}
\label{norm:19}
     {\rm N}(\FOU{\rho},\epsilon) \DEF
     \sqrt{ \frac{1}{(2\pi)^3}
            \iiint \frac{d^3p}{2E_p}
            |\FOU{\rho}(\epsilon\vec{p}\,)|^2 }
            = \OOO(\frac{1}{\epsilon}).
\end{align}
Therefore the function $\Delta_\epsilon(\FOU{\rho},\xi-x)$ is normalizable in the sense that, defining
\begin{align}
\label{norm:20}
      \Delta_{{\rm N}(\FOU{\rho},\epsilon)}(\xi-x)
         \DEF  \frac{\Delta_\epsilon(\FOU{\rho},\xi-x)}
                         {{\rm N}(\FOU{\rho},\epsilon)},\\
\intertext{then}
\label{norm:21}
      \| \Delta_{{\rm N}(\FOU{\rho},\epsilon)}(\xi-x) \|_\mathcal{G}
         =  1 + \OOO(\epsilon^{q-1}).
\end{align}
\end{proposition}
Proof: Let us write the norm of $\Delta_\epsilon(\FOU{\rho},\xi-x)$ in the developed form \eqref{norm:15}
\begin{align}
\label{norm:22}
   \| \Delta_\epsilon(\FOU{\rho},\xi-x) \|^2_\mathcal{G} 
   = \frac{1}{(2\pi)^3} \iiint \frac{d^3p}{2 E_p} \FOU{\rho}^*(\epsilon\vec{p})
     \Bigl( \FOU{\rho}(\epsilon\vec{p})
          + \epsilon^{q+1} R^{q+1}(\vec{p},\theta,\epsilon) \Bigr),
\end{align}
where the $x$ and $t_\xi$ dependence in the remainder $R^{q+1}$ has been ignored because $|h(\vec{p}_1,\vec{p}_2)|^2$ given by \eqref{norm:14} does not depend on $x_1, x_2,$ and $t_\xi$.  Thus, making the change of variable $\epsilon \vec{p} = \vec{s}$,
\begin{align}
\label{norm:23}
     \| \Delta_\epsilon(\FOU{\rho},\xi-x) \|^2_\mathcal{G}
   =  \frac{\epsilon^{-3}}{ {(2\pi)^3}}
            \iiint \frac{d^3s}{2E_p}
            |\FOU{\rho}(\vec{s}\,)|^2
   +  \frac{\epsilon^{q-2}}{(2\pi)^3}
            \iiint \frac{d^3s}{2E_p}
            \FOU{\rho}^*(\vec{s}\,)R^{q+1}(\vec{s},\theta,\epsilon),
\end{align}
which, as both integrals are bounded because of the dampers, and since
\begin{align}
\label{norm:24}
     E_p = \sqrt{m^2 + \frac{|\vec{s}\,|^2}{\epsilon^2}}
         > \frac{|\vec{s}\,|}{\epsilon},
         \qquad \forall |\vec{s}\,|, \forall \epsilon,
\end{align}
enables to write,
\begin{align}
\label{norm:25}
     \| \Delta_\epsilon(\FOU{\rho},\xi-x) \|^2_\mathcal{G}
   =   \OOO(\frac{1}{\epsilon^2})
   +   \OOO(\epsilon^{q-2}).
\end{align}
However, the first integral in \eqref{norm:23} is strictly positive whereas the second one can have either sign. Thus, for $\epsilon > 0$ small enough the first integral dominates and \eqref{norm:23} is positive $\forall q \in \mathbb{N}$.  Consequently, the square root of \eqref{norm:23} can be written as (\ref{norm:18}-\ref{norm:19}), and $\Delta_\epsilon(\FOU{\rho},\xi-x)$ can be normalized as in \eqref{norm:20}. \END

   This proposition implies that the single particle state $f_1 = f_0 \Delta_\epsilon(\FOU{\rho},\xi-x)$ is normalizable.  The generalization to multiparticle states is immediate:  Replacing all classical functions $\Delta_+(\xi_j-x_j)$ in the Fock-space basis elements \eqref{stat:13} by \eqref{norm:20} implies that all multiparticle states $(f_n) \in \mathbb{D}$ are systematically normalized.  The Hilbertian norm \eqref{fock:14} is then finite for all states we a finite number of particles.  In practice, however, we will continue using $\Delta_\epsilon$ for the embedded $\Delta_+$ functions, and employ the normalized states only when considering the impact of normalization.

\section{$\mathcal{G}$-embedding of free-field operators}
%--------------------------------------------------------
\label{embo:0}
\setcounter{equation}{0}
\setcounter{definition}{0}
\setcounter{axiom}{0}
\setcounter{conjecture}{0}
\setcounter{lemma}{0}
\setcounter{theorem}{0}
\setcounter{corollary}{0}
\setcounter{proposition}{0}
\setcounter{example}{0}
\setcounter{remark}{0}
\setcounter{problem}{0}

Let us consider the free-field operator \eqref{stat:14} written in the form  \eqref{stat:21}
\begin{align}
\label{embo:1}
       {\bphi}_0(x) = \mathbf{a}^+(x) + \mathbf{a}^-(x), 
\end{align}
where
\begin{align}
\label{embo:2}
         \mathbf{a}^+(x) = \mathbf{a}^+\Bigl( \Delta_+(\xi-x) \Bigr),
       \qquad \text{and} \qquad
         \mathbf{a}^-(x) = \mathbf{a}^-\Bigl( \Delta_+(\xi-x) \Bigr).
\end{align}
As an operator acting on Fock-space states ${\bphi}_0(x)$ is a function of the variable $x$, while $\xi$ is a parameter associated to its argument $\bigl(\xi \mapsto\Delta_+(\xi-x) \bigr)$.  Thus, we interpret
\begin{align}
\label{embo:3}
       \vec{x} \mapsto {\bphi}_0(x), 
\end{align}
as a distribution in $\vec{x}$, and define its embedding in $\mathcal{G}$ as
\begin{equation}
\label{embo:4}
   \bphi_0(\rho_\epsilon,t,\vec{x}\,) \DEF \iota(\bphi_0)_\epsilon
                    = \bphi_0(x) \ast \rho_\epsilon^\vee(\vec{x}\,).
\end{equation}
Since $\rho \in \mathcal{A}_\infty$ is real by definition, and convolution a linear operation commuting with the creation and annihilation operators, this embedding is simply,
\begin{align}
\label{embo:5}
        \bphi_0(\rho_\epsilon,t,\vec{x}\,) =
           \mathbf{a}^+(\rho_\epsilon,x) + \mathbf{a}^-(\rho_\epsilon,x), 
\end{align}
where
\begin{align}
\label{embo:6}
       \mathbf{a}^\pm(\rho_\epsilon,x) =
      \mathbf{a}^\pm\Bigl( \Delta_+(\xi-x)
                      \ast \rho_\epsilon^\vee(\vec{x}\,) \Bigr),
\end{align}
i.e.,
\begin{align}
\label{embo:7}
       \mathbf{a}^\pm(\rho_\epsilon,x) =
      \mathbf{a}^\pm\Bigl( \Delta_\epsilon(\FOU{\rho},\xi-x) \Bigr),
\end{align}
where $\Delta_\epsilon$ is defined by \eqref{embs:7}.

   We are now going to show that $\bphi_0(\rho_\epsilon,t,\vec{x}\,)$ so-defined is an element of $\mathcal{E}_{\text{M}}\bigl(\mathbb{R}^4,\mathsf{L}(\mathbb{D})\bigr)$, that is a moderate operator-valued function in $\mathcal{G}$ according to Definition \ref{oper:defi:4}.  We have thus to prove:
\begin{conjecture}
%.................
\label{embo:conj:1}
Let $\bphi_0(\rho_\epsilon,t,\vec{x}\,)$ be defined by \eqref{embo:5}. Then,
\begin{align}
\nonumber
  &\forall K \text{~compact in~} \mathbb{R}^4, 
   \forall \rmD^\alpha \text{~derivation with respect to~}\vec{x} \in \mathbb{R}^3,
%% \text{~and~} \forall \alpha,n \in \mathbb{N}_0,
   \\
\nonumber
  &\exists \mathfrak{B} \text{~bounded in~} \mathsf{L}(\mathbb{D}) \text{~and~}
   \exists N \in \mathbb{N} \text{~ such that~}\\
\label{embo:8}
  & \forall \alpha,n \geq 0 \qquad \Rightarrow \qquad
       \rmD^\alpha \frac{\partial^n}{\partial t^n}
       \bphi_0(\rho_\epsilon,t,\vec{x}\,)
       \in \frac{1}{\epsilon^N} \mathfrak{B}, \\
\nonumber
  &\text{if~} \{t,\vec{x}\} \in K \text{~and~} \epsilon >0 \text{~small enough}.
\end{align}
\end{conjecture}
We proceed by first proving two lemmas:
\begin{lemma}
%..............
\label{embo:lemm:1}
Let $\Delta_\epsilon(\FOU{\rho},\xi-x)$ be defined by \eqref{embs:7}. Then
$\forall |\alpha|, \forall n \in \mathbb{N}$, $\exists N \in \mathbb{N}$ such that
\begin{align}
\label{embo:9}
       \|   \rmD^\alpha \frac{\partial^n}{\partial t^n}
            \Delta_\epsilon(\FOU{\rho},\xi-x) \|_{\mathsf{L}^2_\mathcal{G}}
       = \OOO(\frac{1}{\epsilon^N}),
\end{align}
if $\{t,\vec{x}\} \in K$ compact in $\mathbb{R}^4$ and $\epsilon > 0$ small enough. Thus, the mapping
\begin{align}
\label{embo:10}
   (\epsilon,t,\vec{x}\,) \rightarrow  \Delta_\epsilon(\FOU{\rho},\xi-x),
\end{align}
is an element of  $\mathcal{E}_{\text{\emph{M}}}\bigl(\mathbb{R}^4,\mathbb{C}\bigr)$, i.e., a moderate generalized function in $\mathcal{G}$.
\end{lemma}
Proof: We first consider $\Delta_\epsilon$, and then any $\rmD^\alpha \frac{\partial^n}{\partial t^n} \Delta_\epsilon$ with $\alpha,n \neq 0$.  From the norm squared of $\Delta_\epsilon(\FOU{\rho},\xi-x)$, i.e., equation \eqref{norm:18}, it immediately follows that for $\epsilon > 0$ small enough
\begin{align}
\label{embo:11}
     \| \Delta_\epsilon(\FOU{\rho},\xi-x) \|_{\mathsf{L}^2_\mathcal{G}}
                      = \OOO(\frac{1}{\epsilon}).
\end{align}
Therefore the function $\Delta_\epsilon(\FOU{\rho},t,\vec{x}\,)$ satisfies \eqref{embo:9} when $\alpha=n=0$.   In the case of the derivatives of the field, i.e., $\rmD^\alpha \frac{\partial^n}{\partial t^n} \Delta_\epsilon(\FOU{\rho},t,\vec{x}\,)$ with $\alpha>0$ and/or $n>0$, the expression under the integral sign in the definition \eqref{embs:7} of $\Delta_\epsilon(\FOU{\rho},\xi-x)$ will be multiplied by products of powers of the components of the four-vector $p=\{E_p,\vec{p}\}$.  Because $E_p=\sqrt{m^2 + {\vec{p}\,}^2}$, an upper bound on such products is provided by $E_p^k \DEF (E_p)^k$ with $k \in \mathbb{N}$.  Going through the calculations leading to the norm \eqref{norm:19} it is readily seen that an upper bound on the norm of $\rmD^\alpha \frac{\partial^n}{\partial t^n} \Delta_\epsilon(\FOU{\rho},\xi-x)$ is then given by \eqref{norm:19} in which $1/E_p$ is replaced by $E_p^{2k-1}$ with $k=|\alpha|+n$, i.e., 
\begin{align}
\label{embo:12}
     \| \rmD^\alpha \frac{\partial^n}{\partial t^n}
                 \Delta_\epsilon(\xi-x) \|_{\mathsf{L}^2_\mathcal{G}}
         < \sqrt{ \frac{1}{ {(2\pi)^3}}
            \iiint \frac{d^3p}{2E_p} E_p^{2k}
            |\FOU{\rho}(\epsilon\vec{p}\,)|^2 },
\end{align}
However, as $\FOU{\rho} \in \mathcal{S}(\mathbb{R}^3)$, it follows that this bound is finite $\forall k \in \mathbb{N}$. Thus, making the change of variable $\epsilon \vec{p} = \vec{s}$ and replacing $E_p$ by $|\vec{s}\,|/\epsilon$ we conclude that for $\epsilon$ small enough and $\forall k \in \mathbb{N}$ 
\begin{align}
\label{embo:13}
       \|   \rmD^\alpha \frac{\partial^n}{\partial t^n}
            \Delta_\epsilon(\FOU{\rho},\xi-x) \|_{\mathsf{L}^2_\mathcal{G}}
       = \OOO(\frac{1}{\epsilon^{k+1}}),
\end{align}
so that \eqref{embo:9} is satisfied and the lemma is proved.  \END
\begin{lemma} 
%............
\label{embo:lemm:2}
The mapping
\begin{align}
\nonumber
  \mathsf{L}^2(\mathbb{R}^3) &\rightarrow \mathsf{L}(\mathbb{D}),\\
\label{embo:14}
   \psi  &\mapsto \mathbf{a}_p^\pm(\psi),
\end{align}
is well defined and linear-bounded on bounded sets. That is:
\begin{enumerate}
\item Well defined: Let $f \in B$ where $B$ is any bounded set in $\mathbb{D}$.  Then the set $\{\mathbf{a}^\pm(\psi).f\}_{f\in B}$ is again a bounded set in $\mathbb{D}$ for any $\psi \in \mathsf{L}^2(\mathbb{R}^3)$;
\item Linear bounded: Let $\psi \in B_\psi$ where $B_\psi$ is any bounded set in $\mathsf{L}^2(\mathbb{R}^3)$. Then the set $\{\mathbf{a}^\pm(\psi).f\}_{\psi \in B_\psi}$ is again a bounded set in $\mathbb{D}$ for any $f \in B$, that is, the set $\mathfrak{B} = \{\mathbf{a}^\pm(\psi).f\}_{f\in B, \psi \in B_\psi}$ is a bounded set in $\mathsf{L}(\mathbb{D})$.
\end{enumerate}
\end{lemma}
Proof: From Definition \ref{oper:defi:1} any bounded set $B \subset \mathbb{D}$ is contained in the a set of elements $f \in \mathbb{D}$ of the form
\begin{align}
\label{embo:15}
     f =
\begin{pmatrix}
      b_0\\
      b_1\\
       \vdots\\
      b_N\\
       0
\end{pmatrix} ,
\quad\text{where}\quad b_n = 0, \forall n> N,
\end{align}
and where $b_n$ is bounded in $\mathsf{L}_S^2\bigl((\mathbb{R}^3)^n;\mathbb{C} \bigr)$ while $b_0$ is bounded in $\mathbb{C}$. Thus, operating with $\mathbf{a}^+$ defined by \eqref{fock:7},
\begin{equation}\label{embo:16}
     \mathbf{a}^+(\psi) f
      =
\begin{pmatrix}
      b_0\\
      b_1\\
       \vdots\\
      b_N\\
       0\\
       0
\end{pmatrix} 
     =
\begin{pmatrix}
     0 \\
     b_0\psi\\
     \sqrt{2} ~{\rm Sym}( \psi \otimes b_1)\\
     \vdots\\
     \sqrt{n} ~{\rm Sym}( \psi \otimes b_{N})\\
     0
\end{pmatrix} ,
\end{equation}
which is clearly another bounded set in $\mathbb{D}$.  This proves point one, and proving point two simply consists of letting $\psi$ range in $B_\psi$.  Since the proof is similar for $\mathbf{a}^-(\psi)$, the lemma is proved.  \END

\begin{theorem}
%..............
\label{embo:theo:1}
Let $\Delta_\epsilon(\FOU{\rho},\xi-x)$ be defined by \eqref{embs:7}, and let $\epsilon >0$. Then the mapping
\begin{align}
\label{embo:17}
   (\epsilon,t,\vec{x}\,) \rightarrow \bphi_0(\rho_\epsilon,t,\vec{x}\,)
           = \mathbf{a}^+ \bigl( \Delta_\epsilon(\FOU{\rho},\xi-x)  \bigr)
           + \mathbf{a}^- \bigl( \Delta_\epsilon(\FOU{\rho},\xi-x)  \bigr),
\end{align}
is an element of  $\mathcal{E}_{\text{\emph{M}}}\bigl(\mathbb{R}^4,\mathsf{L}(\mathbb{D})\bigr)$, i.e., a moderate operator valued $\mathcal{G}$-function.
\end{theorem}
Proof: From Lemma \ref{embo:lemm:1} it immediately follows that
\begin{align}
\label{embo:18}
     \Bigl( x \rightarrow \rmD^\alpha \frac{\partial^n}{\partial t^n}
                         \Delta_\epsilon(\FOU{\rho},\xi-x) \Bigr)
             \in \frac{1}{\epsilon^N} B,
\end{align}
where $B$ is bounded in $\mathsf{L}^2_\mathcal{G}(\mathbb{R}^3)$.  Thus, from Lemma \ref{embo:lemm:2},
\begin{align}
\label{embo:19}
 \mathbf{a}^\pm \Bigl( x \rightarrow \rmD^\alpha \frac{\partial^n}{\partial t^n}
                                     \Delta_\epsilon(\FOU{\rho},\xi-x) \Bigr)
                          \in \frac{1}{\epsilon^N} \mathfrak{B},
\end{align}
where $\mathfrak{B}$ is bounded in $\mathsf{L}(\mathbb{D})$. \END

   In conclusion, for fixed $\rho \in \mathcal{S}(\mathbb{R}^3)$  the function $(\epsilon,t,\vec{x}\,) \mapsto \bphi_0(\rho_\epsilon,t,\vec{x}\,)$ is a \emph{representative} of an object $\bphi_0(\rho_\epsilon,\cdot,\cdot)$ such that
\begin{align}
\nonumber
       \bphi_0(\rho_\epsilon,\cdot,\cdot) ~: \qquad
       \mathbb{R}^4 &\rightarrow \mathcal{G}\bigl(\mathbb{R}^4,
                                              \mathsf{L}(\mathbb{D}) \bigr)\\
\label{embo:20}
        t, \vec{x}  &\mapsto \bphi_0(\rho_\epsilon,t,\vec{x}\,),
\end{align}
which is a nonlinear generalized function in the variable $\{t,\vec{x}\}$ belonging to $\mathcal{G}\bigl( \mathbb{R}^4,\mathsf{L}(\mathbb{D}) \bigr)$.  Moreover, for fixed $t$, the function $\bphi_0(\epsilon,\cdot,\vec{x}\,)$ is a generalized function belonging to $\mathcal{G}\bigl( \mathbb{R}^3,\mathsf{L}(\mathbb{D}) \bigr)$ so that $\bphi_0$ can also be considered as a map
\begin{align}
\nonumber
       \bphi_0(\rho_\epsilon,\cdot,\vec{x}\,) ~: \quad
       \mathbb{R} &\rightarrow \mathcal{G}\bigl(\mathbb{R}^3,
                                              \mathsf{L}(\mathbb{D}) \bigr),\\
\label{embo:21}
        t  &\mapsto \bphi_0(\rho_\epsilon,t,\vec{x}\,).
\end{align}

   For maps such as (\ref{embo:20}--\ref{embo:21}) one can then define suitable structures and the concept of \emph{differentiability} in $\mathcal{G}$.  But we do not enter into these formal developments because they are straightforward.  For instance, in the $\mathcal{G}$-context, the operator ${\bpi}_0(\rho_\epsilon,t,\vec{x}\,)$ defined by \eqref{stat:16} canonically conjugated to ${\bphi}_0(\rho_\epsilon,t,\vec{x}\,)$ is according to \eqref{embo:5} and \eqref{embo:7} given by
\begin{align}
\nonumber
     {\bpi}_0(\rho_\epsilon,t,\vec{x}\,)
    &= \frac{\partial}{\partial t}{\bphi}_0(\rho_\epsilon,t,\vec{x}\,)\\
\label{embo:22}
    &= \mathbf{a}^+\Bigl( \frac{\partial}{\partial t}
                          \Delta_\epsilon(\FOU{\rho},\xi-x) \Bigr)
     + \mathbf{a}^-\Bigl( \frac{\partial}{\partial t}
                          \Delta_\epsilon(\FOU{\rho},\xi-x) \Bigr).
\end{align}
%

%   Finally, we note that having embedded the operators ${\bphi}_0(\rho_\epsilon,t,\vec{x}\,)$ and ${\bpi}_0(\rho_\epsilon,t,\vec{x}\,)$, the embedding of composite operator is straight forward.  

%%\noindent{\bf !!}     Finally, as $\rho$ is real valued by definition, the operators ${\bphi}_0(\rho_\epsilon,t,\vec{x}\,)$ and ${\bpi}_0(\rho_\epsilon,t,\vec{x}\,)$ are densely defined symmetric operators (of domain $\mathbb{D}$) on $\mathbb{F}$.  They map $\mathbb{D}$ into $\mathbb{D}$ and satisfy on $\mathbb{D}$ the canonical commutation relations (\ref{summ:1}--\ref{summ:3}): 

\section{Commutation relations: formal calculations}
%---------------------------------------------------
\label{comf:0}
\setcounter{equation}{0}
\setcounter{definition}{0}
\setcounter{axiom}{0}
\setcounter{conjecture}{0}
\setcounter{lemma}{0}
\setcounter{theorem}{0}
\setcounter{corollary}{0}
\setcounter{proposition}{0}
\setcounter{example}{0}
\setcounter{remark}{0}
\setcounter{problem}{0}

Before verifying that the free-field operators embedded in $\mathcal{G}$ satisfy the canonical commutation relations (\ref{summ:1}--\ref{summ:3}) it is useful to first recall the standard proof for the corresponding classical operators:

\begin{proposition}[Formal commutation relations of the free-field operators]
%............................................................................
\label{comf:prop:1} 
The classical free-field operators $\bphi_0(x)$ and $\bpi_0(x)= \partial_t{\bphi}_0(x)$ satisfy the equal-time canonical commutation relations
\begin{align}
\label{comf:1}
        [\bphi_0(x_1), \bphi_0(x_2)]\Bigr|_{t_1=t_2=t_\xi} &= 0,\\
\label{comf:2}
        [ \bpi_0(x_1),  \bpi_0(x_2)]\Bigr|_{t_1=t_2=t_\xi} &= 0,\\
\label{comf:3}
        [\bphi_0(x_1),  \bpi_0(x_2)]\Bigr|_{t_1=t_2=t_\xi} &=
                                   i \delta^3(\vec{x_1}-\vec{x_2}),
\end{align}

\end{proposition}
Proof: Let us calculate the commutator  $[\bphi(x_1), \bphi(x_2)]$ using the abbreviated form \eqref{embo:1} for the operators $\bphi(x_1)$ and  $\bphi(x_2)$.  It comes
\begin{align}
\nonumber
        [\bphi_0(x_1) , \bphi_0(x_2)] 
      &=[\mathbf{a}^+(x_1) + \mathbf{a}^-(x_1) ~~,~~
         \mathbf{a}^+(x_2) + \mathbf{a}^-(x_2)]\\
\label{comf:4}
      &=[\mathbf{a}^-(x_1) ~,~ \mathbf{a}^-(x_2)]
       +[\mathbf{a}^+(x_1) ~,~ \mathbf{a}^+(x_2)] \\
\label{comf:5}
      &+[\mathbf{a}^-(x_1) ~,~ \mathbf{a}^+(x_2)]
       +[\mathbf{a}^+(x_1) ~,~ \mathbf{a}^-(x_2)] ,
\end{align}
where, by (\ref{fock:10}--\ref{fock:11}) the two commutators \eqref{comf:4} are identically zero.  On the other hand, \eqref{fock:12} and \eqref{stat:11} imply that the two commutators \eqref{comf:5} give
\begin{align}
\label{comf:6}
    [\bphi(x_1) , \bphi(x_2)]
        =      \Delta_{+}(x_1-x_2)-\Delta_{+}(x_2-x_1)
        =  \Delta_{\text{JP}}(x_1-x_2),
\end{align}
where $\Delta_{\text{JP}}$ is known has the Jordan-Pauli function, and the invariant function $\Delta_+$ is given by the integral \eqref{stat:2}, i.e.,
\begin{align}
\label{comf:7}
    \Delta_+(t,\vec{x}\,) =
\frac{1}{(2\pi)^3}   \iiint \frac{d^3p}{2 E_p}
        \exp( -iE_p t +i\vec{p}\cdot \vec{x} ).
\end{align}
Therefore,
\begin{align}
\label{comf:8}
        [\bphi_0(x_1), \bphi_0(x_2)] &= \Delta_{\text{JP}}(x_1-x_2),\\
\label{comf:9}
        [\bpi_0 (x_1),  \bpi_0(x_2)] &=
            \frac{\partial}{\partial t_1}
            \frac{\partial}{\partial t_2} \Delta_{\text{JP}}(x_1-x_2)
        = - \frac{\partial^2}{\partial t_1^2} \Delta_{\text{JP}}(x_1-x_2),\\
\label{comf:10}
        [\bphi_0(x_1),  \bpi_0(x_2)] &= 
            \frac{\partial}{\partial t_2} \Delta_{\text{JP}}(x_1-x_2)
        = - \frac{\partial}{\partial t_1} \Delta_{\text{JP}}(x_1-x_2).
\end{align}
For $t_1 = t_2$ and $\vec{x} = \vec{x}_1 -\vec{x}_2$ the Jordan-Pauli function is
\begin{align}
\label{comf:11}
    \Delta_{\text{JP}}(0,\vec{x}\,) =
\frac{1}{(2\pi)^3}   \iiint \frac{d^3p}{2 E_p}  \Bigl(
        \exp(+i\vec{p}\cdot \vec{x}\,) 
      - \exp(-i\vec{p}\cdot \vec{x}\,)  \Bigr),
\end{align}
so that the commutators \eqref{comf:8} and \eqref{comf:9} are zero because their right-hand side is the integral of an odd function.  The commutator \eqref{comf:10}, being the integral of an even function, is however non-zero.  Thus, differentiating \eqref{comf:7} in \eqref{comf:6},
\begin{align}
\label{comf:12}
     - \frac{\partial}{\partial t_1} \Delta_{\text{JP}}(0,\vec{x}\,) 
     = i \frac{1}{(2\pi)^3} \iiint d^3p ~ \exp ( \pm i \vec{p} \cdot \vec{x} )
     = i \delta^3(\vec{x}\,),
\end{align}
immediately gives the third equal-time commutation relation in the set (\ref{comf:1}--\ref{comf:3}), which is therefore proved. \END

\section{Commutation relations: rigorous calculations}
%-----------------------------------------------------
\label{comr:0}
\setcounter{equation}{0}
\setcounter{definition}{0}
\setcounter{axiom}{0}
\setcounter{conjecture}{0}
\setcounter{lemma}{0}
\setcounter{theorem}{0}
\setcounter{corollary}{0}
\setcounter{proposition}{0}
\setcounter{example}{0}
\setcounter{remark}{0}
\setcounter{problem}{0}

\begin{proposition}[Rigorous commutation relations of the free-field operators]
%..............................................................................
\label{comr:prop:1}
Let the free-field operators $\bphi_0(\rho_\epsilon,x)$ and $\bpi_0(\rho_\epsilon,x)= \partial_t{\bphi}_0(\rho_\epsilon,x)$ be defined by \eqref{embo:5} and \eqref{embo:19}, and let $\rho \in \mathcal{A}_\infty$ and $\FOU{\chi} \in \mathcal{B}_\infty$.  Then :
\begin{enumerate}
\item[(A)] The equal-time commutation relations are 
\begin{align}
\label{comr:1}
    [{\bphi}_0(\rho_\epsilon,x_1),
     {\bphi}_0(\rho_\epsilon,x_2)]\Bigr|_{t_1=t_2=t_\xi} &=0,\\
\label{comr:2}
    [{\bpi }_0(\rho_\epsilon,x_1),
     {\bpi }_0(\rho_\epsilon,x_2)]\Bigr|_{t_1=t_2=t_\xi} &=0,\\
\label{comr:3}
    [{\bphi}_0(\rho_\epsilon,x_1),
     {\bpi }_0(\rho_\epsilon,x_2)]\Bigr|_{t_1=t_2=t_\xi} &=
    i \delta^3(\rho_\epsilon,\chi_\epsilon,\vec{x}_1,\vec{x}_2)~\mathbf{1}, 
\end{align} 
where the nonlinear generalized function replacing the usual Dirac $\delta$-function is real and identically given by
\begin{align}
\nonumber
     \delta^3(\rho_\epsilon,\chi_\epsilon,\vec{x}_1,\vec{x}_2)
 &=       \iiint d^3p_1~\FOU{\rho}^*(\epsilon \vec{p}_1)
          \iiint d^3p_2~\FOU{\rho}  (\epsilon \vec{p}_2)\\
\nonumber
 &\times  \frac{1}{(2\pi)^3} \frac{(E_{p_1}+E_{p_2})}{2E_{p_1}}
          \exp i\bigl(\vec{p}_1\cdot \vec{x}_1 -\vec{p}_2\cdot \vec{x}_2\bigr)\\
\label{comr:4}
 &\times  \frac{1}{\epsilon^3}
          \chi\bigl(\frac{\vec{p}_2-\vec{p}_1}{\epsilon}\bigr).
\end{align}
\item[(B)] These commutation relations can be written
\begin{align}
\label{comr:5}
    [{\bphi}_0(\rho_\epsilon,x_1),
     {\bphi}_0(\rho_\epsilon,x_2)]\Bigr|_{t_1=t_2=t_\xi} &= 0,\\
\label{comr:6}
    [{\bpi }_0(\rho_\epsilon,x_1),
     {\bpi }_0(\rho_\epsilon,x_2)]\Bigr|_{t_1=t_2=t_\xi} &= 0,\\
\label{comr:7}
    [{\bphi}_0(\rho_\epsilon,x_1),
     {\bpi }_0(\rho_\epsilon,x_2)]\Bigr|_{t_1=t_2=t_\xi} &=
         i \delta^3_{[\bphi_0,\bpi_0]}(\vec{x}_1-\vec{x}_2) ~\mathbf{1}, 
\end{align}
where
\begin{align}
\label{comr:8}
    \delta^3_{[\bphi_0,\bpi_0]}(\vec{x}_1-\vec{x}_2)
     \DEF  \frac{1}{\epsilon^3}(\rho^\vee \ast \rho)
           \bigl(\frac{\vec{x}_1-\vec{x}_2}{\epsilon}\bigr)
         + \OOO(\epsilon^{q-1}),
\end{align}
and $\rho^\vee\ast\rho\in \mathcal{A}_\infty$.
\end{enumerate}
\end{proposition}
Proof:  (A) Let us calculate the commutator  $[\bphi(\rho_\epsilon,x_1), \bphi(\rho_\epsilon,x_2)]$ using the abbreviated form \eqref{embo:5} for the field operators.  As with the formal calculations we get a sum of four commutators
\begin{align}
\label{comr:9}
        [\bphi_0(\rho_\epsilon,x_1) ,
         \bphi_0(\rho_\epsilon,x_2)] 
          &=  [\mathbf{a}^-(\rho_\epsilon,x_1)
          ~,~ \mathbf{a}^-(\rho_\epsilon,x_2)]\\
\label{comr:10}
      &+  [\mathbf{a}^+(\rho_\epsilon,x_1)
     ~,~ \mathbf{a}^+(\rho_\epsilon,x_2)] \\
\label{comr:11}
      &+ [\mathbf{a}^-(\rho_\epsilon,x_1)
      ~,~ \mathbf{a}^+(\rho_\epsilon,x_2)]\\
\label{comr:12}
      &+ [\mathbf{a}^+(\rho_\epsilon,x_1)
      ~,~ \mathbf{a}^-(\rho_\epsilon,x_2)] ,
\end{align}
and the two commutators (\ref{comr:9}--\ref{comr:10}) are again identically zero because of (\ref{fock:10}--\ref{fock:11}).  On the other hand, \eqref{fock:12} and \eqref{norm:12} imply that the commutators (\ref{comr:11}--\ref{comr:12}) give
\begin{align}
\nonumber
        [\bphi_0(\rho_\epsilon,x_1) ,
         \bphi_0(\rho_\epsilon,x_2)]
         \DEF \Delta&_{\text{JP},\mathcal{G}}(x_1,x_2)\\
\label{comr:13}
          = \Delta&_\mathcal{G}(x_1,x_2) - \Delta_\mathcal{G}(x_2,x_1),
\end{align}
where  
\begin{align}
\nonumber
       \Delta_\mathcal{G}(x_1,x_2) 
 &\DEF       \iiint d^3p_1~\FOU{\rho}^*(\epsilon \vec{p}_1)
          \iiint d^3p_2~\FOU{\rho}  (\epsilon \vec{p}_2)\\
\nonumber
 &\times  \frac{1}{(2\pi)^3} \frac{(E_{p_1}+E_{p_2})}{4E_{p_1}E_{p_2}}
   ~ \exp \Bigl( i(E_{p_1}-E_{p_2})t_\xi  -i(E_{p_1}t_1-E_{p_2}t_2) \Bigr)\\
\label{comr:14}
 &\times  \exp i(\vec{p}_1\cdot \vec{x}_1 - \vec{p}_2\cdot \vec{x}_2 \,)
        ~ \frac{1}{\epsilon^3}
          \chi \bigl(\frac{\vec{p}_1-\vec{p}_2}{\epsilon}\bigr).
\end{align}
This function has the important property that if we set $t_1=t_2=t_\xi$ its time-dependence vanishes.  In that case the assumptions $\rho \in \mathcal{A}_\infty$ and $\FOU{\chi} \in \mathcal{B}_\infty$, which imply that $\FOU{\rho}^* = \FOU{\rho}^\vee$ and $\chi^* = \chi^\vee$, have the consequence that \eqref{comr:14} is a real function.  Indeed, taking its complex conjugate and making the change of variables $\vec{p}_1 \rightarrow -\vec{p}_1$ and $ \vec{p}_2 \rightarrow -\vec{p}_2$ is then an identity.  Moreover, interchanging $\vec{x}_1$ and  $\vec{x}_2$ and making the same change of variables is also an identity.  Thus
\begin{align}
\label{comr:15}
      \Delta_\mathcal{G}(x_1,x_2)\Bigr|_{t_1=t_2=t_\xi}
    - \Delta_\mathcal{G}(x_2,x_1)\Bigr|_{t_1=t_2=t_\xi} =0,
\end{align}
and consequently \eqref{comr:13} implies that
\begin{align}
\label{comr:16}
        [\bphi_0(\rho_\epsilon,x_1) ,
         \bphi_0(\rho_\epsilon,x_2)]\Bigr|_{t_1=t_2=t_\xi} = 0,
\end{align}
which proves the first commutation relation, i.e., \eqref{comr:1}.  Similarly, differentiating \eqref{comr:14} with respect to $t_1$ and $t_2$ we find that
\begin{align}
\label{comr:17}
      \frac{\partial}{\partial t_1}\frac{\partial}{\partial t_2}
      \Delta_\mathcal{G}(x_1,x_2)\Bigr|_{t_1=t_2=t_\xi}
    = \frac{\partial}{\partial t_1}\frac{\partial}{\partial t_2}
      \Delta_\mathcal{G}(x_2,x_1)\Bigr|_{t_1=t_2=t_\xi} ,
\end{align}
and thus
\begin{align}
\nonumber
        [\bpi_0(\rho_\epsilon,&x_1) ,
         \bpi_0(\rho_\epsilon,x_2)]\Bigr|_{t_1=t_2=t_\xi}\\
\label{comr:18}
    &=   \frac{\partial}{\partial t_1}\frac{\partial}{\partial t_2}
       [\bphi_0(\rho_\epsilon,x_1) ,
         \bphi_0(\rho_\epsilon,x_2)]\Bigr|_{t_1=t_2=t_\xi} = 0,
\end{align}
which proves the second commutation relation, i.e., \eqref{comr:2}.  To prove the third commutation relation we need the $t_2$ derivative of \eqref{comr:14}, i.e.,
\begin{align}
\nonumber
      \frac{\partial}{\partial t_2}
      \Delta_\mathcal{G}(x_1,x_2)\Bigr|_{t_1=t_2=t_\xi}
 &=  i \iiint d^3p_1~\FOU{\rho}^*(\epsilon \vec{p}_1)
       \iiint d^3p_2~\FOU{\rho}  (\epsilon \vec{p}_2)\\
\nonumber
 &\times  \frac{1}{(2\pi)^3} \frac{(E_{p_1}+E_{p_2})}{4E_{p_1}}
          \exp i (\vec{p}_1\cdot \vec{x}_1 -\vec{p}_2\cdot \vec{x}_2)\\
\label{comr:19}
 &\times  \frac{1}{\epsilon^3}
          \chi\bigl(\frac{\vec{p}_1-\vec{p}_2}{\epsilon}\bigr).
\end{align}
This derivative is pure imaginary, as can be seen by making the substitutions $\vec{p}_1 \rightarrow -\vec{p}_1$ and $\vec{p}_2 \rightarrow -\vec{p}_2$ in its complex conjugate, and using $\FOU{\rho}^* = \FOU{\rho}^\vee$ and $\chi^* = \chi^\vee$.  Thus
\begin{align}
\label{comr:20}
      \frac{\partial}{\partial t_2}
      \Delta_\mathcal{G}(x_1,x_2)\Bigr|_{t_1=t_2=t_\xi}
      = \frac{i}{2}~ \delta^3(\rho_\epsilon,\chi_\epsilon,\vec{x}_1,\vec{x}_2),
\end{align}
where $\delta^3(\rho_\epsilon,\chi_\epsilon,\vec{x}_1,\vec{x}_2) \in \mathbb{R}$ corresponds to \eqref{comr:4}, and therefore
\begin{align}
\nonumber
        [\bphi_0(\rho_\epsilon,&x_1) ,
         \bpi_0  (\rho_\epsilon,x_2)]\Bigr|_{t_1=t_2=t_\xi}\\
\label{comr:21}
    &=   \frac{\partial}{\partial t_2}
        [\bphi_0(\rho_\epsilon,x_1) ,
         \bphi_0(\rho_\epsilon,x_2)]\Bigr|_{t_1=t_2=t_\xi}\\
\label{comr:22}
    &= 2 \frac{\partial}{\partial t_2}
         \Delta_\mathcal{G}(x_1,x_2)\Bigr|_{t_1=t_2=t_\xi}
     = i ~ \delta^3(\rho_\epsilon,\chi_\epsilon,\vec{x}_1,\vec{x}_2),
\end{align}
 so that \eqref{comr:3} is proved.

(B) Since $\rho \in \mathcal{A}_\infty$ and $\FOU{\chi} \in \mathcal{B}_\infty$ we can use Proposition \ref{moll:prop:2} to do one of the $p$-integrations in \eqref{comr:14}.  This has been done in Section \ref{norm:0} to reduce \eqref{norm:12} to \eqref{norm:18}, which in the present notation is
\begin{align}
\label{comr:23}
    \Delta_\mathcal{G}(x_1,x_2)
   =  \Delta_\epsilon(\FOU{\rho^\vee\ast\rho}, x_1-x_2)
   +  \OOO_{t_\xi,x_1,x_2}(\epsilon^{q-1}),
\end{align}
where we have written $\OOO_{t_\xi,x_1,x_2}(\epsilon^{q-1})$ to remind that taking into account the discussion below \eqref{comr:14} the remainder is symmetric in an interchange of $x_1$ and $x_2$ if  $t_1=t_2=t_\xi$.  Therefore,  we still have
\begin{align}
\label{comr:24}
      \Delta_\mathcal{G}(x_1,x_2)\Bigr|_{t_1=t_2=t_\xi}
    - \Delta_\mathcal{G}(x_2,x_1)\Bigr|_{t_1=t_2=t_\xi}
    = 0,
\end{align}
and consequently \eqref{comr:13} implies that
\begin{align}
\label{comr:25}
        [\bphi_0(\rho_\epsilon,x_1) ,
         \bphi_0(\rho_\epsilon,x_2)]\Bigr|_{t_1=t_2=t_\xi}
       = 0,
\end{align}
which proves the first commutation relation, i.e., \eqref{comr:5}.  Similarly, differentiating \eqref{comr:23} with respect to $t_1$ and $t_2$ we find that
\begin{align}
\label{comr:26}
        [\bpi_0(\rho_\epsilon,x_1) ,
         \bpi_0(\rho_\epsilon,x_2)]\Bigr|_{t_1=t_2=t_\xi}
       = 0,
\end{align}
which proves the second commutation relation, i.e., \eqref{comr:6}.  To prove the third one we need the $t_2$ derivative of \eqref{comr:23} at $t_1=t_2=t_\xi$.  Thus we calculate
\begin{align}
\nonumber
      \frac{\partial}{\partial t_2}
      \Delta_\epsilon(\FOU{\rho^\vee\ast\rho}, x_1-x_2)\Bigr|_{t_1=t_2=t_\xi}
     &=   \frac{i}{(2\pi)^3} \iiint \frac{d^3p}{2} \FOU{\rho^\vee\ast\rho}(\epsilon\vec{p})
     \exp i\vec{p}\cdot (\vec{x}_1 - \vec{x}_2\,)\\
\label{comr:27}
     &= \frac{i}{2}  \delta^3_\epsilon(\rho^\vee\ast\rho,\vec{x}_1-\vec{x}_2),
\end{align}
where by inverse Fourier transform
\begin{align}
\label{comr:28}
   \delta^3_\epsilon(\rho^\vee\ast\rho,\vec{x}_1-\vec{x}_2)
     \DEF  \frac{1}{\epsilon^3}(\rho^\vee \ast \rho)
         \bigl(\frac{\vec{x}_2-\vec{x}_1}{\epsilon}\bigr).
\end{align}
in which by Proposition \ref{moll:prop:1} the convoluted mollifier $\rho^\vee\ast\rho\in \mathcal{A}_\infty$.  Proceeding as in \eqref{comr:22} we get finally
\begin{align}
\label{comr:29}
        [\bphi_0(\rho_\epsilon,&x_1) ,
         \bpi_0  (\rho_\epsilon,x_2)]\Bigr|_{t_1=t_2=t_\xi}
     = i \delta^3_\epsilon(\rho^\vee\ast\rho,\vec{x}_1-\vec{x}_2)
     + \OOO(\epsilon^{q-1}),
\end{align}
so that \eqref{comr:7} is proved.  \END

This leads to two remarks:
\begin{itemize}

\item It is remarkable that despite their dependence upon $\rho$ and $\FOU{\chi}$ the first two commutation relations are satisfied identically.  Otherwise there would be serious problems with the physical interpretation of QFT in the $\mathcal{G}$-setting.

\item In contradistinction to the formal commutation relations (\ref{comf:1}--\ref{comf:3}) one has here to explicitly formulate the equal-time condition as $t_1=t_2=t_\xi$, where $t_\xi$ is the unassigned time-variable of the states (which in the Heisenberg picture do not dependent on time).  This is however not a supplementary condition:  It simply turns out that in the classical formulation the statement $t_1=t_2$ is sufficient because the $\delta$-function replacing the mollifier $\chi$ in \eqref{comr:14} implies that there is no $t_\xi$ dependence even if $t_1\neq t_2$.

\end{itemize}

\section{Hamiltonian: formal calculations}
%-----------------------------------------
\label{hamf:0}
\setcounter{equation}{0}
\setcounter{definition}{0}
\setcounter{axiom}{0}
\setcounter{conjecture}{0}
\setcounter{lemma}{0}
\setcounter{theorem}{0}
\setcounter{corollary}{0}
\setcounter{proposition}{0}
\setcounter{example}{0}
\setcounter{remark}{0}
\setcounter{problem}{0}

The Hamiltonian of a free field, defined by setting $g=0$ in \eqref{summ:14}, is
\begin{align}
\notag
       \mathbf{H}_0(t) = \iiint_{\mathbb{R}^3} \Bigl\{
         \frac{1}{2}& \bigl({\bpi}_0(t,\vec{x}\,)\bigr)^2 
       + \frac{1}{2} \sum_{1 \leq \mu \leq 3}
         \bigl(\partial_\mu {\bphi}_0(t,\vec{x}\,)\bigr)^2 \\
\label{hamf:1}  
       + \frac{1}{2}& m^2
         \bigl({\bphi}_0(t,\vec{x}\,)\bigr)^2  \Bigr\} ~d^3x. 
\end{align}
In this section we calculate this operator using the definition \eqref{stat:18} of the field operator, which for the sake of clarity we rewrite using the abbreviations
\begin{equation}\label{hamf:2}
  A^+ =  \psi_p^-(x) \mathbf{a}_p^+(\xi),
        \qquad \text{and} \qquad 
  A^- =  \psi_p^+(x) \mathbf{a}_p^-(\xi), 
\end{equation}
where $\psi_p^-=\psi_p^*(x)$ and $\psi_p^+=\psi_p(x)$, so that
\begin{align}
\label{hamf:3}
    \bphi(x) &= \iiint d^3p~     
       ( A^+ + A^- )  ,\\
\label{hamf:4}
    \bpi (x) &= \iiint d^3p~      
       ( A^+ - A^- ) (+iE_p),\\
\label{hamf:5}
   \vec{\nabla} \bphi(x) &= \iiint d^3p~    
       ( A^+ - A^- ) (-i\vec{p}\,).
\end{align}
The Hamiltonian operator is then
\begin{align}
\nonumber
  \mathbf{H}_0 = \frac{1}{2}& \iiint d^3x  \iiint d^3p_1 \iiint d^3p_2 \\
  \nonumber 
  \times \Bigl( ~~~ 
   &(+A_1^+A_2^+ + A_1^+A_2^- + A_1^-A_2^+ + A_1^-A_2^-) ~ m^2\\
\nonumber 
  +&(-A_1^+A_2^+ + A_1^+A_2^- + A_1^-A_2^+ - A_1^-A_2^-) ~ E_{p_1} E_{p_2}\\
\label{hamf:6}
  +&(-A_1^+A_2^+ + A_1^+A_2^- + A_1^-A_2^+ - A_1^-A_2^-)
                                          ~ \vec{p}_1\cdot\vec{p}_2 ~~~\Bigr).
\end{align}
{\emph{Assuming} that the integration over $x$ can be done first, we get integrals of the type
\begin{equation}\label{hamf:7}
   \iiint d^3x~A_1^{\pm}A_2^{\pm} = \mathbf{a}_1^{\pm}\mathbf{a}_2^{\pm}
   \iiint d^3x~ \psi_{p_1}^{\mp}( x)\psi_{p_2}^{\mp}( x),
\end{equation}
where with the definition \eqref{stat:7} of the $\delta$-function we have
\begin{equation}\label{hamf:8}
   \iiint d^3x~\psi_{p_1}^{\pm}( x)\psi_{p_2}^{\pm}( x)
   = \frac{\exp i(\mp E_{p_1} \mp E_{p_2} )t }
          {2\sqrt{E_{p_1} E_{p_2}}} \delta^3(\vec{p}_1+\vec{p}_2),
\end{equation}
and 
\begin{equation}\label{hamf:9}
   \iiint d^3x~\psi_{p_1}^{\pm}( x)\psi_{p_2}^{\mp}( x)
   = \frac{\exp i(\mp E_{p_1} \pm E_{p_2} )t}
          {2\sqrt{E_{p_1} E_{p_2}}} \delta^3(\vec{p}_1-\vec{p}_2).
\end{equation}
The $\delta$-functions enable to perform one of the momentum integrals trivially, so that
\begin{align}
\nonumber
  \mathbf{H}_0 = \frac{1}{2} \iiint  \frac{d^3p}{2E_p}       \Bigl( ~~~
   &(+e^+\mathbf{a}_p^+\mathbf{a}_p^+ + \mathbf{a}_p^+\mathbf{a}_p^- +
      \mathbf{a}_p^-\mathbf{a}_p^+ + e^-\mathbf{a}_p^-\mathbf{a}_p^-) ~ m^2\\
\nonumber 
  +&(-e^+\mathbf{a}_p^+\mathbf{a}_p^+ + \mathbf{a}_p^+\mathbf{a}_p^- +
      \mathbf{a}_p^-\mathbf{a}_p^+ - e^-\mathbf{a}_p^-\mathbf{a}_p^-)
                                                          ~ E_p^2\\
\label{hamf:10}
  +&(+e^+\mathbf{a}_p^+\mathbf{a}_p^+ + \mathbf{a}_p^+\mathbf{a}_p^- + 
      \mathbf{a}_p^-\mathbf{a}_p^+ + e^-\mathbf{a}_p^-\mathbf{a}_p^-)
                                          ~ \vec{p}\cdot\vec{p} ~~~\Bigr),
\end{align}
where $e^\pm = \exp(\pm2iE_pt)$, and where the two sign changes in the third line are due to \eqref{hamf:8} implying that $\vec{p}_1 = -\vec{p}_2$ whereas \eqref{hamf:9} implies that $\vec{p}_1 = \vec{p}_2$.  We now use the definition of $E_p$ to remark that
\begin{equation}\label{hamf:11}
     m^2 - E_p^2 + \vec{p}\cdot\vec{p} = 0,
     \qquad \text{and} \qquad
     m^2 + E_p^2 + \vec{p}\cdot\vec{p} = 2 E_p^2,
\end{equation}
so that the sums of like-sign terms are zero, whereas the opposite-sign terms give the final result
\begin{align}
\nonumber
  \mathbf{H}_0 &= \frac{1}{2} \iiint d^3p ~ E_p
          (\mathbf{a}_p^+\mathbf{a}_p^- + \mathbf{a}_p^-\mathbf{a}_p^+)\\
 \label{hamf:12}
             &=  \iiint d^3p ~ E_p\Bigl(\mathbf{a}_p^+\mathbf{a}_p^-
               + \frac{1}{2}[\mathbf{a}_p^-,\mathbf{a}_p^+] \Bigr),
 \end{align}
where the commutator is the infinite number $\delta^3(0) = (2\pi)^{-3} \iiint d^3\xi$ because
from \eqref{fock:12} and \eqref{stat:6}
\begin{align}
\label{hamf:13}
    [\mathbf{a}_{p}^-(\xi), \mathbf{a}_{p'}^+(\xi)]
  = \BRA \psi_{p}^{ }(\xi) \| \psi_{p'}^{ }(\xi) \KET
  = \delta^3(\vec{p}-\vec{p'}).
%  = \frac{1}{} \iiint d^3\xi \e^{ i\vec{\xi} \cdot (\vec{p}-\vec{p'})}.
\end{align}
The total energy operator $\mathbf{H}_0$ is therefore badly divergent.  More precisely, if $\mathbf{H}_0$ is operating on any element of the Fock space its eigenvalue will be a sum containing at least the contribution from its action on the vacuum state $\mho$, that is the infinite `zero-point' energy (also called `null-point' or `vacuum' energy) 
\begin{align}
\label{hamf:14}
 E_\mho =\frac{1}{2} \iiint d^3p~ E_p [\mathbf{a}_p^-,\mathbf{a}_p^+]
        =\frac{1}{2} \iiint d^3p~ E_p \frac{1}{(2\pi)^3} \iiint d^3\xi = \infty,
% more:
%  = \frac{4\pi}{2(2\pi)^3} \iiint_0^\infty dp~ p^2 \sqrt{m^2+p^2}.
\end{align}
which is twice infinite because both the $\xi$ and $p$ integrals are diverging.  In fact, looking at the details of the above calculation, if neither $p$-integration is made after the $x$-integration, one sees that the zero-point energy can be put in the form
\begin{align}
\label{hamf:15}
 E_\mho =\frac{1}{2} \iiint d^3p_1 \iiint d^3p_2 
    ~ \delta^3(\vec{p}_1-\vec{p}_2)
    ~ \delta^3(\vec{p}_1-\vec{p}_2)
    ~        E(\vec{p}_1,\vec{p}_2) = \infty,
\end{align}
where the first $\delta$-function comes from the $x$-integration \eqref{hamf:9} and the second one from the $\xi$-integration in the commutator \eqref{hamf:13}, and where\footnote{In this expression a factor $\exp i(E_{p_1} - E_{p_2} )(t - t_\xi)$ coming from \eqref{stat:6} and \eqref{hamf:9} is ignored because in the standard formulation such exponentials are set equal to one since they disappear when the corresponding expressions are evaluated on test functions.} 
\begin{align}
\label{hamf:16}
E(\vec{p}_1,\vec{p}_2) &= \frac{ (E_{p_1}E_{p_2}
             + \vec{p}_1\cdot\vec{p}_2 + m^2)(E_{p_1}  + E_{p_2}) }
                               { 4 E_{p_1} E_{p_2} }
%%             \exp i(- E_{p_1} + E_{p_2} )(t_\xi - t),
\end{align}
reduces to $E_p$ when $\vec{p}_1=\vec{p}_2=\vec{p}$.
\begin{remark}[Elimination of zero-point energy]
%...............................................
\label{hamf:rema:1}
In practice, the zero-point energy contribution (which is an infinite but constant energy shift) is simply ignored, and the Hamiltonian operator of the free field is redefined as
\begin{align}
\label{hamf:17}
  \mathbf{H}_0 \DEF \iiint d^3p ~ E_p \mathbf{a}_p^+(\xi)\mathbf{a}_p^-(\xi).
 \end{align}
\end{remark}
It remains to verify that $\mathbf{H}_0$ does correspond to the quantum-mechanical total energy operator.  We therefore calculate its effect on a multiparticle state $f_n(\xi_1,...,\xi_n)$, which by the definitions of the creation and annihilation operators is\footnote{The symbol $^\times$ means that this label is skipped in the symmetrization process.}
\begin{align}
\label{hamf:18}
  \mathbf{H}_0~f_n =  \iiint d^3p ~ E_p  \sum_{j=1}^n \psi_p(\xi_j)
     \BRA \psi_p(\xi) \| f_n(\xi,...,\xi_j^\times,...,\xi_n) \KET.
 \end{align}
Considering to begin with the single particle state $f_1(\xi_1)=f_0\Delta_+(\xi_1-x_1)$ and using formula \eqref{stat:9}, i.e.,
\begin{align}
\label{hamf:19}
       \BRA \psi_p(\xi) \| \Delta_+(\xi-x_1) \KET 
      &= \psi_p(-x_1),
\end{align}
it follows from the definition of $\Delta_+$ that
\begin{align}
\label{hamf:20}
 \mathbf{H}_0~f_1 &=  \iiint d^3p ~ E_p   \psi_p(\xi_1)
     \BRA \psi_p(\xi) \| f_0\Delta_+(\xi-x_1) \KET\\
\label{hamf:21}
    & = \iiint d^3p ~ E_p f_0 \psi_p(\xi_1)\psi_p(-x_1)\\
 \label{hamf:22}
    & = i\frac{\partial}{\partial t_{\xi_1}} f_0\Delta_+(\xi_1-x_1).
\end{align}
Therefore, the single particle state $f_1(\xi_1)$ has been `reconstructed' and $\mathbf{H}_0~f_1 = i{\partial}/{\partial ~t_{\xi_1}} ~ f_1$.  Because multiparticle states are simply symmetrized linear super\-positions of products of single-particle states this immediately generalizes to the identity
\begin{align}
\label{hamf:23}
 \mathbf{H}_0(t) ~f_n = i \sum_{j=1}^n\frac{\partial}{\partial t_{\xi_j}} ~f_n,
\end{align}
which confirms that $\mathbf{H}_0$ is indeed the quantum-mechanical total energy operator.

\section{Hamiltonian: rigorous calculations}
%-------------------------------------------
\label{hamr:0}
\setcounter{equation}{0}
\setcounter{definition}{0}
\setcounter{axiom}{0}
\setcounter{conjecture}{0}
\setcounter{lemma}{0}
\setcounter{theorem}{0}
\setcounter{corollary}{0}
\setcounter{proposition}{0}
\setcounter{example}{0}
\setcounter{remark}{0}
\setcounter{problem}{0}

Using the rigorous definitions (\ref{embo:5}--\ref{embo:7}) of the field operator, we define the $\mathcal{G}$-embedding of the free field Hamiltonian \eqref{hamf:1} as the distribution
\begin{align}
\notag
    \mathbf{H}_0(\rho_\epsilon,\FOU{\chi}_\epsilon,t)
  = \iiint_{\mathbb{R}^3} \Bigl\{
    \frac{1}{2}& \bigl({\bpi}_0(\rho_\epsilon,t,\vec{x}\,)\bigr)^2 
  + \frac{1}{2} \sum_{1 \leq \mu \leq 3}
    \bigl(\partial_\mu {\bphi}_0(\rho_\epsilon,t,\vec{x}\,)\bigr)^2 \\
\label{hamr:1}  
       + \frac{1}{2}& m^2
         \bigl({\bphi}_0(\rho_\epsilon,t,\vec{x}\,)\bigr)^2  \Bigr\}
         \FOU{\chi}(\epsilon \vec{x}\,) ~d^3x, 
\end{align}
where the suitable damper $\FOU{\chi}(\epsilon \vec{x}\,) \in \mathcal{B}_\infty$ corresponds to the test function.  Mathematically $\FOU{\chi}(\epsilon \vec{x}\,)$ insures that integrating over the whole of $\mathbb{R}^3$ makes sense so that $\mathbf{H}_0(\rho_\epsilon,\FOU{\chi}_\epsilon,t)$ is a \emph{generalized linear operator} on $\mathbb{D}$ for fixed $t$.  Form a physical point of view, $\FOU{\chi}(\epsilon \vec{x}\,)$ is restricting integration over the spatial coordinates in space-time to a large but finite volume as in the scalar product \eqref{norm:1}, which is why consistency requires that $\FOU{\chi}(\epsilon \vec{x}\,)$ in \eqref{hamr:1} is the same as in \eqref{norm:1}.\footnote{A restriction to a finite volume is generally also introduced in the standard formulation of QFT.  In the present formulation this volume is defined by the support of $\FOU{\chi}$.}

   To follow the formal calculations as closely as possible we include the $\FOU{\rho}$ dampers in the abbreviations \eqref{hamf:2}, i.e.,
\begin{equation}
\nonumber%\label{hamr:2??}
  A^+ =  \FOU{\rho}  (\epsilon \vec{p}\,) \psi_p^-(x) \mathbf{a}_p^+(\xi),
        \qquad \text{and} \qquad 
  A^- =  \FOU{\rho}^*(\epsilon \vec{p}\,) \psi_p^+(x) \mathbf{a}_p^-(\xi), 
\end{equation}
so that we still have
\begin{align}
\label{hamr:2}
    \bphi(x) &= \iiint d^3p~  
                            ( A^+ + A^- ),\\
\label{hamr:3}
    \bpi (x) &= \iiint d^3p~  
                            ( A^+ - A^- ) (+iE_p),\\
\label{hamr:4}
   \vec{\nabla} \bphi(x) &= \iiint d^3p~  
                            ( A^+ - A^- ) (-i\vec{p}\,),
\end{align}
while because of $\FOU{\chi}$ the Hamiltonian operator becomes
\begin{align}
\nonumber
  \mathbf{H}_0 = \frac{1}{2}&
               \iiint d^3x  ~\FOU{\chi} (\epsilon \vec{x}\,)
               \iiint d^3p_1
               \iiint d^3p_2 \\
  \nonumber 
  \times \Bigl( ~~~ 
   &(+A_1^+A_2^+ + A_1^+A_2^- + A_1^-A_2^+ + A_1^-A_2^-) ~ m^2\\
\nonumber 
  +&(-A_1^+A_2^+ + A_1^+A_2^- + A_1^-A_2^+ - A_1^-A_2^-) ~ E_{p_1} E_{p_2}\\
\label{hamr:5}
  +&(-A_1^+A_2^+ + A_1^+A_2^- + A_1^-A_2^+ - A_1^-A_2^-)
                                          ~ \vec{p}_1\cdot\vec{p}_2 ~~~\Bigr).
\end{align}
Since we are in the $\mathcal{G}$-setting the integrations can be done in any  order.  Thus, doing the integration over $x$ first, we get integrals of the type
\begin{equation}\label{hamr:6}
   \iiint d^3x~\FOU{\chi} (\epsilon \vec{x}\,)A_1^{\pm}A_2^{\pm}
 = \FOU{\rho}^\pm(\epsilon \vec{p}_1)
   \FOU{\rho}^\pm(\epsilon \vec{p}_2)
   \mathbf{a}_1^{\pm}\mathbf{a}_2^{\pm}
   \iiint d^3x~\psi_{p_1}^{\mp}( x)\psi_{p_2}^{\mp}( x)
   \FOU{\chi} (\epsilon \vec{x}\,),
\end{equation}
where for convenience $\FOU{\rho}^+=\FOU{\rho}$ and $\FOU{\rho}^-=\FOU{\rho}^*$, and where using \eqref{norm:3} we have
\begin{equation}\label{hamr:7}
   \iiint d^3x~\psi_{p_1}^{\pm}( x)\psi_{p_2}^{\pm}( x)
               \FOU{\chi} (\epsilon \vec{x}\,)
   = \frac{\exp i(\mp E_{p_1} \mp E_{p_2} )t }
          {2\sqrt{E_{p_1} E_{p_2}}} ~
     \frac{1}{\epsilon^3} \chi(\frac{\pm\vec{p}_1\pm\vec{p}_2}{\epsilon}),
\end{equation}
and 
\begin{equation}\label{hamr:8}
   \iiint d^3x~\psi_{p_1}^{\pm}( x)\psi_{p_2}^{\mp}( x)
              ~\FOU{\chi} (\epsilon \vec{x}\,)
   = \frac{\exp i(\mp E_{p_1} \pm E_{p_2} )t}
          {2\sqrt{E_{p_1} E_{p_2}}} ~
     \frac{1}{\epsilon^3} \chi(\frac{\pm\vec{p}_1 \mp\vec{p}_2}{\epsilon}).
\end{equation}
Thus, contrary to the corresponding formal equations (\ref{hamf:8}--\ref{hamf:9}), we do not get the $\delta$-functions which enabled to make one of the momentum integrations trivially.  Nevertheless, using the formal calculations as a guide, we expect that we can make use of the kinematical constraints \eqref{hamf:11} in some appropriate limit.  We therefore rewrite the Hamiltonian \eqref{hamr:1} as
\begin{align}
\label{hamr:9}
    \mathbf{H}_0 = \mathbf{H}_{++} + \mathbf{H}_{+-}
                 + \mathbf{H}_{-+} + \mathbf{H}_{--} ,
\end{align}
where each term corresponds to the sum of a column of the matrix defined by \eqref{hamr:5}, that is:
\begin{align}
\nonumber
    \mathbf{H}_{++} &= \frac{1}{2} \iiint d^3p_1 \iiint d^3p_2
              ~\FOU{\rho}(\epsilon \vec{p}_1)
              ~\FOU{\rho}(\epsilon \vec{p}_2) 
          ~\frac{1}{\epsilon^3} \chi(\frac{+\vec{p}_1+\vec{p}_2}{\epsilon})\\
\label{hamr:10} 
    &\times \bigl(-E_{p_1}E_{p_2} - \vec{p}_1\cdot\vec{p}_2 +m^2\bigr)
            \frac{\exp i(+ E_{p_1} + E_{p_2} )t }
                 {2\sqrt{E_{p_1} E_{p_2}}}
              ~\mathbf{a}_{p_1}^+(\xi)
              ~\mathbf{a}_{p_2}^+(\xi),
\end{align}
\begin{align}
\nonumber
    \mathbf{H}_{--} &= \frac{1}{2} \iiint d^3p_1 \iiint d^3p_2
              ~\FOU{\rho}^*(\epsilon \vec{p}_1)
              ~\FOU{\rho}^*(\epsilon \vec{p}_2) 
          ~\frac{1}{\epsilon^3} \chi(\frac{-\vec{p}_1-\vec{p}_2}{\epsilon})\\
\label{hamr:11} 
    &\times \bigl(-E_{p_1}E_{p_2} - \vec{p}_1\cdot\vec{p}_2 +m^2\bigr)
            \frac{\exp i(- E_{p_1} - E_{p_2} )t }
                 {2\sqrt{E_{p_1} E_{p_2}}}
              ~\mathbf{a}_{p_1}^-(\xi)
              ~\mathbf{a}_{p_2}^-(\xi),
\end{align}
\begin{align}
\nonumber
    \mathbf{H}_{+-} &= \frac{1}{2} \iiint d^3p_1 \iiint d^3p_2
              ~\FOU{\rho}  (\epsilon \vec{p}_1)
              ~\FOU{\rho}^*(\epsilon \vec{p}_2) 
          ~\frac{1}{\epsilon^3} \chi(\frac{+\vec{p}_1-\vec{p}_2}{\epsilon})\\
\label{hamr:12} 
    &\times \bigl(+E_{p_1}E_{p_2} + \vec{p}_1\cdot\vec{p}_2 +m^2\bigr)
            \frac{\exp i(+ E_{p_1} - E_{p_2} )t }
                 {2\sqrt{E_{p_1} E_{p_2}}}
              ~\mathbf{a}_{p_1}^+(\xi)
              ~\mathbf{a}_{p_2}^-(\xi),
\end{align}
\begin{align}
\nonumber
    \mathbf{H}_{-+} &= \frac{1}{2} \iiint d^3p_1 \iiint d^3p_2
              ~\FOU{\rho}^*(\epsilon \vec{p}_1)
              ~\FOU{\rho}  (\epsilon \vec{p}_2) 
          ~\frac{1}{\epsilon^3} \chi(\frac{-\vec{p}_1+\vec{p}_2}{\epsilon})\\
\label{hamr:13} 
    &\times \bigl(+E_{p_1}E_{p_2} + \vec{p}_1\cdot\vec{p}_2 +m^2\bigr)
            \frac{\exp i(- E_{p_1} + E_{p_2} )t }
                 {2\sqrt{E_{p_1} E_{p_2}}}
              ~\mathbf{a}_{p_1}^-(\xi)
              ~\mathbf{a}_{p_2}^+(\xi).
\end{align}
In the following we calculate these terms one after the other.  It will be seen that the kinematical constraints play the same role as in the formal calculations, and that with the help of Proposition \ref{moll:prop:2} the mollifiers $\chi_\epsilon$ will take the role of the $\delta$-functions.

\subsection{Calculation of $\mathbf{H}_{++}$}
%...............................................
The operator $\mathbf{a}_{p_1}^+\mathbf{a}_{p_2}^+$ acting on the state $f_n$ gives a state $f_{n+2}$ with $n+2$ particles.  The definition \eqref{fock:7} gives
\begin{align}
  \mathbf{a}_{p_1}^+(\xi)\mathbf{a}_{p_2}^+(\xi)~ f_n
   =  \sqrt{(n+2)(n+1)} \Bigl( \sum_{k=1}^{n+2}   
      \psi_{p_1}(\xi_k) \Bigl( \sum_{j=1}^{n+1} \psi_{p_2}(\xi_j) ~ f_n
                        \Bigr)\Bigr),
\label{hamr:14} 
\end{align}
where the big parentheses suggests that the two sums are nested in a way that is not simple to write down.  But since we are only interested in the form of a general term in that expression we further simplify it by defining
\begin{align}
  \sum =  \sqrt{(n+2)(n+1)} \sum_{k=1}^{n+2}  \sum_{j=1}^{n+1}. 
\label{hamr:15} 
\end{align}
Then, since
\begin{align}
\nonumber
  \psi_{p_1}(\xi_k) \psi_{p_2}(\xi_j) &= 
  \frac{1}{2(2\pi)^3\sqrt{E_{p_1}E_{p_2}}}
             \exp i(-E_{p_1}t_{\xi_k}-E_{p_2}t_{\xi_j})\\
\label{hamr:16} 
     &\times \exp i(\vec{p}_1\cdot\vec{p}_k +\vec{p}_2\cdot\vec{p}_j),
\end{align}
the $\mathbf{H}_{++}$ operator \eqref{hamr:10} is
\begin{align}
\nonumber
    \mathbf{H}_{++} &= \frac{1}{2} \sum \iiint d^3p_1 \iiint d^3p_2
              ~\FOU{\rho}(\epsilon \vec{p}_1)
              ~\FOU{\rho}(\epsilon \vec{p}_2)\\
\nonumber
    &\times \exp i\Bigl( E_{p_1}(t-t_{\xi_k}) + E_{p_2}(t-t_{\xi_j})
         +(\vec{p}_1\cdot\vec{p}_k +\vec{p}_2\cdot\vec{p}_j) \Bigr)\\
\label{hamr:17} 
    &\times \frac{-E_{p_1}E_{p_2} - \vec{p}_1\cdot\vec{p}_2 +m^2}
                 {4(2\pi)^3E_{p_1} E_{p_2}}
     \frac{1}{\epsilon^3} \chi(\frac{\vec{p}_1+\vec{p}_2}{\epsilon}),
\end{align}
which in order to use Proposition \ref{moll:prop:2} we rewrite as
\begin{align}
\nonumber
  \mathbf{H}_{++} = \frac{1}{2} \sum
                          &\iiint d^3p_1 ~\FOU{\rho}(\epsilon \vec{p}_1)\\
\label{hamr:18}
                   \times &\iiint d^3p_2 ~ \FOU{\rho}(\epsilon \vec{p}_2)~
                               h(\vec{p}_1,\vec{p}_2)
     \frac{1}{\epsilon^3} \chi(\frac{\vec{p}_1+\vec{p}_2}{\epsilon}),
\end{align}
where
\begin{align}
\nonumber
    h(\vec{p}_1,\vec{p}_2) 
    &=     \frac{-E_{p_1}E_{p_2} - \vec{p}_1\cdot\vec{p}_2 +m^2}
                {4(2\pi)^3E_{p_1} E_{p_2}}\\
\label{hamr:19} 
    &\times \exp i\Bigl( E_{p_1}(t-t_{\xi_k}) + E_{p_2}(t-t_{\xi_j})
         +(\vec{p}_1\cdot\vec{p}_k +\vec{p}_2\cdot\vec{p}_j) \Bigr).
\end{align}
Since $E_p = \sqrt{m^2 + |\vec{p}|^2}$ it follows that $h(\vec{p}_1,\vec{p}_2) \in \mathcal{O}_{\text{M}}(\mathbb{R}^3,\mathbb{C})$ in both variables $\vec{p}_1$ and $\vec{p}_2$.  We can therefore apply Proposition \ref{moll:prop:2} to the $p_2$-integral for fixed $\vec{p}_1$.  Since $h(\vec{p}_1,\vec{p}_1)=0$ we get 
\begin{align}
\label{hamr:20}
     \iiint d^3p_2 ~ \FOU{\rho}(\epsilon \vec{p}_1)~h(\vec{p}_1,\vec{p}_2)
     \frac{1}{\epsilon^3} \chi(\frac{\vec{p}_1+\vec{p}_2}{\epsilon})
     = \epsilon^{q+1}  R^{q+1} (\vec{p}_1),
\end{align}
where the remainder $R^{q+1}(\vec{p}_1)$ is in $\mathcal{O}_{\text{M}}(\mathbb{R}^3,\mathbb{C})$.  Thus, after the $p_1$-integration, as the damper $\FOU{\rho}(\epsilon \vec{p}_1) \in \mathcal{S}$, we finally obtain
\begin{align}
\label{hamr:21}
               \mathbf{H}_{++} = \OOO(\epsilon^{q-1})
               \qquad \text{i.e.,} \qquad
               \mathbf{H}_{++} = 0,
\end{align}
because $\mathbf{H}_{++}f_n$ is negligible in $\mathcal{G}$.

\subsection{Calculation of $\mathbf{H}_{--}$}
%............................................
The operator $\mathbf{a}_{p_1}^-\mathbf{a}_{p_2}^-$ transforms the state $f_n$ into a state $f_{n-2}$ with $n-2$ particles.  To see exactly how this transformation occurs we begin to do it step by step in the customary formulation.  Thus, for the first annihilation the definition \eqref{fock:8} gives
\begin{align}
\label{hamr:22}
   \mathbf{a}_{p_1}^-(\xi)\mathbf{a}_{p_2}^-(\xi) f_n
 = \mathbf{a}_{p_1}^-(\xi) ~ \sqrt{n} 
   \BRA \psi_{p_2}(\xi_j) \| f_n(\xi_j,\xi_1,...,\xi_{n-1}) \KET,
\end{align}
where we relabeled $\xi_n$ as $\xi_j$ although there is no summation on $j$.  As $f_n$ is just a properly symmetrized polynominal of $\Delta_+$ functions we only need formula  \eqref{stat:9}, i.e.,
\begin{align}
\label{hamr:23}
   \BRA \psi_{p_2}(\xi_j) \| \Delta_+(\xi_j-x_j) \KET = \psi_{p_2}(-x_j),
\end{align}
to obtain
\begin{align}
\label{hamr:24}
\mathbf{a}_{p_1}^-(\xi)\mathbf{a}_{p_2}^-(\xi) f_n
 = \mathbf{a}_{p_1}^-(\xi) ~ \sqrt{n}
   \psi_{p_2}(-x_j) f_{n-1}(\xi_1,...,\xi_{n-1}),
\end{align}
where $f_{n-1}$ is the properly symmetrized $(n-1)$-particles state function. Thus, repeating the same procedure with $\mathbf{a}_{p_1}^-(\xi)$, we get
\begin{align}
\label{hamr:25}
\mathbf{a}_{p_1}^-(\xi)\mathbf{a}_{p_2}^-(\xi) f_n =  \sqrt{(n-1)n} 
   \psi_{p_1}(-x_k)\psi_{p_2}(-x_j) f_{n-2}(\xi_1,...,\xi_{n-2}).
\end{align}
In $\mathcal{G}$ the function $\Delta_+(\xi-x)$ becomes $\Delta_\epsilon(\FOU{\rho},\xi-x)$ and according to \eqref{norm:10} formula \eqref{hamr:23} becomes
\begin{align}
\label{hamr:26}
   \BRA \psi_p(\xi) \| \Delta_\epsilon(\FOU{\rho},\xi-x) \KET_\mathcal{G}
 = \FOU{\rho}(\epsilon\vec{p}\,)\psi_p(-x) + \OOO(\epsilon^{q+1}).
\end{align}
Consequently, using this formula twice, the embedded form of \eqref{hamr:25} is simply
\begin{align}
\label{hamr:27}
\mathbf{a}_{p_1}^-(\xi)\mathbf{a}_{p_2}^-(\xi) f_n
% \bigr|_{\mathsf{L}^2_\mathcal{G}}
  =  \Bigl( \sqrt{(n-1)n} 
    \FOU{\rho}(\epsilon\vec{p}_1) \FOU{\rho}(\epsilon\vec{p}_2)
   \psi_{p_1}(-x_k)              \psi_{p_2}(-x_j)
            + \OOO(\epsilon^{q+1}) \Bigr) f_{n-2}.
\end{align}
Comparing with \eqref{hamr:14} we remark that ignoring the dampers in  \eqref{hamr:27} and the summations in \eqref{hamr:14} these two expressions have the same general form, modulo a negligible term, because they are basically just a product of two $\psi_p$ functions.  We therefore expect the calculations in the present $\mathbf{H}_{--}$ case to be very similar to those in the $\mathbf{H}_{++}$ case.  Thus, we write 
\begin{align}
\nonumber
  \mathbf{H}_{--} = \frac{1}{2}\sum
                          &\iiint d^3p_1~\FOU{\rho}(\epsilon \vec{p}_1)\\
\label{hamr:28}
                   \times &\iiint d^3p_2~\FOU{\rho}(\epsilon \vec{p}_2)
                                              ~ h(\vec{p}_1,\vec{p}_2)
     \frac{1}{\epsilon^3} \chi(\frac{-\vec{p}_1-\vec{p}_2}{\epsilon}),
\end{align}
where because of various differences the function $h(\vec{p}_1,\vec{p}_2)$ is not identical to \eqref{hamr:19}, but still has the properties required to apply Proposition \ref{moll:prop:2} again, which leads us to conclude that 
\begin{align}
\label{hamr:29}
               \mathbf{H}_{--} = \OOO(\epsilon^{q-1}),
               \qquad \text{i.e.,} \qquad
               \mathbf{H}_{--} = 0,
\end{align}
because $\mathbf{H}_{--}f_n$ is also negligible in $\mathcal{G}$.

\subsection{Calculation of $\mathbf{H}_{+-}$}
%............................................
The operator $\mathbf{H}_{+-}$ transforms a state with $n$ particles into a state with $n$ particles again, first by an annihilation, then by a creation.  Let us first consider the formal calculation.  Inspecting the definitions (\ref{fock:7}--\ref{fock:8}) we see that due to the symmetrization \eqref{fock:3} there is an overall statistical factor of $\sqrt{n}\sqrt{n}/n = 1$.  We therefore get, as in (\ref{hamf:17}--\ref{hamf:18}),
\begin{align}
\label{hamr:30}
   \mathbf{a}_{p_1}^+(\xi)\mathbf{a}_{p_2}^-(\xi) f_n
 = \sum_{k=1}^n \psi_{p_1}(\xi_k) 
   \BRA \psi_{p_2}(\xi_j) \| f_n(\xi_j,...,x_k^\times,...,\xi_n) \KET.
\end{align}
Embedded in $\mathcal{G}$, the state function $f_n$ is of the form \eqref{embs:9} and the scalar product is no more given by \eqref{hamf:19} but by \eqref{hamr:26}.  Thus
\begin{align}
\label{hamr:31}
   \mathbf{a}_{p_1}^+(\xi)\mathbf{a}_{p_2}^-(\xi) f_n
 = \sum \psi_{p_1}(\xi_j)
   \Bigl( \FOU{\rho}(\epsilon \vec{p}_2)\psi_{p_2}(-x_j)
        + \OOO(\epsilon^{q+1}) \Bigr)
    f_{n-1},
\end{align}
where the details of the summation are ignored and a symbolic $\sum$ with $k=j$ is written instead.  Introducing this equation in \eqref{hamr:12} we obtain
\begin{align}
\nonumber
               \mathbf{H}_{+-}f_n = \frac{1}{2} \sum \iiint& d^3p_1
              ~\FOU{\rho}(\epsilon \vec{p}_1)\\
\label{hamr:32}
              \times \iiint& d^3p_2~ \FOU{\rho}^*(\epsilon \vec{p}_2)
                                    ~ h(\vec{p}_1,\vec{p}_2)
   \frac{1}{\epsilon^3} \chi(\frac{\vec{p}_1-\vec{p}_2}{\epsilon}) ~ f_{n-1},
\end{align}
where
\begin{align}
\nonumber
    h(\vec{p}_1,\vec{p}_2) 
    &=       \exp i( E_{p_1} - E_{p_2} )t ~
             \frac{E_{p_1}E_{p_2} + \vec{p}_1\cdot\vec{p}_2 +m^2}
                  {2\sqrt{E_{p_1} E_{p_2}}}\\
\label{hamr:33} 
    &\times  \psi_{p_1}(\xi_j)
             \Bigl( \FOU{\rho}(\epsilon \vec{p}_2)\psi_{p_2}(-x_j)
                  + \OOO(\epsilon^{q+1}) \Bigr).
\end{align}
As in the previous similar expressions $h(\vec{p}_1,\vec{p}_2) \in \mathcal{O}_{\text{M}}(\mathbb{R}^3,\mathbb{C})$ as a function of both $\vec{p}_1$ and $\vec{p}_2$.  We can thus use Proposition \ref{moll:prop:2} which gives for the ${p}_2$ integral
\begin{align}
\nonumber
     \FOU{\rho}^*(\epsilon \vec{p}_1)h(\vec{p}_1,\vec{p}_1)
          &= \FOU{\rho}^*(\epsilon \vec{p}_1)E_{p_1} \psi_{p_1}(\xi_j)
            \Bigl(\FOU{\rho}(\epsilon \vec{p}_1) \psi_{p_1}(-x_j)
                                     + \OOO(\epsilon^{q+1}) \Bigr) \\
\label{hamr:34}
          &+ \OOO(\epsilon^{q+1}).
\end{align}
Inserting this result in \eqref{hamr:32} we remark that just like in the derivation of \eqref{hamr:21} the $p_1$-integration of the negligible contributions poses no problem.  Thus
\begin{align}
\label{hamr:35}
               \mathbf{H}_{+-}f_n = \frac{1}{2} \sum \iiint d^3p_1
              ~\FOU{\rho}(\epsilon \vec{p}_1)
              |\FOU{\rho}(\epsilon \vec{p}_1)|^2
              ~ E_{p_1} \psi_{p_1}(\xi_j)\psi_{p_1}(-x_j) ~ f_{n-1},
\end{align}
where we used the concept of equality in $\mathcal{G}$ to avoid writing the integrated negligible contributions explicitly.  As was the case for the formal expression \eqref{hamf:21} this equation can be rewritten as
\begin{align}
\label{hamr:36}
               \mathbf{H}_{+-}f_n = 
               \frac{1}{2} \sum
               i \frac{\partial }{\partial t_{\xi_j}} \iiint d^3p
              ~\FOU{\rho}(\epsilon \vec{p}\,)
              |\FOU{\rho}(\epsilon \vec{p}\,)|^2
              ~\psi_{p}(\xi_j)\psi_{p}(-x_j) ~ f_{n-1},
\end{align}
where we recognize the embedded form \eqref{embs:7} of the $\Delta_+$ function in the form \eqref{stat:3}. Therefore,
\begin{align}
\label{hamr:37}
  \mathbf{H}_{+-} f_n = 
         \frac{1}{2} \sum i \frac{\partial }{\partial t_{\xi_j}}
     \Delta_\epsilon( \FOU{\rho}\,|\FOU{\rho}\,|^2 ,\xi_j-x_j)
                ~ f_{n-1},
\end{align}
where the damper is not $\FOU{\rho}$ as in \eqref{embs:7} but $\FOU{\rho}\,|\FOU{\rho}\,|^2$ instead.  Thus, contrary to the formal calculations, where for example according to \eqref{hamf:22} the single particle state function $f_1(\xi_1)=f_0\Delta_+(\xi_1-x_1)$ is identically reconstructed, this is not the case here because $\Delta_\epsilon( \FOU{\rho}\, |\FOU{\rho}\,|^2, \xi_j-x_j) \neq \Delta_\epsilon( \FOU{\rho} ,\xi_j-x_j)$ in general.  However, from Proposition \ref{embs:prop:2} we know that
\begin{align}
\label{hamr:38}
          \Delta_\epsilon( \FOU{\rho}\,|\FOU{\rho}\,|^2 ,\xi_j-x_j)
        = \Delta_\epsilon( \FOU{\rho} ,\xi_j-x_j)
        + \OOO(\epsilon^{q+1}), \forall q \in \mathbb{N}.
\end{align}
Consequently, the summation in \eqref{hamr:37} is actually reconstructing the $f_n$ state from the $f_{n-1}$ state modulo a null quantity in the $\mathcal{G}$-context, which implies that in $\mathcal{G}$ we rigorously have
\begin{align}
\label{hamr:39}
 \mathbf{H}_{+-} ~f_n = \frac{1}{2}
                 i \sum_{j=1}^n\frac{\partial }{\partial t_{\xi_j}} ~f_n.
\end{align}

\subsection{Calculation of $\mathbf{H}_{-+}$ and of the zero-point energy}
%...................................................................
The operator $\mathbf{H}_{-+}$ starts by creating an additional particle by a symmetrized summation, then this particle is annihilated by integration.  Since this calculation is more complicated than that of $\mathbf{H}_{+-}$ the standard practice in the formal calculations is to use the commutation relations, as we did in \eqref{hamf:12}, to derive the identity
\begin{align}
\label{hamr:40}
   \frac{1}{2} (\mathbf{a}_p^+\mathbf{a}_p^- + \mathbf{a}_p^-\mathbf{a}_p^+)
              =  \mathbf{a}_p^+\mathbf{a}_p^-
               + \frac{1}{2}[\mathbf{a}_p^-,\mathbf{a}_p^+].
 \end{align}
In the rigorous case we do not have $\vec{p}_1 = \vec{p}_2$ in equations (\ref{hamr:12}--\ref{hamr:13}).  There are also differences in signs and a difference in the labeling of $\mathbf{a}_p^\pm$.  However, by interchanging the labels of the dummy variables $\vec{p}_1$ and $\vec{p}_2$ in either \ref{hamr:12} or \eqref{hamr:13} all these differences disappear.  This enables to use the identity
\begin{align}
\label{hamr:41}
  \frac{1}{2} (\mathbf{a}_{p_1}^+\mathbf{a}_{p_2}^-
             + \mathbf{a}_{p_2}^-\mathbf{a}_{p_1}^+)
             =  \mathbf{a}_{p_1}^+\mathbf{a}_{p_2}^-
               + \frac{1}{2}[\mathbf{a}_{p_2}^-,\mathbf{a}_{p_1}^+],
 \end{align}
to write
\begin{align}
\label{hamr:42}
 \mathbf{H}_0  =   \mathbf{H}_{+-} + \mathbf{H}_{-+}
               =  2\mathbf{H}_{+-} + E_\mho,
\end{align}
where from (\ref{hamr:12}--\ref{hamr:13})
\begin{align}
\nonumber
    E_\mho &= \frac{1}{2} \iiint d^3p_1 \iiint d^3p_2
              ~\FOU{\rho}  (\epsilon \vec{p}_1)
              ~\FOU{\rho}^*(\epsilon \vec{p}_2) 
              ~[\mathbf{a}_{p_2}^-,\mathbf{a}_{p_1}^+]\\
\label{hamr:43} 
    &\times \bigl(E_{p_1}E_{p_2} + \vec{p}_1\cdot\vec{p}_2 +m^2\bigr)
            \frac{\exp i(E_{p_1} - E_{p_2} )t }
                 {2\sqrt{E_{p_1} E_{p_2}}}
          \frac{1}{\epsilon^3} \chi(\frac{\vec{p}_1-\vec{p}_2}{\epsilon}).
\end{align}
The commutator $[\mathbf{a}_{p_2}^-,\mathbf{a}_{p_1}^+]  = \BRA \psi_{p_2}^{ }(\xi) \| \psi_{p_1}^{ }(\xi) \KET_\mathcal{G}$ is given by \eqref{norm:4}, i.e.,
\begin{align}
\label{hamr:44}
       [\mathbf{a}_{p_2}^-,\mathbf{a}_{p_1}^+]
           =  (E_{p_2} + E_{p_1})\frac{\exp {i(E_{p_2}-E_{p_1})t_\xi}}
                                      {2\sqrt{E_{p_2} E_{p_1}}}
         \frac{1}{\epsilon^3} \chi(\frac{\vec{p}_2-\vec{p}_1}{\epsilon}).
\end{align}
Thus, since $\chi^\vee=\chi^*$, we get
\begin{align}
\nonumber
    E_\mho &= \frac{1}{2} \iiint d^3p_1 \iiint d^3p_2
              ~\FOU{\rho}  (\epsilon \vec{p}_1)
              ~\FOU{\rho}^*(\epsilon \vec{p}_2)
              ~\exp i(E_{p_1} - E_{p_2} )(t - t_\xi)\\
\label{hamr:45} 
    &\times\frac{ (E_{p_1}E_{p_2} + \vec{p}_1\cdot\vec{p}_2 + m^2)(E_{p_1}
                 + E_{p_2}) } { 4 E_{p_1} E_{p_2} }
\Bigl| \frac{1}{\epsilon^3} \chi(\frac{\vec{p}_1-\vec{p}_2}{\epsilon}) \Bigr|^2.
\end{align}
Taking the complex conjugate and interchanging the integration variables $\vec{p}_1$ and $\vec{p}_2$ one finds that $E_\mho^* = E_\mho$ so that $E_\mho$ is real.  Moreover, interchanging  $\vec{p}_1$ and $-\vec{p}_2$ after taking the complex conjugate shows that $E_\mho$ can be written
\begin{align}
\nonumber
    E_\mho &= \frac{1}{2} \iiint d^3p_1 \iiint d^3p_2
              ~\FOU{\rho}  (\epsilon \vec{p}_1)
              ~\FOU{\rho}^*(\epsilon \vec{p}_2)
              ~\cos \, (E_{p_1} - E_{p_2} )(t - t_\xi)\\
\label{hamr:46} 
    &\times\frac{ (E_{p_1}E_{p_2} + \vec{p}_1\cdot\vec{p}_2 + m^2)(E_{p_1}
                 + E_{p_2}) } { 4 E_{p_1} E_{p_2} }
 \frac{1}{\epsilon^6} |\chi|^2(\frac{\vec{p}_1-\vec{p}_2}{\epsilon}).
\end{align}
Equations \eqref{hamr:45} and \eqref{hamr:46} are rigorous $\mathcal{G}$-expressions replacing the formal expression \eqref{hamf:15} of the zero-point energy in which the square of a $\delta^3$-function appears.  These expressions are finite and well defined $\forall \epsilon \in ]0,1[$ and $\forall t, t_\xi \in \mathbb{R}$.  Moreover, they are symmetrical between $\vec{p}_1$ and $\vec{p}_2$ and the fraction on the second line is equal to $E_p$ when $\vec{p}_1 = \vec{p}_2$.  Hence applying Proposition \ref{moll:prop:4} to either the ${p}_1$ or ${p}_2$ integral we get
\begin{align}
\label{hamr:47} 
    E_\mho = \frac{1}{2}\iiint d^3p~E_p   \Bigl(
             ~|\FOU{\rho}(\epsilon \vec{p}\,)|^2
             \frac{1}{\epsilon^3} M[^2_0]
           + \OOO(\frac{1}{\epsilon^2})   \Bigr),
\end{align}
where, by Parseval's theorem,
\begin{equation}
\label{hamr:48}
    M[^2_0]  = \iiint d^3z~  |\chi|^2(\vec{z}\,)
             = \frac{1}{(2\pi)^3} \iiint d^3z~  |\FOU{\chi}|^2(\vec{z}\,),
\end{equation}
and where it is not possible in general to remove the terms beyond the leading one because the moments
\begin{equation}
\label{hamr:49}
      M[^2_n]  = \iiint d^3z~ \vec{z}\,^n |\chi|^2(\vec{z}),
\end{equation}
with $n_1,n_2,$ and $n_3$ even cannot be zero unless $\chi \equiv 0$.

\begin{remark}[$E_\mho$ as vacuum expectation value of $\mathbf{H}_0$]
%.....................................................................
\label{hamr:rema:1}
The calculation of $E_\mho$ can be greatly simplified by remarking that it is equal to the expectation value $\SCA \mho | \mathbf{H}_0 | \mho \LAR = \SCA \mho | \mathbf{H}_{-+} | \mho \LAR$ because the other three contributions are zero.  Then, since $\SCA \mho | \mho \LAR = 1$, one obtains $\SCA \mho | \mathbf{a}_{p_1}^- \mathbf{a}_{p_2}^+ | \mho \LAR = \BRA \psi_{p_1}^{ }(\xi) \| \psi_{p_2}^{ }(\xi) \KET_\mathcal{G}$ in \eqref{hamr:13},  and \eqref{hamr:45} follows immediately.
\end{remark}
\begin{remark}[Time averaging of $E_\mho$]
%.........................................
\label{hamr:rema:2}
Starting from the Hamiltonian \eqref{hamr:1} all calculations leading to \eqref{hamr:47} were made for fixed $t$, whereas there is in \eqref{hamr:45} an oscillating factor $\exp i (- E_{p_1} + E_{p_2} )(t - t_\xi)$ which in the customary formulation (\ref{hamf:14}--\ref{hamf:15}) is brushed aside because, for any finite time $t - t_\xi$,
\begin{align}
\label{hamr:50} 
      E_{p_1} \rightarrow E_{p_2}
      \qquad  \Rightarrow  \qquad
              \exp i (E_{p_1} - E_{p_2} )(t - t_\xi) \rightarrow 1.
\end{align}
This suggests investigating whether this oscillating factor could have an effect on $E_\mho$, for instance if a time-averaging is made over the variable $t$ for a duration $\Delta t$ before the $p$-integrations
 --- a process that should have no impact on the result since $E_\mho$ is part of the Hamiltonian which is time-invariant.  So, let us rewrite \eqref{hamr:45} as
\begin{align}
\label{hamr:51}
        E_\mho = \iiint d^3p_1 \iiint d^3p_2
               ~ h(\vec{p}_1,\vec{p}_2)
%               ~ |\chi_\epsilon|^2(\vec{p}_1,\vec{p}_2)
               ~\exp i (E_{p_1} - E_{p_2} )t,
\end{align}
where $h(\vec{p}_1,\vec{p}_2)$ contains everything except the exponential factor in the variable $t$. Then, to properly calculate the time-average of $E_\mho$ in $\mathcal{G}$, we define a time-averaging operator $A_{\Delta t} \bigl( \cdot \bigr)$ with the help of a damper $\FOU{\omega}(t/\Delta t)$ whose effect is by definition equivalent to that of a smooth cut-off at some large time $\pm t$ on the order of $\Delta t$.  Thus, by Fourier transform,
\begin{align}
\label{hamr:52}
        \int_{-\infty}^{+\infty} dt ~ \FOU{\omega}\Big(\frac{t}{\Delta t}\Bigr)
               ~\exp i (Et)  = 2 \pi \Delta t ~ \omega(E \Delta t),
\end{align}
where $\omega \in \mathcal{S}$, implies
\begin{align}
\label{hamr:53}
        \int_{-\infty}^{+\infty} dt ~ \FOU{\omega}\Big(\frac{t}{\Delta t}\Bigr)
                = 2 \pi \Delta t ~ \omega(0),
\end{align}
so that
\begin{align}
\label{hamr:54}
       A_{\Delta t} \bigl( \cdot \bigr)
 \DEF \frac{1}{2 \pi \omega(0)}
      \int_{-\infty}^{+\infty} \frac{dt}{\Delta t}~
               \FOU{\omega}\Big(\frac{t}{\Delta t}\Bigr)\bigl( \cdot \bigr),
\end{align}
is a normalized and smooth time-averaging operator over a duration of order $\Delta t$.  Applying this operator to \eqref{hamr:51} before making the $p$-integrations, and using \eqref{hamr:52}, we get
\begin{align}
\label{hamr:55}
    A_{\Delta t} \bigl(E_\mho\bigr) = \iiint d^3p_1 \iiint d^3p_2
               ~ h(\vec{p}_1,\vec{p}_2)
%               ~ |\chi_\epsilon|^2(\vec{p}_1,\vec{p}_2)
               ~\frac{\omega\bigl((E_{p_1} - E_{p_2})\Delta t\bigr)}{\omega(0)},
\end{align}
which applying Proposition \ref{moll:prop:4} as in the derivation of \eqref{hamr:47} confirms that an averaging over $t$ has no effect on the value of $E_\mho$, at least as long as $\Delta t$ is finite.  On the other hand, since $\omega \in \mathcal{S}$, 
\begin{align}
\label{hamr:56}
    \lim_{\Delta t \rightarrow \infty} A_{\Delta t} \bigl(E_\mho\bigr) = 0,
\end{align}
if the limit $\Delta t \rightarrow \infty$ is taken \emph{before} any of the $\vec{p}$ integrations is made.  However, similar oscillating factors appear in all four terms of the Hamiltonian \eqref{hamr:9}, and in particular in $\mathbf{H}_{+-}$, so that everything would be zero in that limit, which implies that it is not possible to let $\Delta t \rightarrow \infty$.
\end{remark}

\subsection{Interpretation of zero-point energy}
%...............................................

We have rigorously shown that in $\mathcal{G}$ the Hamiltonian operator can be written 
\begin{align}
\label{hamr:57}
   \mathbf{H}_{0}(\rho_\epsilon,\FOU{\chi}_\epsilon,t) = i\sum_{j=1}^n \frac{\partial }{\partial t_{\xi_j}}
                 + E_\mho(\rho_\epsilon,\FOU{\chi}_\epsilon),
\end{align}
i.e., as the sum of the quantum mechanical total energy operator \eqref{hamf:23} and of the zero-point energy \eqref{hamr:45} whose leading term after performing one of the momentum integrations \eqref{hamr:47} can be written
\begin{align}
\label{hamr:58} 
    E_\mho (\rho_\epsilon,\FOU{\chi}_\epsilon)
       = \frac{1}{2(2\pi)^3} \iiint d^3x  ~|\FOU{\chi}|^2(\epsilon \vec{x}\,) 
                         \iiint d^3p~E_p  ~|\FOU{\rho}|^2(\epsilon \vec{p}\,).
\end{align}
When $\epsilon=0$ this expression yields the customary formula \eqref{hamf:14}, so that $E_\mho \rightarrow \infty$ as $\epsilon \rightarrow 0$.  On the other hand, when $\epsilon\neq 0$, and if neither $\rho$ nor $\chi$ are identically zero, $E_\mho$ is definite positive because $E_p \geq m > 0$.

  As will be seen in Sec.\,\ref{canc:0}, a non-zero value for $E_\mho$ has no measurable consequencies since it drops out of the expression of the scattering operator.  However, like any energy, $E_\mho$ contributes to the gravitational field.  In fact, $E_\mho$ is interpreted as `dark energy' in contemporary cosmological models.  According to this interpretation, the zero-point energy divided by the $d^3x$ volume-integral appearing in factor in the customary expression \eqref{hamf:14} defines an energy density
\begin{align}
\label{hamr:59} 
    \Lambda_{\mho}  = \frac{1}{2(2\pi)^3} \iiint d^3p~E_p ,
\end{align}
which is assumed to contribute to the cosmological constant $\Lambda \approx (2 \times 10^{-3} \text{ eV})^4$ whose non-zero value is required to account for the observation that the expansion of the Universe is accelerating.  But this interpretation is problematic because \eqref{hamr:59} is divergent so that a cut-off is required.  For instance, assuming that the cut-off is at a large energy, one can replace $E_p = \sqrt{m^2 + p^2}$ by $p$ to get an estimate.  Then
\begin{align}
\label{hamr:60} 
    \Lambda_{\mho}  = \frac{1}{2(2\pi)^3} 4\pi \frac{1}{4} p_{\text{cut}}^4,
\end{align}
yields a contribution that is very much larger than $\Lambda$ if $p_{\text{cut}}$ is set to a value that appears reasonable from the point of view of particle physics, i.e., larger than the proton mass but smaller than the Planck energy: $ 10^{9} < p_{\text{cut}} < 10^{28} \text{ eV}$.

  What about the $\mathcal{G}$-formulation? Could expression \eqref{hamr:58} be compatible with a cosmological constant as small as $\Lambda \approx (2 \times 10^{-3} \text{ eV})^4$?  Or could there be a possibility such that $E_\mho$ is actually zero in $\mathcal{G}$?  Indeed, while the very large estimate of $\Lambda_{\mho}$ based on Eq.\,\eqref{hamr:60} is disturbing, the physical meaning of $E_\mho$ and of the cut-off $p_{\text{cut}}$ are already unclear in many respects \cite[footnote 19]{PEEBL2003-}.  Moreover, while it has long been conjectured that the zero-point energy was `physically real,' and that the Casimir effect was experimental evidence for that, it has been proved that this is actually not the case \cite{JAFFE2005-}.  It is therefore plausible that  $E_\mho$ could be zero.

% In such as case eqref{hamr:57} would reduce to the quantum mechanical energy operator, and $\mathbf{H}_{0}$ would properly yield an eigenvalue rather then  of the normalization of the states.

  In order to have $E_\mho = 0$ some of the basic assumptions made in the calculation of $\mathbf{H}_0$ have to be changed.  One such assumption is that the dampers $\FOU{\chi}$ in the definitions of the scalar product \eqref{norm:1} and of the Hamiltonian \eqref{hamr:1} are the same.  This insures that the Hamiltonian is real, but it is not an absolute necessity.  Indeed, even if the Hamiltonian were not real the eigenvalues of $\mathbf{H}_0$ could still be real and given by \eqref{hamr:57} if $E_\mho=0$.  Writing $\chi_\xi$ and $\chi_x$ for the $\xi$- and $x$-integration dampers, this would be the case if the moments
\begin{equation}
\label{hamr:61}
  M[^2_n] = \iiint d^3z~ \vec{z}\,^n \chi_\xi(\vec{z}) \chi_x^*(\vec{z}), 
\end{equation}
are zero for all $n = 0, 1, 2,...$  In fact, this is mathematically possible, for example if $\chi_\xi = \chi_x^*$, as shown in Ref.\,\cite{COLOM2008X}.  As is well known, the implication of a Hamiltonian that contains complex numbers is violation of time-reversal invariance, and consequently by the CPT theorem violation of CP invariance.  But a non-real Hamiltonian such that the vacuum energy is zero could be consistent with the fact that CP-violation affects only fields which have quantum numbers that differ from those of the vacuum by non-strictly conserved quantum numbers (e.g., the $K^0$ particles).  Whether or not a pair $\chi_\xi$, $\chi_x$ yielding such a Hamiltonian exists remains to be investigated.

 %(Such a cancellation mechanism would also remind us of neutrino oscillations which like the cosmological constant are characterized by a milli-eV scale.)

  The idea of putting specific constraints on the mollifiers/dampers to remove the zero-point energy has first been put forward in \cite[p.,\,304]{COLOM1984-},  and the possibility that $\chi_\xi \neq \chi_x$ can be seen as a more sophisticated version of that idea.  It remains however to verify that such constraints do not spoil any other aspects of the theory.  In particular, one has to confirm that this method is compatible with the introduction of interactions, and is applicable at any order in perturbation theory.

  Another possibility of a different kind could be to consider $\epsilon$ as a complex number. Then the infinite looking quantities are holomorphic in the variable $\epsilon$ around $\epsilon=0$, that is $\epsilon=0$ is a pole. The removal of divergences would consist in subtracting the irregular part at $\epsilon=0$, i.e., the terms with negative powers of $\epsilon$. This could lead to a nonambiguous renormalization prescription, and its justification would be an average taking place on complex values of $\epsilon$.  But this is only a conjecture at this point, and one would have to find a procedure to deal with divergences in $\log(\epsilon)$.

  Finally, the consequencies of working with normalized states, as defined in Sec.\,\ref{norm:0}, should also be addressed.  Indeed, if such states are used the divergence of $\Lambda_\mho$ is reduced from $p^4$ to $p^2$.  But working with normalized states requires to redefine the states or the Hamiltonian by introducing dimensional factors, and the physical meaning of introducing such factors must be thoroughly investigated. 

  These examples illustrate that the interpretation of the results obtained by formulating quantum field theory in $\mathcal{G}$ is open to many possibilities.  Let us now return to the less speculative option that $E_\mho$ is non-zero but finite.

% For this reason one has also to investigate the implications of associating $E_\mho$ to a finite value in  $\mathcal{G}$.

   In the formal calculations the zero-point energy is given by an ill-defined integral over the square of a $\delta$-function, i.e., Eq.~\eqref{hamf:15}, which yields the infinite expression \eqref{hamf:14}.  In $\mathcal{G}$ quantum field theory the fields are interpreted as nonlinear generalized functions and quantities such as the zero-point energy as generalized numbers.  This implies that \eqref{hamr:58} should be compared to the data for $\epsilon$ finite rather than in the limit $\epsilon \rightarrow 0$ in which quantum field theory becomes mathematically inconsistent.  Moreover, as $\epsilon$ on its own has no particular physical meaning it can be set to any convenient value.  This value can even be chosen close to $1$ since `zero,' and thus `equality,' in $\mathcal{G}$ is defined as $\OOO(\epsilon^q)$ with $q \in \mathbb{N}$ as large as we please.  

   Hence we take for $\epsilon$ a positive real number such that $\epsilon \lesssim 1$, and rewrite \eqref{hamr:58} in dimensional form in order to facilitate its physical interpretation as a non-zero quantity.  To do that we define two lengths $\lambda_r$ and $\lambda_c$ associated to the mollifier/damper $\rho$ and $\FOU{\chi}$ so that \eqref{hamr:58} becomes
\begin{align}
\label{hamr:63} 
    E_\mho (\rho_\epsilon,\FOU{\chi}_\epsilon)
       = \frac{1}{2(2\pi\hbar)^3}
   \iiint d^3x      ~|\FOU{\chi}|^2(\epsilon \frac{\vec{x}}{\lambda_c}) 
   \iiint d^3p~E_p  ~|\FOU{\rho}|^2(\epsilon \frac{\lambda_r \vec{p}}{\hbar}),
\end{align}
where $\hbar$ is Planck's constant.  Of course, changing the value of $\epsilon$ is equivalent to rescaling $\lambda_r$ and $\lambda_c$, although they can only be changed in such a way that the product $\lambda_r \lambda_c$ remains constant.\footnote{The property might be related to the well-known relation between the Compton, Schwarzschild, and Planck lengths.  That is, if $\lambda_{\text{C}} = \hbar/mc^2$ and $\lambda_{\text{S}}=2mG_{\text{N}}/c^2$ then $\lambda_{\text{C}} \lambda_{\text{S}} = 2 \lambda_{\text{P}}^2$.}  Then, to get an order of magnitude estimate, we approximate $E_p= \sqrt{m^2 + c^2p^2}$ by $cp$ and make two obvious change of integration variables to obtain
\begin{align}
\label{hamr:64} 
    E_\mho (\rho_\epsilon,\FOU{\chi}_\epsilon)
       \approx \frac{1}{2(2\pi)^3}  \frac{1}{\epsilon^7}
       \Bigl( \frac{\lambda_c}{\lambda_r} \Bigr)^3
        \frac{\hbar c} {\lambda_r}
   \iiint d^3y    ~|\FOU{\chi}|^2(y) 
   \iiint d^3s~s  ~|\FOU{\rho}|^2(s),
\end{align}
where $\hbar c = 197 \times 10^{-9}$ eV m.  
%  For example, if $\lambda_c \ll \lambda_r$, we see that $E_\mho$ can be very small --- that is possibly negligible in comparison to $\partial_{t_\xi}$ in \eqref{hamr:57} --- even when the $p$-integral is cut-off at a reasonably large value.
   To further simplify \eqref{hamr:64} we assume that beyond the neighborhoods of zero over which $\FOU{\chi}$ and $\FOU{\rho}$ are identical to $1$ the functions $\FOU{\chi}$ and $\FOU{\rho}$ rapidly drop to $0$ so that both integrals will be roughly equal to $1$.  An approximation, neglecting all numerical factors, is thus
\begin{align}
\label{hamr:65} 
      E_\mho (\lambda_c,\lambda_r) 
     \approx \Bigl( \frac{\lambda_c}{\epsilon} \Bigr)^3
     \Lambda_\mho (\lambda_r),
     \quad \text{where} \quad
     \Lambda_\mho (\lambda_r)
     \approx \Bigl( \frac{1}{\epsilon\lambda_r} \Bigr)^3
     \frac{\hbar c} {\epsilon\lambda_r}.
\end{align}

   Therefore, the zero-point energy-density scale is set by $\lambda_r$, the cut-off in the momentum integral, and the volume by $\lambda_c$, the cut-off in the volume integral --- just like in the customary formulation.  But beyond these similarities there are important differencies between the two formulations. This is because in the customary formulation the effect of setting $\lambda_r$ and $\lambda_c$ to some numbers is limited to giving a value to $\Lambda_\mho$ and $E_\mho$.  On the other hand, in the $\mathcal{G}$-formulation, the mollifier/damper $\rho$ and $\FOU{\chi}$ are related by Fourier transformation to the complex functions $\FOU{\rho}$ and ${\chi}$ whose `characteristic widths' are set by  $\lambda_r/\hbar$ and $\lambda_c/\hbar$.  

   For instance, chosing a certain $\lambda_r$ to dampen the $p$-integral in $\Lambda_\mho$ will at the same time define the width of the $\delta$-function representative $\rho^\vee \ast \rho$ appearing on the right-hand side of the commutation relation \eqref{comr:8}.  In other words, in the $\mathcal{G}$-formulation, chosing $\lambda_r$ and $\lambda_c$  has \emph{physical} implications which go beyond the problem of fitting $E_\mho$ or $\Lambda_\mho$ to some non-zero value.  In particular, whereas the $\delta$-functions are `all the same' in the customary formulation, the small-distance structure at the infinitesimal level of the $\delta$-function represented in $\mathcal{G}$ by the scaled mollifier $\rho_\epsilon$ is by Fourier transform connected to the large-energy structure at the cosmological level by means of the scaled damper $\FOU{\rho}_{\epsilon/\hbar}$.  Similarly, the large-distance and the small-energy structures are connected by means of $\FOU{\chi}_\epsilon$ and $\chi_{\epsilon/\hbar}$.

   The conclusion, therefore, is that it is possible in principle to fit $E_\mho$ and $\Lambda_\mho$ to any finite value, and that $E_\mho$ may possibly even be set to zero, but that any option, as well as any choice of the parameters $\lambda_r$ and $\lambda_c$, have implications which have to be evaluated in the full context of the theory.  This means that it is not possible to chose $\lambda_r$ and $\lambda_c$ on the basis of obtaining a certain value of $E_\mho$ or $\Lambda_\mho$ alone:  It is necessary to go beyond the free-field theory and to study self-interaction phenomena, perturbation theory, etc., before a definite conclusion can be reached.

%File: self.29.tex           arXiv version 2       Date: 6 September 2008
%=====                    
%                         

\chapter{The self-interacting field}
%===================================
\label{self:0}

\section{Self-interacting field: formal calculations}
%----------------------------------------------------
\label{sefo:0}
\setcounter{equation}{0}
\setcounter{definition}{0}
\setcounter{axiom}{0}
\setcounter{conjecture}{0}
\setcounter{lemma}{0}
\setcounter{theorem}{0}
\setcounter{corollary}{0}
\setcounter{proposition}{0}
\setcounter{example}{0}
\setcounter{remark}{0}
\setcounter{problem}{0}

Let us recall Definition \ref{summ:defi:1}:

\begin{definition}[Interacting field equation]
%.............................................
\label{sefo:defi:1}
The operator equation
\begin{align}
\label{sefo:1}
   \frac{\partial}{\partial t} \bphi(t,\vec{x}\,) &= \bpi(t,\vec{x}\,),\\
\label{sefo:2}
   \frac{\partial}{\partial t} \bpi (t,\vec{x}\,) &= 
   \sum_{1\leq\mu\leq 3} \frac{\partial^2}{\partial{x_\mu}^2} \bphi(t,\vec{x}\,)
   - m^2 \bphi(t,\vec{x}\,)
   - g \bigl(\bphi(t,\vec{x}\,)\bigr)^N,
\end{align}
completed by the initial conditions at the time $t=\tau$
\begin{align}
\label{sefo:3}
      \bphi(\tau,\vec{x}\,) = 
      \bphi_{\text{\rm ini}}(\tau,\vec{x}\,),
      \qquad \text{and} \qquad
      \bpi                 (\tau,\vec{x}\,) =
      \bpi_{\text{\rm ini}}(\tau,\vec{x}\,), 
\end{align}
as well as the equal-time commutations relations
\begin{align}
\label{sefo:4}
[\bphi(t,\vec{x}_1),\bphi(t,\vec{x}_2)] &=0,\\
\label{sefo:5}
[ \bpi(t,\vec{x}_1), \bpi(t,\vec{x}_2)] &=0,\\
\label{sefo:6}
[\bphi(t,\vec{x}_1), \bpi(t,\vec{x}_2)] &=
            i \delta^3(\vec{x}_1-\vec{x}_2)~\mathbf{1}, 
\end{align}
is the field equation of a quantized self-interacting neutral spin-0 field.
\end{definition}

The standard method for solving this equation is to use the Hamiltonian formalism.  One introduces the Hamiltonian operator
\begin{align}
\label{sefo:7}  
    \mathbf{H}(\bphi,\bpi,t) &= \iiint_{\mathbb{R}^3}
                     \pmb{\mathcal{H}}
       \Bigl(  \bphi(t,\vec{x}\,),\bpi(t,\vec{x}\,)  \Bigr) ~d^3x, 
\end{align}
where $\pmb{\mathcal{H}}(t,\vec{x}\,)$ is the Hamiltonian density
\begin{align}
\notag
       \pmb{\mathcal{H}}(t,\vec{x}\,) &=
         \frac{1}{2} \bigl(\bpi(t,\vec{x}\,)\bigr)^2 
       + \frac{1}{2} \sum_{1 \leq \mu \leq 3}
         \bigl(\partial_\mu \bphi(t,\vec{x}\,)\bigr)^2 \\
\label{sefo:8}  
       &+ \frac{1}{2} m^2
         \bigl(\bphi(t,\vec{x}\,)\bigr)^2 
       + \frac{g}{N+1} \bigl(\bphi(t,\vec{x}\,)\bigr)^{N+1}. 
\end{align}
We now recall and prove Conjecture \ref{summ:conj:1}:
\begin{proposition}[Heisenberg equations of motion]
%..................................................
\label{sefo:prop:1}
Any pair of field operators $\bphi(t,\vec{x}\,)$ and $\bpi(t,\vec{x}\,)$ satisfying the commutation relations \emph{(\ref{sefo:4}--\ref{sefo:5})}, as well as the Heisenberg equations of motion
\begin{align}
\label{sefo:9}
     \frac{\partial}{\partial t} \bphi(t,\vec{x}\,)   
     &= i[ \mathbf{H}, \bphi(t,\vec{x}\,) ] ,\\
\label{sefo:10}
     \frac{\partial}{\partial t} \bpi (t,\vec{x}\,) 
     &= i[ \mathbf{H}, \bpi (t,\vec{x}\,) ] ,
\end{align}
with the initial conditions \eqref{sefo:3}, where $\mathbf{H}$ is given by \emph{(\ref{sefo:7}--\ref{sefo:8})}, is a solution of the interacting-field equation \emph{(\ref{sefo:1}--\ref{sefo:6})}.

\end{proposition}
Proof: We first calculate the right-hand of \eqref{sefo:9} after replacing $\mathbf{H}$ by $\pmb{\mathcal{H}}(t,\vec{y}\,)$.  It comes, using the commutation relations,
\begin{align}
\nonumber
      [ \pmb{\mathcal{H}}(t,\vec{y}\,), \bphi(t,\vec{x}\,) ]
      &= \frac{1}{2} [ \bigl(\bpi(t,\vec{y}\,)\bigr)^2, \bphi(t,\vec{x}\,) ],\\
\nonumber
      &= \frac{1}{2} \bpi(t,\vec{y}\,)[ \bpi(t,\vec{x}\,), \bphi(t,\vec{x}\,) ]
       + \frac{1}{2} [ \bpi(t,\vec{y}\,), \bphi(t,\vec{x}\,) ]\bpi(t,\vec{y}\,)\\
\label{sefo:11}
      &= -i \delta^3(\vec{x}-\vec{y}\,)\bpi(t,\vec{y}\,).
\end{align}
Therefore \eqref{sefo:9} gives
\begin{align}
\label{sefo:12}  
    \frac{\partial}{\partial t} \bphi(t,\vec{x}\,)
        = \iiint\delta^3(\vec{x}-\vec{y}\,)\bpi(t,\vec{y}\,) ~d^3y
        = \bpi(t,\vec{x}\,), 
\end{align}
which is just \eqref{sefo:1}.   We similarly calculate the right-hand of \eqref{sefo:10} and get
\begin{align}
\nonumber
      [ \pmb{\mathcal{H}}(t,\vec{y}\,), \bpi(t,\vec{x}\,) ]
      &= \frac{1}{2} [ \sum_{1 \leq \mu \leq 3}
         \bigl(\partial_\mu \bphi(t,\vec{y}\,)\bigr)^2  , \bpi(t,\vec{x}\,) ]
\label{sefo:13}
       + \frac{1}{2} [  m^2
         \bigl(\bphi(t,\vec{y}\,)\bigr)^2               , \bpi(t,\vec{x}\,) ]\\
      &+ \frac{g}{N+1} [  
         \bigl(\bphi(t,\vec{y}\,)\bigr)^{N+1}           , \bpi(t,\vec{x}\,) ]
\end{align}
Using the identity
\begin{align}
\label{sefo:14}
   [\bphi^{N+1},\bpi] =  \bphi^{N}[\bphi,\bpi] +  \bphi^{N-1}[\bphi,\bpi]\bphi
                    + ... +  [\bphi,\bpi]\bphi^N,
\end{align}
and the commutations relations we get
\begin{align}
\nonumber
      [ \pmb{\mathcal{H}}(t,\vec{y}\,), \bpi(t,\vec{x}\,) ]
      &= i\sum_{1 \leq \mu \leq 3}
      \Bigl(\frac{\partial}{\partial y_\mu} \delta^3(\vec{x}-\vec{y}\,)\Bigr)
      \Bigl(\frac{\partial}{\partial y_\mu} \bphi(t,\vec{y}\,) \Bigr)\\
\label{sefo:15}
      &+i m^2 \delta^3(\vec{x}-\vec{y}\,)\bphi(t,\vec{y}\,) 
       +i g \delta^3(\vec{x}-\vec{y}\,)
         \bigl(\bphi(t,\vec{y}\,)\bigr)^N  .         
\end{align}
We now integrate over $\vec{y} \in\mathbb{R}^3$ and use for the first line the formula
\begin{align}
\label{sefo:16}  
         \iiint 
      \Bigl(\frac{\partial}{\partial y_\mu} \delta^3(\vec{x}-\vec{y}\,)\Bigr)
      \Bigl(\frac{\partial}{\partial y_\mu} \bphi(t,\vec{y}\,) \Bigr)
      ~d^3y= - \frac{\partial^2}{\partial x_\mu^2} \bphi(t,\vec{x}\,). 
\end{align}
The right-hand side of \eqref{sefo:10} is then
\begin{align}
\label{sefo:17}  
    \frac{\partial}{\partial t} \bpi(t,\vec{x}\,)
   =\sum_{1\leq\mu\leq 3} \frac{\partial^2}{\partial{x_\mu}^2} \bphi(t,\vec{x}\,)
   - m^2 \bphi(t,\vec{x}\,)
   - g \bigl(\bphi(t,\vec{x},)\bigr)^N,
\end{align}
which is just \eqref{sefo:2}. \END

\section{Rigorous calculations: the method}
%------------------------------------------
\label{rigo:0}
\setcounter{equation}{0}
\setcounter{definition}{0}
\setcounter{axiom}{0}
\setcounter{conjecture}{0}
\setcounter{lemma}{0}
\setcounter{theorem}{0}
\setcounter{corollary}{0}
\setcounter{proposition}{0}
\setcounter{example}{0}
\setcounter{remark}{0}
\setcounter{problem}{0}

Our goal is to show that Definition \ref{sefo:defi:1} makes sense and to provide a rigorous proof of Proposition \ref{sefo:prop:1} in the context of nonlinear generalized functions.  Unfortunately, if we examine the formal proof given in the previous section, and try to transpose it into the $\mathcal{G}$ setting, we are immediately confronted with the problem that the Hamiltonian $\mathbf{H}(\bphi,\bpi,t)$ is not yet defined since it depends on the self-interacting fields $\bphi$ and $\bpi$, which are the unknowns of the problem.  Moreover, the only thing that we can say \emph{a priori} about $\bphi$ and $\bpi$ is that, contrary to the free-fields $\bphi_0$ and $\bpi_0$, the coefficients of their expansion in creation and annihilation operators are certainly not simple, and thus possibly not such that $\bphi$ and $\bpi$ are moderate operator-valued functions in $\mathcal{G}$.

   In order to proceed we therefore need a constructive proof, i.e., a method such that we only use and build upon objects that are already known and mathematically well defined.  As it turns out such a method exists, but of course it is not without having its own limitations and difficulties.  In order define it, and to discuss its main shortcomings, we will begin by revisiting some basic aspects of quantum theory, starting with systems with a finite number of degrees of freedom, and then moving to field theory.\footnote{These aspects are likely to be well known to physicists familiar with QFT.  We nevertheless reformulate them here in a rather elementary language to render them accessible to physicists and mathematicians who do not share this familiarity.  Doing so will also motivate the unusual notations $\mathbf{H}^{(0)}, \mathbf{H}^{<0>}$, etc., that will be used in this chapter and the next.}

   Let us consider the variational problem\footnote{The following is adapted from the introduction of the book by Daniel Kastler on quantum electrodynamics \cite{KASTL1961-}.  This textbook has been used by J.-F.~C.~as the main reference for his early work on the formulation of QFT in the $\mathcal{G}$-framework \cite{COLOM1984-}.}
\begin{align}
\label{rigo:1}
  \delta \int L\Bigl( q_j(t), \dot{q}_j(t),t \Bigr) dt = 0, \qquad j = 1,..., d,
\end{align}
relative to the given classical Lagrangian $L(q_j,\dot{q}_j,t)$.  Its solution is given by Hamilton's equations
\begin{equation}
\label{rigo:2}
\left.
\begin{array}{l}
  \dfrac{dq_j}{dt}  = ~~~\dfrac{\partial H(q_j,p_j,t)}{\partial p_j} ,\\
\rule{0mm}{7mm}
  \dfrac{dp_j}{dt}  =   -\dfrac{\partial H(q_j,p_j,t)}{\partial q_j},
\end{array}
\quad \right\}
\end{equation}
where $H(q_j,p_j,t)$ is obtained by the elimination of the $\dot{q}_j$ between the relations
\begin{align}
\label{rigo:3}
  p_j  = \frac{\partial L(q_j,\dot{q}_j,t)}{\partial \dot{q}_j} ,
\qquad \text{and} \qquad
   H  = \sum_{j=1}^d p_j\dot{q}_j -  L(q_j,\dot{q}_j,t).
\end{align}
In quantum theory the problem defined by \eqref{rigo:1} is reformulated by replacing the canonical variables $q_j,p_j, j = 1,..., d$ with linear operators $\mathbf{q}_j,\mathbf{p}_j, j = 1,..., d$ operating on a Hilbert space $\mathcal{H}$, and subject to commutation relations.  Thus, instead of solving Hamilton's equation \eqref{rigo:2}, one is led to the problem:
\begin{problem}[Heisenberg]
%..........................
\label{rigo:probl:1}
 Let $\mathbf{H}(\mathbf{q}_j,\mathbf{p}_j,t)$ be given, and for simplicity let $\mathbf{H}(\mathbf{q}_j,\mathbf{p}_j,t)$ by a polynomial in $\mathbf{q}_j,\mathbf{p}_j, j = 1,..., d$ with $d \in \mathcal{N}$.  Find linear operators $\mathbf{q}_j$ and $\mathbf{p}_j$, operating on a Hilbert space $\mathcal{H}$, such that
\begin{equation}
\label{rigo:4}
\left.
\begin{array}{l}
     [\mathbf{q}_j, \mathbf{q}_k] = [\mathbf{p}_j, \mathbf{p}_k] =0,\\
\rule{0mm}{7mm}
     [\mathbf{q}_j, \mathbf{p}_k] = i \delta_{jk}~\mathbf{1},
\end{array}
\quad \right\}
\end{equation}
where $j,k = 1,..., d$, and
\begin{equation}
\label{rigo:5}
\left.
\begin{array}{l}
     \dfrac{\partial}{\partial t} \mathbf{q}_j   
      = i[ \mathbf{H}, \mathbf{q}_j ] ,\\
\rule{0mm}{7mm}
     \dfrac{\partial}{\partial t} \mathbf{p}_j 
      = i[ \mathbf{H}, \mathbf{p}_j ] .
\end{array}
\quad \right\}
\end{equation}
\end{problem}
This problem is solved in two steps, which can be qualified as `kinematical' and `dynamical:'

    {\bf (H-1)}  One begins by looking for a set of operators $\mathbf{q}_j^0,\mathbf{p}_j^0, j = 1,..., d$ operating on $\mathcal{H}$ satisfying the commutation relations
\begin{equation}
\label{rigo:6}
\left.
\begin{array}{l}
[\mathbf{q}_j^0, \mathbf{q}_k^0] = [\mathbf{p}_j^0, \mathbf{p}_k^0] =0,\\
\rule{0mm}{7mm}
[\mathbf{q}_j^0, \mathbf{p}_k^0] = i \delta_{jk}~\mathbf{1}.
\end{array}
\quad \right\}
\end{equation}
This problem has been solved by von~Neumann \cite{VONNE1931-}, who proved that there is essentially one solution (up to unitary equivalence): $\mathcal{H}$ is the space $\mathsf{L}^2(\mathbb{R}^d,\mathbb{C})$ of square integrable functions $\psi(q_1,q_2,...,q_d)$, and the operators $\mathbf{q}_j^0,\mathbf{p}_j^0$ are given by
\begin{equation}
\label{rigo:7}
\left.
\begin{array}{l}
(\mathbf{q}_j^0 \psi) (q_j) = ~~~~~ q_j \cdot \psi (q_j),\\
\rule{0mm}{7mm}
(\mathbf{p}_j^0 \psi) (q_j) = -i \dfrac{\partial}{\partial q_j}\psi (q_j).
\end{array}
\quad \right\}
\end{equation}
This realization of the $\mathbf{q}_j$ and $\mathbf{p}_j$ is that of the wave mechanics of de Broglie and Schr\"odinger.  It is called the \emph{Schr\"odinger representation of the canonical commutation relations}, and by the Stone-von~Neumann uniqueness theorem \cite{VONNE1931-}, \cite[p.\,6]{SCHWE1961-}, all other irreducible representations are unitarily equivalent to it provided $d < \infty$.  This means that one can go from one representation to another without changing the physical description of the system, which is why such a unitary transformation is referred to as a `change of picture.'

    {\bf (H-2)} One then considers the operator 
\begin{align}
\label{rigo:8}
     \mathbf{H}^{(0)}(\mathbf{q}_j^0,\mathbf{p}_j^0,t) \DEF
     \mathbf{H}(\mathbf{q}_j^0,\mathbf{p}_j^0,t),
\end{align}
i.e., the Hamiltonian in which the $\mathbf{q}_j$ and $\mathbf{p}_j$ have been replaced by $\mathbf{q}_j^0$ and $\mathbf{p}_j^0$.  In the physically interesting cases $\mathbf{H}^{(0)}$ is self-adjoint and thus defines, by the equation
\begin{equation}
\label{rigo:9}
\left.
\begin{array}{l} 
\dfrac{\partial \mathbf{U}(t,\tau)}{\partial t}
   = -i\mathbf{H}^{(0)}(\mathbf{q}_j^0,\mathbf{p}_j^0,t) ~ \mathbf{U}(t,\tau),\\
\rule{0mm}{7mm}
\mathbf{U}(\tau,\tau) = \mathbf{1},
\end{array}
\quad \right\}
\end{equation}
a unitary time-evolution operator $\mathbf{U}(t,\tau)$  which by means of
\begin{equation}
\label{rigo:10}
\left.
\begin{array}{l} 
  \mathbf{q}_j(t) = \mathbf{U}^{-1}(t,\tau) ~\mathbf{q}_j^0 ~\mathbf{U}(t,\tau),\\
\rule{0mm}{7mm}
  \mathbf{p}_j(t) = \mathbf{U}^{-1}(t,\tau) ~\mathbf{p}_j^0 ~\mathbf{U}(t,\tau),
\end{array}
\quad \right\}
\end{equation}
provides the solution of Problem \ref{rigo:probl:1}.  Of course, expressed in that form, the solution is simply an instance of Heisenberg's picture.  To switch to Schr\"odinger's picture one sets 
\begin{align}
\label{rigo:11}
     \psi(q_j,t) =  \mathbf{U}(t,\tau) ~\psi(q_j,\tau),
\end{align}
and \eqref{rigo:9} yields,
\begin{equation}
\label{rigo:12}
i\dfrac{\partial}{\partial t} \psi(q_j,t)
   = \mathbf{H}^{(0)}(\mathbf{q}_j^0,\mathbf{p}_j^0,t) ~ \psi(q_j,\tau),
\end{equation}
which is simply Schr\"odinger's equation.

  Now, in the quantum field theoretical case, the operators $\mathbf{q}_j$ and $\mathbf{p}_j$ operating on a Hilbert space $\mathcal{H}$ are replaced by the operators $\bphi(t,\vec{x}\,)$ and $\bpi(t,\vec{x}\,)$ operating on a Fock space $\mathbb{F}$, where the finite number of indices $j = 1, ..., d \in \mathbb{N}$ is replaced by the continuous parameters $\{t,\vec{x}\,\} \in \mathbb{R}^4$.  The problem to be solved is then:\footnote{This formulation corresponds to a spin-0 field.  In the case $s \neq 0$ the operators $\bphi$ and $\bpi$ have to be replaced by the collections $\{ \bphi_j \}$ and $\{ \bpi_j \}$ where $j = 1, 2, ..., (2s+1)$, and the commutators by anticommutators when $s$ is a half-integer rather than an integer.}
\begin{problem}[Heisenberg-Pauli]
%................................
\label{rigo:probl:2}
 Let $\mathbf{H}\bigl(\bphi(t,\vec{x}\,),\bpi(t,\vec{x}\,)\bigr)$ be given, and for simplicity let $\mathbf{H}\bigl(\bphi(t,\vec{x}\,),\bpi(t,\vec{x}\,)\bigr)$ by a polynomial in $\bphi(t,\vec{x}\,),\bpi(t,\vec{x}\,), \partial_t\bphi(t,\vec{x}\,),$ and $\partial_{\vec{x}}\bphi(t,\vec{x}\,)$ with $\{t,\vec{x}\,\} \in \mathbb{R}^4$.  Find linear operators $\bphi(t,\vec{x}\,)$ and $\bpi(t,\vec{x}\,)$ operating on a Fock space $\mathbb{F}$, such that
\begin{equation}
\label{rigo:13}
\left.
\begin{array}{l}
     [\bphi(t,\vec{x}_1), \bphi(t,\vec{x}_2)]
   = [\bpi (t,\vec{x}_1), \bpi (t,\vec{x}_2)] =0,\\
\rule{0mm}{7mm}
     [\bphi(t,\vec{x}_1), \bpi (t,\vec{x}_2)]
   = i \delta^3(\vec{x}_1-\vec{x}_2)~\mathbf{1},
\end{array}
\quad \right\}
\end{equation}
and
\begin{equation}
\label{rigo:14}
\left.
\begin{array}{l}
     \dfrac{\partial}{\partial t} \bphi(t,\vec{x}\,)   
      = i[ \mathbf{H}, \bphi(t,\vec{x}\,) ] ,\\
\rule{0mm}{7mm}
     \dfrac{\partial}{\partial t} \bpi(t,\vec{x}\,) 
      = i[ \mathbf{H}, \bpi(t,\vec{x}\,) ] .
\end{array}
\quad \right\}
\end{equation}
\end{problem}
To solve this problem one can try to use a two step procedure similar to the one above.  That is:

    {\bf (H-P-1)} One begins with the `kinematics,' i.e., one looks for a set of operators $\bphi^0(t,\vec{x}\,)$ and $\bpi^0(t,\vec{x}\,)$ such that
\begin{equation}
\label{rigo:15}
\left.
\begin{array}{l}
     [\bphi^0(t,\vec{x}_1), \bphi^0(t,\vec{x}_2)]
   = [\bpi^0 (t,\vec{x}_1), \bpi^0 (t,\vec{x}_2)] =0,\\
\rule{0mm}{7mm}
     [\bphi^0(t,\vec{x}_1), \bpi^0 (t,\vec{x}_2)]
   = i \delta^3(\vec{x}_1-\vec{x}_2)~\mathbf{1}.
\end{array}
\quad \right\}
\end{equation}

    {\bf (H-P-2)}  One solves the `dynamics' induced by the operator $\mathbf{H}^{(0)}\bigl(\bphi^0(t,\vec{x}\,),$ $ \bpi^0(t,\vec{x}\,)\bigr)$, i.e., the Hamiltonian in which the operators  $\bphi(t,\vec{x}\,)$ and $\bpi(t,\vec{x}\,)$ have been replaced by $\bphi^0(t,\vec{x}\,)$ and $\bpi^0(t,\vec{x}\,)$. Thus, provided $\mathbf{H}^{(0)}$ is self-adjoint, the equation 
\begin{equation}
\label{rigo:16}
\left.
\begin{array}{l} 
\dfrac{\partial \mathbf{U}(t,\tau)}{\partial t}
   = -i\mathbf{H}^{(0)}\bigl(\bphi^0(t,\vec{x}\,),\bpi^0(t,\vec{x}\,)\bigr) ~ \mathbf{U}(t,\tau),\\
\rule{0mm}{7mm}
\mathbf{U}(\tau,\tau) = \mathbf{1},
\end{array}
\quad \right\}
\end{equation}
should yield a unitary operator $\mathbf{U}(t,\tau)$ which by means of
\begin{equation}
\label{rigo:17}
\left.
\begin{array}{l} 
  \bphi(t,\vec{x}\,) = \mathbf{U}^{-1}(t,\tau) ~\bphi^0(\tau,\vec{x}\,)
                      ~\mathbf{U}(t,\tau),\\
\rule{0mm}{7mm}
  \bpi (t,\vec{x}\,) = \mathbf{U}^{-1}(t,\tau) ~\bpi^0 (\tau,\vec{x}\,)
                       ~\mathbf{U}(t,\tau),
\end{array}
\quad \right\}
\end{equation}
provides the solution of Problem \ref{rigo:probl:2}.

   The trouble with this approach is that for $d=\infty$, as is the case here because the operators $\bphi(t,\vec{x}\,)$ and $\bpi(t,\vec{x}\,)$ are labeled by the continuous indices $\{t,\vec{x}\,\} \in \mathbb{R}^4$, the Stone-von~Neumann theorem does not apply (see, e.g., \cite[p.\,163]{SCHWE1961-}).  Any two irreducible representations of the canonical commutation relations \eqref{rigo:15} are then in general unitarily inequivalent:  Different inequivalent representations will give rise to different physical pictures with different physical implications.  However, as is well known since the early days of QFT, there is at least one representation, called the \emph{Fock representation of the canonical commutation rules}, which has many (if not all) of the good properties that are necessary to get the proper physical implications of QFT.  This representation is provided by the free fields, that is
\begin{equation}
\label{rigo:18}
\left.
\begin{array}{l} 
  \bphi^0(t,\vec{x}\,) \DEF \bphi_0(t,\vec{x}\,),\\
\rule{0mm}{7mm}
  \bpi^0 (t,\vec{x}\,) \DEF \bpi_0 (t,\vec{x}\,).
\end{array}
\quad \right\}
\end{equation}
%
%where $\bphi_0(\rho_\epsilon,t,\vec{x}\,)$ defined by \eqref{embo:5} and $\bpi_0(\rho_\epsilon,t,\vec{x}\,)$ by \eqref{embo:22}
%
Indeed, as was seen in Chapter \ref{free:0} where these field were embedded in $\mathcal{G}$, this representation satisfies the canonical commutation relations (\ref{comr:1}--\ref{comr:3}) and it leads to a \emph{free-field} Hamiltonian $\mathbf{H}_0^{(0)}\bigl(\bphi^0(t,\vec{x}\,),$ $ \bpi^0(t,\vec{x}\,)\bigr) \equiv \mathbf{H}_0\bigl(\bphi_0(\rho_\epsilon,t,\vec{x}\,),$ $ \bpi_0(\rho_\epsilon,t,\vec{x}\,)\bigr)$ which is positive, bounded from below, and even finite in $\mathcal{G}$.  Moreover, the Fock representation is characterized by the fact that a vacuum state exists (which in general is not the case for other possible irreducible representations) and this representation is devoid of the pathologies associated with those which do not satisfy the hypotheses leading to the Stone-von~Neumann theorem.  We can therefore hope that in the case of an \emph{interacting-field} the corresponding Hamiltonian $\mathbf{H}^{(0)}$ will also turn out to be physically acceptable.

   But, of course, the operators $\bphi_0$ and  $\bpi_0$, and thus $\mathbf{H}^{(0)}$, are unbounded --- an intrinsic problem which has its origin in the definitions (\ref{fock:7}--\ref{fock:8}) of the creation and annihilation operators.  Nevertheless, in view of the extraordinary predictive power of QFT, which in its standard formulation is based on the systematic use of the Fock representation, we will admit as a \emph{postulate} the uniqueness of the representation \eqref{rigo:18} in $\mathcal{G}$, and we will continue to cope with the unboundness of the operators by restricting their action to the states $\Psi \in \mathbb{D}$, i.e., to the states with a finite number of particles \eqref{fock:6}:\footnote{This conjecture may possibly be proved with the help of the Weyl form of the commutation relations as in the original proof of von~Neumann \cite{VONNE1931-}, or with the methods used in the recent generalizations of the Stone-von~Neumann theorem, e.g., \cite{CAVAL1999-}.}
\begin{conjecture}[Unicity of ~the ~Fock representation]
%.....................................................
\label{rigo:conj:1}
~ Let $(\bphi^0, \bpi^0, \mathbb{D})$ where $\bphi^0(t,\vec{x}\,), \bpi^0(t,\vec{x}\,) \in \mathcal{E}_{\text{\rm M}}$ are defined by \eqref{rigo:18} be the Fock representation of the canonical commutation relations \eqref{rigo:15} in $\mathbb{D}$.  Then any other irreducible representation $(\bphi, \bpi, \mathbb{D})$ where $\bphi(t,\vec{x}\,), \bpi(t,\vec{x}\,) \in \mathcal{E}_{\text{\rm M}}$ is unitarily equivalent to $(\bphi^0, \bpi^0, \mathbb{D})$, i.e., $\exists \mathbf{V}$ unitary such that $\bphi = \mathbf{V}^{-1} ~\bphi^0~\mathbf{V}$ and  $\bpi = \mathbf{V}^{-1} ~\bpi^0~\mathbf{V}$.
\end{conjecture}

  Consequently, the Hamiltonian $\mathbf{H}^{(0)}$ of the self-interacting field corresponding to Definition \ref{sefo:defi:1}, i.e., (\ref{sefo:7}--\ref{sefo:8}), will be
\begin{align}  % summ:21
\notag
       \mathbf{H}^{(0)}(\tau) &= \iiint_{\mathbb{R}^3} \Bigl\{
         \frac{1}{2} \bigl(\bpi_0(\tau,\vec{x}\,)\bigr)^2 
       + \frac{1}{2} \sum_{1 \leq \mu \leq 3}
         \bigl(\partial_\mu \bphi_0(\tau,\vec{x}\,)\bigr)^2 \\
\label{rigo:19}
       &+ \frac{1}{2} m^2
         \bigl(\bphi_0(\tau,\vec{x}\,)\bigr)^2 
 + \frac{g}{N+1} \bigl(\bphi_0(\tau,\vec{x}\,)\bigr)^{N+1} \Bigr\} ~d^3x. 
\end{align}
The solution of \eqref{rigo:16} is then 
\begin{equation}
\label{rigo:20}
    \mathbf{U}(t,\tau) = \exp\bigl(-i(t-\tau)\mathbf{H}^{(0)}\bigr),
\end{equation}
so that the solution \eqref{rigo:17} of the Heisenberg-Pauli problem can be written as
\begin{align}
\label{rigo:21}
     \bphi(t,\tau,\vec{x}\,)   =
          \rme^{ i(t-\tau)\mathbf{H}^{(0)}(\tau) }
                 ~ \bphi_0(\tau,\vec{x}\,) ~
          \rme^{-i(t-\tau)\mathbf{H}^{(0)}(\tau) },\\
\label{rigo:22}
     \bpi(t,\tau,\vec{x}\,)   =
          \rme^{ i(t-\tau)\mathbf{H}^{(0)}(\tau) }
                 ~ \bpi_0(\tau,\vec{x}\,) ~
          \rme^{-i(t-\tau)\mathbf{H}^{(0)}(\tau) }.
\end{align}
These are the Heisenberg equations for the interacting fields operators $\bphi(t,\tau,\vec{x}\,)$ and $\bpi(t,\tau$, $\vec{x}\,)$ that will be considered in the following.  As is manifest in their form, where $\mathbf{H}^{(0)}(\tau) \equiv \mathbf{H}\bigl(\bphi_0(\tau,\vec{x}\,)$, $ \bpi_0(\tau,\vec{x}\,)\bigr)$, they can be seen as formulas expressing these field-operators in terms of some initial free-field operators.  They are therefore suitable for a `constructive' proof of Proposition \ref{sefo:prop:1} in the context of nonlinear generalized functions.\footnote{Furthermore, their form will turn out to be well suited to the applications to scattering problems that will be discussed in Chapter \ref{scat:0}.  The initial conditions \eqref{sefo:3} will then correspond to asymptotic fields (that is defined at times $\tau \rightarrow -\infty$) which will be postulated to correspond to free-field operators, i.e., $\bphi_{\text{\rm ini}} = \bphi_0$ and $\bpi_{\text{\rm ini}} = \bpi_0$.}

  In practice, to go through this proof, we will have to follow a somewhat tortuous path.  This is because we will encounter technical problems due to the domains of the operators (which create difficulties when composing them) as well as to the time-dependences of these domains (which create difficulties when calculating their time-derivatives).

Our procedure will thus be as as follows:

\begin{enumerate}

\item The Hamiltonian $\mathbf{H}^{(0)}(\tau)$ will be embedded as $\mathbf{H}^{(0)}(\rho_\epsilon,\FOU{\chi}_\epsilon,\tau)$ and shown to have a self-adjoint extension denoted by $\mathbf{H}^{<0>}(\rho_\epsilon,\FOU{\chi}_\epsilon,\tau)$;

\item The domains $\mathsf{D}$ and ranges $\mathsf{R}$ of the operators will be analyzed;

\item The commutation relations of the $\mathcal{G}$-embedded self-interacting fields $\bphi$ and $\bpi$ will be proved;

\item A definition of time-differentiation consistent with time-dependent domains will be provided;

\item The equivalence of the integral and differential forms of the Heisenberg equations will be proved;

\item That, in their $\mathcal{G}$-embedded forms, $\bpi$ defined by \eqref{rigo:22} is the $t$-derivative of $\bphi$ defined by \eqref{rigo:21} will be proved;

\item Finally, that the rigorous Heisenberg equations of motion (\ref{diff:10}--\ref{diff:11}) solve the rigorous interacting-field equation (\ref{seri:18}--\ref{seri:19}) will be demonstrated.

\end{enumerate}

\section{Hamiltonian: self-adjoint extension}
%--------------------------------------------
\label{hami:0}
\setcounter{equation}{0}
\setcounter{definition}{0}
\setcounter{axiom}{0}
\setcounter{conjecture}{0}
\setcounter{lemma}{0}
\setcounter{theorem}{0}
\setcounter{corollary}{0}
\setcounter{proposition}{0}
\setcounter{example}{0}
\setcounter{remark}{0}
\setcounter{problem}{0}

The $\mathcal{G}$-embedding of the full Hamiltonian \eqref{rigo:19} is obtained by adding the self-interaction term to the free-field Hamiltonian \eqref{hamr:1} and setting $t=\tau$, i.e.,
\begin{align}
\notag
     \mathbf{H}^{(0)}(\rho_\epsilon,\FOU{\chi}_\epsilon,\tau)
    &= \iiint_{\mathbb{R}^3} \Bigl\{
      \frac{1}{2} \bigl(\bpi_0(\rho_\epsilon,\tau,\vec{x}\,)\bigr)^2 
    + \frac{1}{2} \sum_{1 \leq \mu \leq 3}
      \bigl(\partial_\mu \bphi_0(\rho_\epsilon,\tau,\vec{x}\,)\bigr)^2 \\
\label{hami:1}  
    &+ \frac{1}{2} m^2
      \bigl(\bphi_0(\rho_\epsilon,\tau,\vec{x}\,)\bigr)^2 
    + \frac{g}{N+1} \bigl(\bphi_0(\rho_\epsilon,\tau,\vec{x}\,)\bigr)^{N+1}
               \Bigr\}~\FOU{\chi}(\epsilon \vec{x}\,) ~d^3x, 
\end{align}
where $\FOU{\chi}$ is the usual damper insuring that integrating over the whole of $\mathbb{R}^3$ makes sense.  Obviously, $\mathbf{H}^{(0)}(\rho_\epsilon,\FOU{\chi}_\epsilon,\tau)$ maps $\mathbb{D}$ into $\mathbb{D}$. Moreover,  $\mathbf{H}^{(0)}(\rho_\epsilon,\FOU{\chi}_\epsilon,\tau)$ is symmetric because $\rho_\epsilon$ and $\FOU{\chi}_\epsilon$ are real, i.e.,
\begin{align}
\label{hami:2}  
  \forall \Phi_1, \Phi_2 \in \mathbb{D}  \qquad 
   \SCA \Phi_1 , \mathbf{H}^{(0)}(\rho_\epsilon,\FOU{\chi}_\epsilon,\tau)
        \Phi_2 \LAR_\mathbb{F} =
   \SCA \mathbf{H}^{(0)}(\rho_\epsilon,\FOU{\chi}_\epsilon,\tau) \Phi_1
      , \Phi_2 \LAR_\mathbb{F}.
\end{align}

   We are now going to prove that $\mathbf{H}^{(0)}(\rho_\epsilon,\FOU{\chi}_\epsilon,\tau)$ admits a self-adjoint extension denoted by $\mathbf{H}^{<0>}(\rho_\epsilon,\FOU{\chi}_\epsilon,\tau)$.   That extension is needed to prove that the operator $\exp\bigl(i(t-\tau) \mathbf{H}^{<0>}(\rho_\epsilon,\FOU{\chi}_\epsilon,\tau) \bigr)$, which will replace $\exp\bigl(i(t-\tau) \mathbf{H}^{(0)}(\rho_\epsilon,\FOU{\chi}_\epsilon,\tau) \bigr)$ in  the embedded forms of (\ref{rigo:21}--\ref{rigo:22}), is defined mathematically and is unitary:  
\begin{theorem} [Self-adjoint extension]
%.......................................
\label{hami:theo:1}
The Hamiltonian $\mathbf{H}^{(0)}(\rho_\epsilon,\FOU{\chi}_\epsilon,\tau) : \mathbb{D} \rightarrow \mathbb{D} \subset \mathbb{F}$ admits the self-adjoint extension $\mathbf{H}^{<0>}(\rho_\epsilon,\FOU{\chi}_\epsilon,\tau)$, on a domain containing $\mathbb{D}$ denoted by $\mathbb{D}^{<0>}$, constructed below.\footnote{The original version of this proof is due to B.\ Perrot (unpublished), see also, \cite[p.\,311--313]{COLOM1984-}. For brevity we do not always write the arguments $\rho_\epsilon$, $\FOU{\chi}_\epsilon$, and $\tau$ of $\mathbf{H}^{<0>}$ and $\mathbf{H}^{(0)}$ anymore.}
\end{theorem}
Proof:  Let us consider the infinite product of Hilbert space
$\mathbf{F} = \prod_{n=0}^\infty\mathsf{L}_S^2\bigl((\mathbb{R}^3)^n \bigr)$, i.e.,  the collection of sequences $\mathbf{F} = \bigl\{ (...,f_n,...) \bigr\}$, without any information on $\sum_{n=1}^{+\infty}\|(f_n)\|^2_{\mathsf{L}_S^2\bigl((\mathbb{R}^3)^n \bigr)}$ which may be a divergent series.\footnote{We write $\SCA~|~\LAR_\mathbb{F}$ and $\|~\|_{\mathsf{L}^2_S}$ for the scalar product and norm since the embedded forms $\BRA~\|~\KET_\mathcal{G}$ and $\|~\|_{\mathsf{L}^2_\mathcal{G}}$ are not explicitly needed in the proof.  }  (Contrary to $\mathbb{F} \subset \mathbf{F}$, no norm exist on $\mathbf{F}$.)   Because the operator $\mathbf{H}^{<0>}$ has a finite number of creation and annihilation operators, it defines a map $\mathbf{H}^{<0>} : \mathbf{F} \rightarrow \mathbf{F}$ which, just like $\mathbf{H}^{(0)} : \mathbb{D} \rightarrow \mathbb{D}$, gives non-zero contributions on only a finite number of indices $n$, i.e., $p-N-1 \leq n \leq p+N+1$, when acting on a state $(...,f_n,...)$ with $f_i = 0$ for $i\neq n$. This is why $\mathbf{H}^{<0>}$ can be defined as an operator from $\mathbf{F}$ into $\mathbf{F}$.  As $\mathbb{F} \subset \mathbf{F}$, we therefore define the range
\begin{align}
\label{hami:3}  
 \mathbb{D}^{<0>} \DEF \Bigl\{ \Phi \in \mathbb{F} 
                  \quad\text{such that}\quad
    \mathbf{H}^{<0>}(\rho_\epsilon,\FOU{\chi}_\epsilon,\tau)\Phi \in \mathbb{F}
                       \Bigr\},
\end{align}
i.e., $\mathbb{D}^{<0>} \subset \mathbb{F}$ is the dense subspace of states $\Phi$ such that $\mathbf{H}^{<0>}\Phi$ has the property that $\sum_{n=1}^{+\infty}\|(f_n)\|^2_{\mathsf{L}_S^2\bigl((\mathbb{R}^3)^n \bigr)}$ converges.   One obviously has $\mathbb{D} \subset \mathbb{D}^{<0>}$.

Let us use the notation $\bigl(g; \mathsf{D}(g)\bigr)=\bigl(\text{linear map};\text{domain of map}\bigr)$.  We want to prove that  $\bigl(\mathbf{H}^{<0>}|_{\mathbb{D}^{<0>}};\mathbb{D}^{<0>}\bigr)$, i.e., the restriction $\mathbf{H}^{<0>}|_{\mathbb{D}^{<0>}}$ of $\mathbf{H}^{<0>}$ to ${\mathbb{D}^{<0>}}$, is self-adjoint on $\mathbb{F}$, i.e., that the domain of its adjoint $\bigl(\mathbf{H}^{<0>}|_{\mathbb{D}^{<0>}};\mathbb{D}^{<0>}\bigr)^\ADJ$ is $\mathbb{D}^{<0>}$ again.  Since $\rho_\epsilon$ and  $\FOU{\chi}_\epsilon$ are real $\mathbf{H}^{<0>}|_{\mathbb{D}^{<0>}}$ is obviously symmetric, i.e., 
\begin{align}
\label{hami:4}  
  \forall \Phi_1, \Phi_2 \in \mathbb{D}^{<0>},  \qquad 
   \SCA \Phi_1 , \mathbf{H}^{<0>}|_{\mathbb{D}^{<0>}} \Phi_2 \LAR_\mathbb{F} =
   \SCA \mathbf{H}^{<0>}|_{\mathbb{D}^{<0>}} \Phi_1 , \Phi_2 \LAR_\mathbb{F}.
\end{align}

   Let us recall the meaning of the formula $\SCA \mathbf{A} x, y\LAR = \SCA  x,\mathbf{A}^\ADJ y\LAR$ for an unbounded operator $\bigl(\mathbf{A}; \mathsf{D}(\mathbf{A})\bigr)$ on the Hilbert space $\mathbb{F}$.  By definition, the domain $\mathsf{D}(\mathbf{A}^\ADJ) \DEF \bigl\{ ~ y \in \mathbb{F}$ such that the linear map $l_y : x \mapsto \SCA \mathbf{A} x, y\LAR $ from $\mathsf{D}(\mathbf{A}) \rightarrow \mathbb{C}$ can be extended on $\mathbb{F}$ as a continuous linear map, i.e., $\exists C(y)$ such that $|\SCA \mathbf{A} x, y\LAR| \leq C(y) \| x \|_\mathbb{F}, \forall x \in \mathsf{D}(\mathbf{A}) ~ \bigr\}$.  Then, by Riesz's theorem, $\exists z \in \mathbb{F}$ such that $l_y(x) = \SCA x, z \LAR, \forall x$, so that by definition $z=\mathbf{A}^\ADJ y$.

    Let $\mathbb{D}^{<0>\ADJ}$ be the domain of the adjoint $\bigl(\mathbf{H}^{<0>}|_{\mathbb{D}^{<0>}};\mathbb{D}^{<0>}\bigr)^\ADJ$.  Obviously, $\mathbb{D}^{<0>} \subset \mathbb{D}^{<0>\ADJ}$.  So let us prove that $\mathbb{D}^{<0>\ADJ} \subset \mathbb{D}^{<0>}$, i.e., we take $\Phi \in \mathbb{D}^{<0>\ADJ}$ and prove that $\Phi \in \mathbb{D}^{<0>}$.  Let $s_n \in  \mathsf{L}_S^2\bigl((\mathbb{R}^3)^n \bigr) \subset \mathbb{D}$ such that $s_n\neq 0$ for some index $n$ and zero otherwise.  Then, since $s_n \in  \mathbb{D}$ and $\mathbf{H}^{(0)}: \mathbb{D} \rightarrow \mathbb{D}$, it follows that $\mathbf{H}^{<0>}|_{\mathbb{D}^{<0>}}(s_n) = \mathbf{H}^{(0)}(s_n) \in  \mathbb{D} \subset \mathbb{F}$.  Thus $s_n \in \mathbb{D}^{<0>}$ by definition of $\mathbb{D}^{<0>}$.  Therefore, using  $\mathbf{H}^{<0>}|_{\mathbb{D}^{<0>}}(s_n) = \mathbf{H}^{(0)}(s_n)$ and Eq.~\eqref{hami:4} we have
\begin{align}
\label{hami:5}  
       \SCA \mathbf{H}^{(0)} s_n, \Phi \LAR_\mathbb{F} 
     = \SCA s_n, (\mathbf{H}^{<0>}|_{\mathbb{D}^{<0>}})^\ADJ \Phi \LAR_\mathbb{F} 
     = \SCA s_n, \bigl((\mathbf{H}^{<0>}|_{\mathbb{D}^{<0>}})^\ADJ \Phi \bigr)_n \LAR_\mathbb{F},
\end{align}
because only the component $n$ of $s_n$ is non zero.  On the other hand, as $\SCA~,~\LAR$ can evidently be extended from $\mathbb{D}\times\mathbb{F}\rightarrow\mathbb{C}$ to $\mathbb{F}\times\mathbb{F}\rightarrow\mathbb{C}$, using  Eq.~\eqref{hami:4} and $\mathbf{H}^{(0)}(s_n) \in \mathbb{D}$, we have 
\begin{align}
\label{hami:6}  
       \SCA \mathbf{H}^{(0)} s_n, \Phi \LAR_\mathbb{F} 
     = \SCA s_n, \mathbf{H}^{<0>} \Phi \LAR_\mathbb{F} 
     = \SCA s_n, \bigl( \mathbf{H}^{<0>} \Phi \bigr)_n \LAR_\mathbb{F},
\end{align}
where $(\mathbf{H}^{<0>} \Phi)_n \in \mathbb{F}$ is equal to the $n$th component of $\mathbf{H}^{<0>}\Phi$ at the index $n$ and zero otherwise.  Consequently, comparing (\ref{hami:5}--\ref{hami:6}),
\begin{align}
\label{hami:7}  
      \SCA s_n, \bigl((\mathbf{H}^{<0>}|_{\mathbb{D}^{<0>}})^\ADJ \Phi \bigr)_n
      \LAR_\mathbb{F}
    = \SCA s_n, \bigl( \mathbf{H}^{<0>} \Phi \bigr)_n \LAR_\mathbb{F}, 
      \quad
      \forall n, \forall s_n \in  \mathsf{L}_S^2\bigl((\mathbb{R}^3)^n \bigr),
\end{align}
and therefore
\begin{align}
\label{hami:8}  
      \bigl((\mathbf{H}^{<0>}|_{\mathbb{D}^{<0>}})^\ADJ \Phi \bigr)_n
    = \bigl( \mathbf{H}^{<0>} \Phi \bigr)_n, 
      \quad
      \forall n.
\end{align}
Since $\Phi \in \mathbb{D}^{<0>\ADJ}$, we have then
\begin{align}
\nonumber
    (\mathbf{H}^{<0>}|_{\mathbb{D}^{<0>}})^\ADJ \Phi \in \mathbb{F}
    \qquad \Rightarrow \qquad  &\sum_{n=1}^{+\infty}
   \| \bigl( (\mathbf{H}^{<0>}|_{\mathbb{D}^{<0>}})^\ADJ \Phi \bigr)_n
   \|^2_{\mathsf{L}_S^2\bigl((\mathbb{R}^3)^n \bigr)}
 < \infty,\\
\label{hami:9}
    \qquad \Rightarrow \qquad  &\sum_{n=1}^{+\infty}
   \| \bigl(  \mathbf{H}^{<0>}                       \Phi \bigr)_n
   \|^2_{\mathsf{L}_S^2\bigl((\mathbb{R}^3)^n \bigr)}
 < \infty,
\end{align}
hence $\mathbf{H}^{<0>}\Phi \in \mathbb{F}$, i.e., $\Phi \in \mathbb{D}^{<0>}$. ~~\END
\begin{remark} [Self-adjoint extension of free-field operators]
%..............................................................
\label{hami:rema:1}
The same proof can be used to demonstrate that $\bphi_0(\rho_\epsilon,\tau,\vec{x}\,)$,  $\bpi_0(\rho_\epsilon,\tau,\vec{x}\,)$, and polynomials in these operators, admit a similar self-adjoint extension.
\end{remark}

\section{Analysis of domain and range of operators}
%--------------------------------------------------
\label{anal:0}
\setcounter{equation}{0}
\setcounter{definition}{0}
\setcounter{axiom}{0}
\setcounter{conjecture}{0}
\setcounter{lemma}{0}
\setcounter{theorem}{0}
\setcounter{corollary}{0}
\setcounter{proposition}{0}
\setcounter{example}{0}
\setcounter{remark}{0}
\setcounter{problem}{0}

  Thanks to Theorem \ref{hami:theo:1} we can use the Hille-Yosida theory which implies that $\{\exp(it \mathbf{H}^{<0>})\}_{t \in \mathbb{R}}$ is a strongly continuous group of unitary operators on $\mathbb{F}$.  We therefore replace the Hamiltonian $\mathbf{H}^{(0)}$ appearing in the formal expression $\exp\bigl( i(t-\tau)\mathbf{H}^{(0)} \bigr)$ by its self-adjoint extension $\mathbf{H}^{<0>}$, i.e., we henceforth make the substitution
\begin{equation}\label{anal:1}
  \exp\bigl( i(t-\tau)\mathbf{H}^{(0)} \bigr)
     \quad \rightarrow \quad
  \exp\bigl( i(t-\tau)\mathbf{H}^{<0>} \bigr),
\end{equation}
which yields an unitary operator $\mathbb{F} \rightarrow \mathbb{F}$ that maps $\mathbb{D}^{<0>} \rightarrow \mathbb{D}^{<0>}$.  Similarly, as $\mathbf{H}^{<0>}$ is self-adjoint, we define the inverse of $\exp\bigl( i(t-\tau) \mathbf{H}^{<0>} \bigr)$ by
\begin{equation}\label{anal:2}
  \exp\bigl(-i(t-\tau)\mathbf{H}^{<0>} \bigr)
     \circ
  \exp\bigl( i(t-\tau)\mathbf{H}^{<0>} \bigr)
   = \mathbf{1},
\end{equation}
which enables us to make the substitution
\begin{equation}\label{anal:3}
  \exp\bigl( -i(t-\tau)\mathbf{H}^{(0)} \bigr)
     \quad \rightarrow \quad
  \exp\bigl( -i(t-\tau)\mathbf{H}^{<0>} \bigr).
\end{equation}
It is also an unitary operator mapping $\mathbb{D}^{<0>} \rightarrow \mathbb{D}^{<0>}$.

   We now define the range
\begin{equation}\label{anal:4}
  \mathbb{D}(\rho_\epsilon,\FOU{\chi}_\epsilon,t,\tau) \DEF
  \rme^{i(t-\tau)\mathbf{H}^{<0>}(\rho_\epsilon,\FOU{\chi}_\epsilon,\tau) }
  ~ \mathbb{D}\subset \mathbb{D}^{<0>}.
\end{equation}
 Since the exponential is unitary, $\mathbb{D}(\rho_\epsilon,\FOU{\chi}_\epsilon,t,\tau)$ is a dense vector subspace of $\mathbb{F}$ depending on $t$, and also on $\rho_\epsilon,\FOU{\chi}_\epsilon$, and $\tau$ as parameters.  (It is a subset of $\mathbb{D}^{<0>}$ because $\rme^{+i(t-\tau)\mathbf{H}^{<0>}}\mathbb{D}^{<0>} \subset \mathbb{D}^{<0>}$, and $\mathbb{D} \subset \mathbb{D}^{<0>}$.) Then,
\begin{align}
\label{anal:5}
  \rme^{+i(t-\tau)\mathbf{H}^{<0>} } ~ &: ~
   ~ \mathbb{D}   \rightarrow
     \mathbb{D}(\rho_\epsilon,\FOU{\chi}_\epsilon,t,\tau),\\
\intertext{and from the definition of $\mathbb{D}(\rho_\epsilon,\FOU{\chi}_\epsilon,t,\tau)$}
\label{anal:6}
  \rme^{-i(t-\tau)\mathbf{H}^{<0>} } ~ &: ~
   ~ \mathbb{D}(\rho_\epsilon,\FOU{\chi}_\epsilon,t,\tau) \rightarrow
     \mathbb{D}.
\end{align}
Therefore, if $\mathbf{A}$ is an operator (such as $\bphi_0$ or $\bpi_0$) which maps $\mathbb{D}$ into $\mathbb{D}$, or which has   $\mathsf{D}(\mathbf{A})=\mathbb{D}$ and $\mathsf{R}(\mathbf{A})=\mathbb{D}$ as possible domain and range, we have
\begin{align}
\label{anal:7}
  \rme^{ i(t-\tau)\mathbf{H}^{<0>} }
   \circ \mathbf{A} \circ
  \rme^{-i(t-\tau)\mathbf{H}^{<0>} }~ : ~
   ~ \mathbb{D}(\rho_\epsilon,\FOU{\chi}_\epsilon,t,\tau) \rightarrow
     \mathbb{D}(\rho_\epsilon,\FOU{\chi}_\epsilon,t,\tau).
\end{align}
Consequently, \emph{the domain and range of the Heisenberg equations} (\ref{rigo:21}--\ref{rigo:22}) \emph{is the time-dependent space defined by} \eqref{anal:4}.\footnote{The domain of the Heisenberg equations is in general time-dependent, even in the case of a finite number of degrees of freedom.}  Moreover
\begin{align}
\label{anal:8}
  \rme^{+i(t-\tau)\mathbf{H}^{<0>} }
  \circ    \mathbf{H}^{<0>} ~ &: ~
   ~ \mathbb{D} \rightarrow \mathbb{D}(\rho_\epsilon,\FOU{\chi}_\epsilon,t,\tau),
\end{align}
because $\mathbf{H}^{<0>}$ maps $\mathbb{D}$ into $\mathbb{D}$ since it extends $\mathbf{H}^{(0)}$, and
\begin{align}
\label{anal:9}
  \mathbf{H}^{<0>} \circ 
  \rme^{+i(t-\tau)\mathbf{H}^{<0>} } ~ &: ~
   ~ \mathbb{D} \rightarrow \mathbb{D}(\rho_\epsilon,\FOU{\chi}_\epsilon,t,\tau),
\end{align}
because $\mathbf{H}^{<0>}$ also maps $\mathbb{D}(\rho_\epsilon,\FOU{\chi}_\epsilon,t,\tau)$ into $\mathbb{D}(\rho_\epsilon,\FOU{\chi}_\epsilon,t,\tau)$.  Indeed
\begin{align}
\label{anal:10} 
  \rme^{+i(t-\tau)\mathbf{H}^{<0>} }
            \circ \mathbf{H}^{<0>} \circ
  \rme^{-i(t-\tau)\mathbf{H}^{<0>} } 
                = \mathbf{H}^{<0>},
\end{align}
maps $\mathbb{D}(\rho_\epsilon,\FOU{\chi}_\epsilon,t,\tau)$ into $\mathbb{D}(\rho_\epsilon,\FOU{\chi}_\epsilon,t,\tau)$ because $\mathbf{H}^{<0>}$ commutes with the exponentials.

\section{Commutation relations of self-interacting field}
%--------------------------------------------------------
\label{comm:0}
\setcounter{equation}{0}
\setcounter{definition}{0}
\setcounter{axiom}{0}
\setcounter{conjecture}{0}
\setcounter{lemma}{0}
\setcounter{theorem}{0}
\setcounter{corollary}{0}
\setcounter{proposition}{0}
\setcounter{example}{0}
\setcounter{remark}{0}
\setcounter{problem}{0}

We define the $\mathcal{G}$-embedding of $\bphi(t,\tau,\vec{x}\,)$ expressed by \eqref{rigo:21} as 
\begin{equation}\label{comm:1}
     \bphi(\rho_\epsilon,\FOU{\chi}_\epsilon,t,\tau,\vec{x}\,)   =
     \rme^{ i(t-\tau)\mathbf{H}^{<0>}(\rho_\epsilon,\FOU{\chi}_\epsilon,\tau) }
                 ~ \bphi_0(\rho_\epsilon,\tau,\vec{x}\,) ~
     \rme^{-i(t-\tau)\mathbf{H}^{<0>}(\rho_\epsilon,\FOU{\chi}_\epsilon,\tau) },
\end{equation}
According to \eqref{anal:7} it maps $\mathbb{D}(\rho_\epsilon,\FOU{\chi}_\epsilon,t,\tau)$ into itself.  Moreover, since $\mathbb{D}(\rho_\epsilon,\FOU{\chi}_\epsilon,t,\tau)$ is a dense vector subspace of $\mathbb{F}$ which does not depend on $\vec{x}$, the composition $\bphi(\rho_\epsilon,\FOU{\chi}_\epsilon,t,\tau,\vec{x}_1) \circ \bphi(\rho_\epsilon,\FOU{\chi}_\epsilon,t,\tau,\vec{x}_2)$ makes sense as an operator from $\mathbb{D}(\rho_\epsilon,\FOU{\chi}_\epsilon,t,\tau)$ into itself.

   Similarly, we define the $\mathcal{G}$-embedding of $\bpi(t,\tau,\vec{x}\,)$ expressed by \eqref{rigo:22} as formula \eqref{comm:1} with  $\bpi_0$ in place of $\bphi_0$ (we do not yet know that $\bpi$ is the time derivative of $\bphi$, although $\bpi_0=\partial_t\bphi_0$).  Therefore
\begin{equation}\label{comm:2}
     \bpi(\rho_\epsilon,\FOU{\chi}_\epsilon,t,\tau,\vec{x}\,)   =
     \rme^{ i(t-\tau)\mathbf{H}^{<0>}(\rho_\epsilon,\FOU{\chi}_\epsilon,\tau) }
                 ~ \bpi_0(\rho_\epsilon,\tau,\vec{x}\,) ~
     \rme^{-i(t-\tau)\mathbf{H}^{<0>}(\rho_\epsilon,\FOU{\chi}_\epsilon,\tau) }.
\end{equation}
The operator $\bpi$ has the same properties as $\bphi$.  In particular, the compositions $\bpi(\rho_\epsilon,\FOU{\chi}_\epsilon,t,\tau,\vec{x}_1) \circ \bpi(\rho_\epsilon,\FOU{\chi}_\epsilon,t,\tau,\vec{x}_2)$ and $\bphi(\rho_\epsilon,\FOU{\chi}_\epsilon,t,\tau,\vec{x}_1) \circ \bpi(\rho_\epsilon,\FOU{\chi}_\epsilon,t,\tau,\vec{x}_2)$ make sense, so that we can calculate the  commutation relations of $\bphi$ and $\bpi$.  Simplification of the exponentials, which eliminates the $t$-dependence, and the equal-time commutation relations (\ref{comr:5} -- \ref{comr:7}) of the free-field operators at the time $\tau$ give
\begin{align}
\label{comm:3}
 [    \bphi(\rho_\epsilon,\FOU{\chi}_\epsilon,t,\tau,\vec{x}_1),
      \bphi(\rho_\epsilon,\FOU{\chi}_\epsilon,t,\tau,\vec{x}_2)  ] &=0,\\
\label{comm:4}
 [    \bpi(\rho_\epsilon,\FOU{\chi}_\epsilon,t,\tau,\vec{x}_1),
      \bpi(\rho_\epsilon,\FOU{\chi}_\epsilon,t,\tau,\vec{x}_2)   ] &=0,\\
\label{comm:5}
 [    \bphi(\rho_\epsilon,\FOU{\chi}_\epsilon,t,\tau,\vec{x}_1),
      \bpi(\rho_\epsilon,\FOU{\chi}_\epsilon,t,\tau,\vec{x}_2)   ] &=
      i \delta^3_\epsilon(\rho^\vee \ast \rho,\vec{x}_1-\vec{x}_2)
      + \OOO(\epsilon^{q+1}), 
\end{align}
which are the rigorous form of the commutation relations (\ref{sefo:4}--\ref{sefo:6}), valid at least on $\mathbb{D}(\rho_\epsilon,\FOU{\chi}_\epsilon,t,\tau)$ where these operators are defined.

\section{Differentiation on time-dependent domains}
%--------------------------------------------------
\label{diff:0}
\setcounter{equation}{0}
\setcounter{definition}{0}
\setcounter{axiom}{0}
\setcounter{conjecture}{0}
\setcounter{lemma}{0}
\setcounter{theorem}{0}
\setcounter{corollary}{0}
\setcounter{proposition}{0}
\setcounter{example}{0}
\setcounter{remark}{0}
\setcounter{problem}{0}

   The field operator $\bphi(\rho_\epsilon,\FOU{\chi}_\epsilon,t,\tau,\vec{x}\,)$ defined by \eqref{comm:1} maps $\mathbb{D}(\rho_\epsilon,\FOU{\chi}_\epsilon,t,\tau)$  into itself.  This creates a serious problem with the $t$-derivative $\partial_t\bphi(\rho_\epsilon,\FOU{\chi}_\epsilon,t,\tau,\vec{x}\,)$ since the domain $\mathsf{D}(\bphi)=\mathbb{D}(\rho_\epsilon,\FOU{\chi}_\epsilon,t,\tau)$ depends on $t$.  In particular, it is not clear how this derivative should be defined and related to the conjugate operator $\bpi(\rho_\epsilon,\FOU{\chi}_\epsilon,t,\tau,\vec{x}\,)$.

   In the absence of a formulation such that $\bphi$ has a time-independent domain we content ourselves with a weaker definition of the $t$-derivative.\footnote{May be one should consider $\bphi$ as a generalized function on space-time for this problem of domain.}  Such a time-derivative can be obtained as follows:
\begin{definition}[Time-derivative]
%..................................
\label{diff:defi:1}
The map $\mathbf{A} : t \mapsto \mathbf{A}(t)$ from $\mathbb{R}$ to the set $L\bigl(\mathbb{D}(t),\mathbb{F}\bigr)$ of linear operators from $\mathbb{D}(t) \rightarrow \mathbb{F}$, where $\mathbb{D}(t)$ is a dense subset of $\mathbb{F}$, is differentiable with derivative $t \mapsto \mathbf{B}(t) \in L\bigl(\mathbb{D}(t),\mathbb{F}\bigr)$ iff: for any $\mathcal{C}^1$ map $\Phi: t \mapsto \Phi(t)$ from $\mathbb{R}$ to $\mathbb{F}$, with $\Phi(t) \in \mathbb{D}(t)$ and $\Phi'(t) = \partial_t \Phi \in \mathbb{D}(t), \forall t$, then the map $t \mapsto \mathbf{A}(t).\Phi(t)$ from $\mathbb{R}$ to $\mathbb{F}$ is differentiable with derivative $t \mapsto \mathbf{B}(t).\Phi(t) + \mathbf{A}(t).\Phi'(t)$.
\end{definition}
We adopt this definition with
\begin{equation}\label{diff:1}
  \mathbb{D}(t) \DEF \mathbb{D}(\rho_\epsilon,\FOU{\chi}_\epsilon,t,\tau).
\end{equation}
given by \eqref{anal:4} and the maps $\Phi$ of the form
\begin{equation}\label{diff:2}
  \Phi(t) = \rme^{i(t-\tau)\mathbf{H}^{<0>}}\Psi(t),
\end{equation}
with $\Psi$ a $\mathcal{C}^\infty$ (or $\mathcal{C}^1$: it does not matter) map from $\mathbb{R}$ into $\mathbb{D}$, where $\mathbb{D}$ has the structure of union of normed spaces defined in Sec.~\ref{oper:0}.1. (The definition of $\mathcal{C}^\infty$ maps is the same as the one given in definition~\ref{oper:defi:3} in which $\mathsf{L}(\mathbb{D})$ is replaced by $\mathbb{D}$.)  Then
\begin{align}
\label{diff:3}
  \Phi'(t) =  i \rme^{i(t-\tau)\mathbf{H}^{<0>}}
                \mathbf{H}^{<0>}(\rho_\epsilon,\FOU{\chi}_\epsilon,\tau)~\Psi(t)
              + \rme^{i(t-\tau)\mathbf{H}^{<0>}}\Psi'(t),
\end{align}
is in $\mathbb{D}(t)$ since $\Psi(t)$ and $\Psi'(t) \in \mathbb{D}$ and $\mathbf{H}^{<0>}\bigr|_{\mathbb{D}}=\mathbf{H}^{(0)}$.
\begin{proposition}[Unicity of time-derivative]
%..............................................
\label{diff:prop:1}
~ The time-derivative $\partial_t \mathbf{A}(t) \DEF \mathbf{B}(t)$ defined by Definition \ref{diff:defi:1} is unique on $\mathbb{D}(t)$.
\end{proposition}
Proof: Let $\mathbf{B}_1$ and  $\mathbf{B}_2$ be two such maps. Then, $\mathbf{B}_1(t).\Phi(t) = \partial_t\bigl( \mathbf{A}(t).\Phi(t) \bigr) - \mathbf{A}(t).\Phi'(t)$, and the same for $\mathbf{B}_2$.  Thus $\mathbf{B}_1(t).\Phi(t) = \mathbf{B}_2(t).\Phi(t)$.  Therefore $\mathbf{B}_1(t) = \mathbf{B}_2(t)$ provided the set $\bigl\{ \Phi(t) \bigr\}$ such that $\Phi \in \mathcal{C}^1(\mathbb{R} \rightarrow \mathbb{F})$, with $\Phi(t) \in \mathbb{D}(t)$ and $\Phi'(t) \in \mathbb{D}(t)$, covers $\mathbb{D}(t), \forall t$, which is the case: Take $\Phi(t) = \rme^{i(t-\tau)\mathbf{H}^{<0>}}\Psi$ with $\Psi$ a constant element of $\mathbb{D}$. ~~\END

 Definition \ref{diff:defi:1} applies to $\mathbf{A} = \bphi$ as well as to $\mathbf{A} = \bpi$.  We therefore start with $\bphi$ given by \eqref{comm:1} and according to Definition~\ref{diff:defi:1} consider
\begin{align}
\notag
\bphi(\rho_\epsilon,\FOU{\chi}_\epsilon,t,\tau,\vec{x}\,).\Phi(t)
         & = \rme^{ i(t-\tau)\mathbf{H}^{<0>} }
           ~ \bphi_0(\rho_\epsilon,\tau,\vec{x}\,) ~
             \rme^{-i(t-\tau)\mathbf{H}^{<0>} }
            .\Phi(t)\\
\label{diff:4}
         & =  \rme^{ i(t-\tau)\mathbf{H}^{<0>} }
            ~ \bphi_0(\rho_\epsilon,\tau,\vec{x}\,).\Psi(t),
\end{align}
where we used \eqref{diff:2}.  Thus, differentiating,
\begin{align}
\notag
    \frac{\partial}{\partial t} \Bigl(
    \bphi(\rho_\epsilon,\FOU{\chi}_\epsilon,t,\tau,\vec{x}\,).\Phi(t) \Bigr)
         & =  i  \rme^{ i(t-\tau)\mathbf{H}^{<0>} }
 ~ \mathbf{H}^{<0>}(\rho_\epsilon,\FOU{\chi}_\epsilon,\tau)
            ~ \bphi_0(\rho_\epsilon,\tau,\vec{x}\,)
              .\Psi(t)\\
\label{diff:5}
         & +  \rme^{ i(t-\tau)\mathbf{H}^{<0>} }
            ~ \bphi_0(\rho_\epsilon,\tau,\vec{x}\,)
              .\Psi'(t),
\end{align}
since $\bphi_0(\rho_\epsilon,\tau,\vec{x}\,).\Psi(t) \in \mathbb{D}$ and from the properties of $\Psi(t)$.  Then, expressing $\Psi(t)$ and $\Psi'(t)$ in terms of $\Phi(t)$ and $\Phi'(t)$, i.e., from \eqref{diff:2},
\begin{align}
\label{diff:6}
  \Psi\,(t) &=  \rme^{-i(t-\tau)\mathbf{H}^{<0>}}\Phi(t),\\
\label{diff:7}
  \Psi'(t) &=  -i \mathbf{H}^{<0>}(\rho_\epsilon,\FOU{\chi}_\epsilon,\tau)
                 \rme^{-i(t-\tau)\mathbf{H}^{<0>}}\Phi(t)
              +  \rme^{-i(t-\tau)\mathbf{H}^{<0>}}\Phi'(t),
\end{align}
we obtain
\begin{align}
\label{diff:8}
    \frac{\partial}{\partial t} \Bigl(
    \bphi(\rho_\epsilon,\FOU{\chi}_\epsilon,t,\tau,\vec{x}\,).\Phi(t) \Bigr)
           =  \mathbf{B}.\Phi(t) 
   + \bphi(\rho_\epsilon,\FOU{\chi}_\epsilon,t,\tau,\vec{x}\,).\Phi'(t),
\end{align}
where $\mathbf{B}$ is the time-derivative defined by Definition \ref{diff:defi:1}, i.e.,
\begin{align}
\notag
     \frac{\partial}{\partial t} \bphi(\rho_\epsilon,\FOU{\chi}_\epsilon,t,\tau,\vec{x}\,)
         & = i  \rme^{ i(t-\tau)\mathbf{H}^{<0>} }
 ~ \mathbf{H}^{<0>}(\rho_\epsilon,\FOU{\chi}_\epsilon,\tau)
 ~ \bphi_0(\rho_\epsilon,\tau,\vec{x}\,) ~
          \rme^{-i(t-\tau)\mathbf{H}^{<0>} }\\
\label{diff:9}
         & - i  \rme^{ i(t-\tau)\mathbf{H}^{<0>} }
 ~ \bphi_0(\rho_\epsilon,\tau,\vec{x}\,)
 ~ \mathbf{H}^{<0>}(\rho_\epsilon,\FOU{\chi}_\epsilon,\tau) ~
          \rme^{-i(t-\tau)\mathbf{H}^{<0>} }.
\end{align}
which clearly maps $\mathbb{D}(\rho_\epsilon,\FOU{\chi}_\epsilon,t,\tau)$ into itself.  Then, rearranging the exponentials (which obviously commute with $\mathbf{H}^{<0>}$) we finally obtain
\begin{align}
\label{diff:10}
     \frac{\partial}{\partial t} \bphi(\rho_\epsilon,\FOU{\chi}_\epsilon,t,\tau,\vec{x}\,)
          = i[ \mathbf{H}^{<0>}(\rho_\epsilon,\FOU{\chi}_\epsilon,\tau),
               \bphi(\rho_\epsilon,\FOU{\chi}_\epsilon,t,\tau,\vec{x}\,) ].
\end{align}
Similarly, for $\bpi$, differentiating \eqref{comm:2}, we obtain 
\begin{align}
\label{diff:11}
     \frac{\partial}{\partial t} \bpi(\rho_\epsilon,\FOU{\chi}_\epsilon,t,\tau,\vec{x}\,)
          = i[ \mathbf{H}^{<0>}(\rho_\epsilon,\FOU{\chi}_\epsilon,\tau),
               \bpi(\rho_\epsilon,\FOU{\chi}_\epsilon,t,\tau,\vec{x}\,)  ].
\end{align}
Equations \eqref{diff:10} and \eqref{diff:11} are the Heisenberg equations of motion in differential form (\ref{sefo:9}--\ref{sefo:10}), which are therefore, in the $\mathcal{G}$-context, rigorously equivalent to the integral form  (\ref{rigo:21}--\ref{rigo:22}).  However, we have still not proved that $\bpi$ is the $t$-derivative of $\bphi$.  This will be proved in the next section, i.e., Proposition~\ref{seri:prop:1}.

\section{The rigorous interacting-field equation}
%================================================
\label{seri:0}
\setcounter{equation}{0}
\setcounter{definition}{0}
\setcounter{axiom}{0}
\setcounter{conjecture}{0}
\setcounter{lemma}{0}
\setcounter{theorem}{0}
\setcounter{corollary}{0}
\setcounter{proposition}{0}
\setcounter{example}{0}
\setcounter{remark}{0}
\setcounter{problem}{0}

Since $\rho_\epsilon$ and $\FOU{\chi}_\epsilon$ can be considered as fixed we simplify the notation by writing the field \eqref{comm:1} and its conjugate \eqref{comm:2} as
\begin{align}
\label{seri:1}
    \bphi(\epsilon,t,\tau,\vec{x}\,)
    \DEF \bphi(\rho_\epsilon,\FOU{\chi}_\epsilon,t,\tau,\vec{x}\,),\\
\label{seri:2}
    \bpi (\epsilon,t,\tau,\vec{x}\,)
    \DEF \bpi (\rho_\epsilon,\FOU{\chi}_\epsilon,t,\tau,\vec{x}\,),
\end{align}
where $\tau$ is also fixed until its dependence will be taken into account in the next chapter when studying the scattering operator. 

   In terms of the rigorous field operators the Hamiltonian density \eqref{sefo:8} is then
\begin{align}
\notag
    \pmb{\mathcal{H}}(\rho_\epsilon,\FOU{\chi}_\epsilon,t,\tau,\vec{y}\,)
    &=  \Bigl\{
      \frac{1}{2}
      \bigl(\bpi(\epsilon,t,\tau,\vec{y}\,)\bigr)^2 
    + \frac{1}{2} \sum_{1 \leq \mu \leq 3}
      \bigl(\partial_\mu
      \bphi(\epsilon,t,\tau,\vec{y}\,)\bigr)^2 \\
\label{seri:3}  
    &+ \frac{1}{2} m^2
      \bigl(\bphi(\epsilon,t,\tau,\vec{y}\,)\bigr)^2 
    + \frac{g}{N+1}
     \bigl(\bphi(\epsilon,t,\tau,\vec{y}\,)\bigr)^{N+1}
               \Bigr\}. 
\end{align}
The compositions of these operators, from $\mathbb{D}(t)$ into $\mathbb{D}(t)$, make sense.  Moreover, since $\mathbb{D}(t)$ is independent of $\vec{y}$, the $\partial_\mu$  derivatives make sense. Replacing $\bphi$ and $\bpi$ by \eqref{comm:1} and \eqref{comm:2} yields 
\begin{align}\label{seri:4}
       \pmb{\mathcal{H}}(\rho_\epsilon,\FOU{\chi}_\epsilon,t,\tau,\vec{y}\,)  =
                 \rme^{ i(t-\tau)\mathbf{H}^{<0>} }
                  \pmb{\mathcal{H}}^{(0)}(\rho_\epsilon,\tau,\vec{y}\,)
                ~\rme^{-i(t-\tau)\mathbf{H}^{<0>} },
\end{align}
where $\pmb{\mathcal{H}}^{(0)}$ is defined like $\pmb{\mathcal{H}}$ but with $\bphi_0$ and $\bpi_0$ instead of $\bphi$ and $\bpi$; $\pmb{\mathcal{H}}^{(0)}$ maps $\mathbb{D}$ into $\mathbb{D}$.  Then, integrating the density \eqref{seri:4} with the damper $\FOU{\chi}$ as in the Hamiltonian \eqref{hami:1} gives 
\begin{align}
\nonumber
  \iiint_{\mathbb{R}^3}  \pmb{\mathcal{H}}(\rho_\epsilon,\FOU{\chi}_\epsilon,t,\tau,\vec{y}\,)
                        ~\FOU{\chi}(\epsilon\vec{y}\,) ~d^3y
\nonumber
             &=   \rme^{ i(t-\tau)\mathbf{H}^{<0>}}
            \mathbf{H}^{(0)}(\rho_\epsilon,\FOU{\chi}_\epsilon,\tau,\vec{y}\,)
                 ~\rme^{-i(t-\tau)\mathbf{H}^{<0>}}\\
\label{seri:5}
         &= \mathbf{H}^{<0>}(\rho_\epsilon,\FOU{\chi}_\epsilon,\tau,\vec{y}\,),
\end{align}
where we could replace $\mathbf{H}^{(0)}$ by its self-adjoint extension $\mathbf{H}^{<0>}$ since these two lines make sense only when acting on $\mathbb{D}(t) = \rme^{i(t-\tau)\mathbf{H}^{<0>}}\mathbb{D}$, because then $\mathbf{H}^{(0)}$ and $\mathbf{H}^{<0>}$ are the same because then they act on $\mathbb{D}$ only.  (This is of course the crucial point which means that the interacting field equation makes sense only on a suitable domain depending on $t$.)

  Thus the Heisenberg equations of motion (\ref{diff:10}--\ref{diff:11}) can be written 
\begin{align}
\label{seri:6}
     \frac{\partial}{\partial t}
     \bphi(\epsilon,t,\tau,\vec{x}\,)
      = i \iiint
   [ \pmb{\mathcal{H}}(\rho_\epsilon,\FOU{\chi}_\epsilon,t,\tau,\vec{y}\,),
     \bphi(\epsilon,t,\tau,\vec{x}\,) ] 
         ~\FOU{\chi}(\epsilon\vec{y}\,) ~d^3y,\\
\label{seri:7}
     \frac{\partial}{\partial t}
     \bpi (\epsilon,t,\tau,\vec{x}\,)
      = i\iiint
   [ \pmb{\mathcal{H}}(\rho_\epsilon,\FOU{\chi}_\epsilon,t,\tau,\vec{y}\,),
     \bpi (\epsilon,t,\tau,\vec{x}\,) ] 
         ~\FOU{\chi}(\epsilon\vec{y}\,) ~d^3y.
\end{align}
From the commutation relations (\ref{comm:3}--\ref{comm:5}) and the formula \eqref{seri:3} for $\pmb{\mathcal{H}}$, we easily obtain 
\begin{align}
\label{seri:8}
     \frac{\partial}{\partial t}
     \bphi(\epsilon,t,\tau,\vec{x}\,)
  &= i\iiint \Bigl\{ 
    \delta^3_\epsilon(\rho^\vee \ast \rho,\vec{y}-\vec{x}\,)
     ~\bpi (\epsilon,t,\tau,\vec{y}\,)
     + \OOO(\epsilon^{q+1}) \Bigr\} ~\FOU{\chi}(\epsilon\vec{y}\,) ~d^3y,\\
\label{seri:9}
     \frac{\partial}{\partial t}
     \bpi(\epsilon,t,\tau,\vec{x}\,)
  &= i\iiint \Bigl\{ 
       \sum_{1 \leq \mu \leq 3}
       \bigl(\partial_\mu \bphi(\epsilon,t,\tau,\vec{y}\,) \bigr)
       \bigl(\partial_\mu \delta^3_\epsilon(\rho^\vee
                               \ast \rho,\vec{y}-\vec{x}\,)\bigr)\\
\label{seri:10}
  &+ m^2 \bphi(\epsilon,t,\tau,\vec{y}\,)
  ~\delta^3_\epsilon(\rho^\vee \ast \rho,\vec{y}-\vec{x}\,)\\
\label{seri:11} 
  &+ g\bigl(\bphi(\epsilon,t,\tau,\vec{y}\,)\bigr)^{N}
     \delta^3_\epsilon(\rho^\vee \ast \rho,\vec{y}-\vec{x}\,)
   + \OOO(\epsilon^{q+1}) \Bigr\} ~\FOU{\chi}(\epsilon\vec{y}\,) ~d^3y,
\end{align}
because we can apply the same algebraic rules as those used in deriving the formal equations \eqref{sefo:11} and \eqref{sefo:15}.

    To calculate the right-hand sides of (\ref{seri:8}--\ref{seri:11}) we remark that if we temporarily ignore $\partial_\mu$ affecting $\delta^3_\epsilon$ in \eqref{seri:9}, and put $N=1$ in \eqref{seri:11}, all four integrals have the general form
\begin{align}
\label{seri:12}
  \mathcal{O} \iiint
  \delta^3_\epsilon(\rho^\vee \ast \rho,\vec{y}-\vec{x}\,)
 ~\bphi(\epsilon,t,\tau,\vec{y}\,)
 ~\FOU{\chi}(\epsilon\vec{y}\,) ~d^3y,
\end{align}
where $\mathcal{O}$ is the operator $\partial_t$ or $\sum \partial_\mu$, or else the constant $m^2$ or $g$.  Thus we are led to evaluate the integral in \eqref{seri:12} where according to \eqref{comm:1} and (\ref{embo:5}--\ref{embo:7}) the interacting field operator is 
\begin{align}
\nonumber
        \bphi(\epsilon,t,\tau,\vec{y}\,) 
    &=    \rme^{ i(t-\tau)\mathbf{H}^{<0>} }
           \mathbf{a}^+\bigl( \Delta_\epsilon(\FOU{\rho},\xi-y) \bigr)
         ~\rme^{-i(t-\tau)\mathbf{H}^{<0>} }\\
\label{seri:13}
    &+    \rme^{ i(t-\tau)\mathbf{H}^{<0>} }
           \mathbf{a}^-\bigl( \Delta_\epsilon(\FOU{\rho},\xi-y) \bigr)
         ~\rme^{-i(t-\tau)\mathbf{H}^{<0>} }. 
\end{align}
However, both $\rho^\vee \ast \rho$ and $\FOU{\rho}$ are real scalar quantities.  Together with the integral sign they can therefore be introduced inside the argument of $\mathbf{a}^\pm$.  What has actually to be calculated is then
\begin{align}
\label{seri:14}
       \iiint 
  \delta^3_\epsilon(\rho^\vee \ast \rho,\vec{y}-\vec{x}\,)
        ~\Delta_\epsilon(\FOU{\rho},\xi-y) 
        ~\FOU{\chi}(\epsilon\vec{y}\,) ~d^3y.
\end{align}
The problem is now to reduce this integral in $\mathcal{G}$.  The integrand consists of three factors, all depending on $\epsilon$, which can be seen as the `$\mathcal{G}$-regularization' of $\Delta_\epsilon(\FOU{\rho},\xi-y)$.  In standard quantum field theory this object is interpreted as a distribution, and in the $\mathcal{G}$-context we have the possibility to go beyond distribution theory because $\delta^3_\epsilon(\rho^\vee \ast \rho,\vec{y}-\vec{x}\,)$ is actually a suitable mollifier.  But, in order to take advantage of this feature we need to integrate \eqref{seri:14} with some arbitrary but $\mathcal{C}^\infty_0$ function $\Xi(\vec{x}\,)$ --- i.e., $\Xi \in \mathcal{C}^\infty$ with compact support in the variable $\vec{x}$ --- on which, after a change of variable, one applies the Taylor formula as usual in the $\mathcal{G}$-context.  Following this idea, and proceeding step by step, we prove the following proposition:

\begin{proposition}[Interpretation of equation \eqref{seri:8}]
%.............................................................
\label{seri:prop:1} 
Let $\bphi(\epsilon,t,\tau,\vec{x}\,)$ be defin\-ed by \eqref{seri:1} and $\bpi(\epsilon,t,\tau,\vec{x}\,)$ by \eqref{seri:2}.  Then $\forall \Xi(\vec{x}\,) \in \mathcal{C}^\infty_0(\mathbb{R}^3)$, and $\forall q \in \mathbb{N}$,
\begin{align}
\label{seri:15}
     \iiint d^3x~ \Xi(\vec{x}\,) \Bigl\{
     \frac{\partial}{\partial t} \bphi(\epsilon,t,\tau,\vec{x}\,) 
 &-  \bpi(\epsilon,t,\tau,\vec{x}\,) \Bigr\} = \OOO(\epsilon^q)~\mathbf{1}, \end{align}
where $\OOO(\epsilon^q)$ means that the left-hand side of the equation is contained in $\Cst(q)~\epsilon^q~\mathcal{B}$, ~where $\mathcal{B}$ is a bounded set independent of $q$ in our space of linear operators.
\end{proposition}
Proof: We start from \eqref{seri:8}.  We note that in this equation one has 
\begin{align}
\label{seri:16}
    \iiint \OOO(\epsilon^{q+1})~\FOU{\chi}(\epsilon\vec{y}\,)~d^3y
  = \OOO(\epsilon^{q+1-3}),
\end{align}
since $\FOU{\chi}$ has compact support, which is in fact a scalar multiplied by the identity operator coming from (\ref{comr:7}--\ref{comr:8}).  Thus it can be considered as $\OOO(\epsilon^{q+1-3})$ in the space of linear operators, i.e., something contained in $\Cst(q)~\epsilon^{q+1-3}~\mathcal{B}$, where  $\mathcal{B}$ is a bounded set in our space of linear operators (here $\mathcal{B}$ is simply the identity).  So the $\OOO(\epsilon^{q+1-3})$ term in \eqref{seri:8} is consistent with \eqref{seri:15} and we may focus on the first term inside the integral.  We want to prove that, $\forall \Xi \in \mathcal{C}^\infty_0(\mathbb{R}^3)$,
\begin{align}
\label{seri:17}
     \iiint \Xi(\vec{x}\,) \Bigl\{
     \frac{\partial}{\partial t} \bphi(\epsilon,t,\tau,\vec{x}\,) 
  -  \bpi(\epsilon,t,\tau,\vec{x}\,) \Bigr\} d^3x
   = \OOO(\epsilon^q)~\mathbf{1},
     \quad \forall q \in \mathbb{N},
\end{align}
where $\OOO(\epsilon^q)$ means that the first member is contained in $\Cst(q)~\epsilon^q~\mathcal{B}$ for some bounded set $\mathcal{B}$ in our space of linear operators.  Taking into account the factors $\rme^{\pm i(t-\tau)\mathbf{H}^{<0>}}$ as in the derivation of \eqref{seri:5}, we are now in the space of linear operators from $\mathbb{D}$ into $\mathbb{D}$.  Therefore, from \eqref{seri:8} and the free-field formulas, as was done when going from \eqref{seri:12} to \eqref{seri:14}, we set
\begin{align}
\nonumber
      J(\xi)= \iiint \Xi(\vec{x}\,) \Bigl\{
    \iiint \delta^3_\epsilon(\rho^\vee \ast \rho,\vec{y}-\vec{x}\,)
        ~&\Delta_\epsilon(\FOU{\rho},\xi-y) 
        ~\FOU{\chi}(\epsilon\vec{y}\,) ~d^3y\\
\label{seri:18}
    - &\Delta_\epsilon(\FOU{\rho},\xi-x) ~\Bigr\} d^3x.
\end{align}
Thus it suffice to prove that
\begin{align}
\label{seri:19}
   J \in \OOO(\epsilon^q) \mathcal{B}
     \subset \mathsf{L}^2(\mathbb{R}^3), \forall q,
    \qquad \text{that is} \qquad
   \| J \|_{\mathsf{L}^2(\mathbb{R}^3)} = \OOO(\epsilon^q), \forall q.
\end{align}

   To simplify the notation we define $\sigma = \rho^\vee \ast \rho$, where $\sigma$ has the properties of a mollifier by Proposition~\ref{moll:prop:1}.  Then
\begin{align}
\nonumber
      J(\xi)= \iiint \Xi(\vec{x}\,) \Bigl\{
    \iiint \frac{1}{\epsilon^3}   
            \sigma\Bigl(\frac{\vec{y}-\vec{x}}{\epsilon}\Bigr)
        ~&\Delta_\epsilon(\FOU{\rho},\xi-y) 
        ~\FOU{\chi}(\epsilon\vec{y}\,) ~d^3y\\
\label{seri:20}
    - ~&\Delta_\epsilon(\FOU{\rho},\xi-x) ~\Bigr\} d^3x.
\end{align}
We make the change of variable\footnote{Note that the argument $\xi - y= \{t_\xi-t_y, \vec{\xi}-\vec{y}\}$ of $\Delta_\epsilon$ is a four-dimensional quantity, and that we leave its time part unchanged.}
\begin{align}
\label{seri:21}
   \vec{z} \DEF \frac{\vec{y}-\vec{x}}{\epsilon}, 
   \quad \Rightarrow \quad
   \epsilon^3 d^3z = d^3y, \quad \vec{y} = \vec{x} + \epsilon\vec{z},
\end{align}
\begin{align}
\nonumber
      J(\xi) &= \int \cdots \int \sigma(\vec{z}\,) 
             ~\Delta_\epsilon(\FOU{\rho},\xi-x-\epsilon z)
              \FOU{\chi}(\epsilon\vec{x} + \epsilon^2\vec{z}\,)
             ~\Xi(\vec{x}\,) ~d^3x ~d^3z\\
\label{seri:22}
             &- \int \cdots \int \sigma(\vec{z}\,) 
             ~\Delta_\epsilon(\FOU{\rho},\xi-x)
             ~\Xi(\vec{x}\,) ~d^3x ~d^3z,
\end{align}
where a $\sigma(\vec{z}\,)$ has been inserted in the second integral 
since $\int \sigma(\vec{z}\,) d^3z = 1$. Then we make another change of variable, but in the first integral only
\begin{align}
\label{seri:23}
   \vec{y} \DEF \vec{x} + \epsilon\vec{z}, 
   \quad \Rightarrow \quad
   d^3x = d^3y, \quad \vec{x} = \vec{y} - \epsilon\vec{z},
\end{align}
\begin{align}
\nonumber
      J(\xi) &= \int \cdots \int \sigma(\vec{z}\,) 
             ~\Delta_\epsilon(\FOU{\rho},\xi-y)
              \FOU{\chi}(\epsilon\vec{y}\,)
             ~\Xi(\vec{y} - \epsilon\vec{z}\,) ~d^3y ~d^3z\\
\label{seri:24}
             &- \int \cdots \int \sigma(\vec{z}\,) 
             ~\Delta_\epsilon(\FOU{\rho},\xi-y)
             ~\Xi(\vec{y}\,) ~d^3y ~d^3z,
\end{align}
where we changed $\vec{x}$ into $\vec{y}$ in the second integral.  Now we remark that we can insert $\FOU{\chi}(\epsilon\vec{y}\,)$ in the second integral because $\FOU{\chi}(\epsilon\vec{y}\,)\Xi(\vec{y}\,) \cong \Xi(\vec{y}\,)$ for $0 <\epsilon < 1$ (from the fact that \,$\operatorname{supp} \Xi$ is compact and $\FOU{\chi} \equiv 1$ in a $0$-neighborhood). Then
\begin{align}
\label{seri:25}
      J(\xi) &= \int \cdot \int \sigma(\vec{z}\,) 
             ~\Delta_\epsilon(\FOU{\rho},\xi-y)
              \FOU{\chi}(\epsilon\vec{y}\,) 
             ~\Bigl( \Xi(\vec{y} - \epsilon\vec{z}\,) - \Xi(\vec{y}\,)
              \Bigr) ~d^3y ~d^3z.
\end{align}
We now recall the definition \eqref{embs:7} of $\Delta_\epsilon$, i.e.,
\begin{align}
\label{seri:26}
     \Delta_\epsilon(\FOU{\rho}, x) \DEF
     \frac{1}{(2\pi)^3}   \iiint \frac{d^3p}{2 E_p}
     \FOU{\rho}(\epsilon\vec{p}\,)
     \exp i (\vec{p}\cdot \vec{x} - E_p t),
\end{align}
so that, writing $t$ for $t_\xi - t_y$, equation \eqref{seri:25} becomes
\begin{align}
\nonumber
      J(\xi) &= \frac{1}{(2\pi)^3}  \int \cdots \int
                \frac{1}{2 E_p} \FOU{\rho}(\epsilon\vec{p}\,)
                \exp i (\vec{p}\cdot (\vec{\xi}-\vec{y}\,) - E_p t)\\
\label{seri:27}
            &\times \sigma(\vec{z}\,)~\FOU{\chi}(\epsilon\vec{y}\,)
             ~\Bigl( \Xi(\vec{y} - \epsilon\vec{z}\,) - \Xi(\vec{y}\,)
              \Bigr) ~d^3y ~d^3z ~d^3p.
\end{align}
Next we use Taylor's theorem with remainder to write
\begin{align}
\label{seri:28}
  \Xi(\vec{y} - \epsilon\vec{z}\,) - \Xi(\vec{y}\,) 
  = \frac{(\epsilon|\vec{z}\,|)^{q+1}}{(q+1)!}
       \bigl(\rmD^{q+1}\Xi\bigr) (\vec{y} - \theta\epsilon\vec{z}\,),
       \quad 0 < \theta <1,
\end{align}
because all terms $ \int |\vec{z}\,|^n \sigma(\vec{z}\,) d^3z = 0$ for $n>1$ in the Taylor development since $\sigma \in \mathcal{A}_\infty$.  Therefore,
\begin{align}
\nonumber
      J(\xi) &= \frac{1}{(2\pi)^3} \frac{\epsilon^{q+1}}{(q+1)!}
                \int \cdots \int
                \frac{1}{2 E_p} \FOU{\rho}(\epsilon\vec{p}\,)
                \exp i (\vec{p}\cdot (\vec{\xi}-\vec{y}\,) - E_p t)\\
\label{seri:29}
            &\times \sigma(\vec{z}\,)~\FOU{\chi}(\epsilon\vec{y}\,)
             ~|\vec{z}\,|^{q+1}
             ~\bigl(\rmD^{q+1}\Xi\bigr) (\vec{y} - \theta\epsilon\vec{z}\,)
             ~d^3y ~d^3z ~d^3p.
\end{align}
What we want to prove is \eqref{seri:19}, that is $\| J \|_{\mathsf{L}^2(\mathbb{R}^3)} = \OOO(\epsilon^q), \forall q$, that is
\begin{align}
\label{seri:30}
     2E_p \iiint | J(t,\vec{\xi}\,)|^2~d^3\xi = \OOO(\epsilon^q),
      \qquad \forall q \in \mathbb{N}.
\end{align}
where the factor $2E_p$ comes from the time-derivative in the norm induced by the relativistically invariant scalar-product \eqref{fock:15}.  However, from the isometric properties of the Fourier transformation in $\mathsf{L}^2$, i.e., Plancherel's theorem, we have (modulo a coefficient)
\begin{align}
\label{seri:31}
      \iiint | J(t,\vec{\xi}\,)|^2~d^3\xi = 
      \iiint | \FOU{J}(t,\vec{p}\,)|^2~d^3p.
\end{align}
where by \eqref{moll:2}
\begin{equation}\label{seri:32}
  J(t,\vec{\xi}\,) \DEF 
  (2\pi)^{-3} \iiint \rme^{-i\vec{p} \cdot \vec{\xi}}
                                     \FOU{J}(t,\vec{p}\,) ~d^3p.
\end{equation}
Therefore, from \eqref{seri:29},
\begin{align}
\nonumber
      J(t,-\vec{p}\,) &= \frac{\epsilon^{q+1}}{(q+1)!}
                \frac{1}{2 E_p} \FOU{\rho}(\epsilon\vec{p}\,)
                \int \cdots \int
                \exp i (-\vec{p}\cdot \vec{y} - E_p t)\\
\label{seri:33}
            &\times \sigma(\vec{z}\,)~\FOU{\chi}(\epsilon\vec{y}\,)
             ~|\vec{z}\,|^{q+1}
             ~\bigl(\rmD^{q+1}\Xi\bigr) (\vec{y} - \theta\epsilon\vec{z}\,)
             ~d^3y ~d^3z.
\end{align}
Since $\FOU{\rho} \equiv 1$ in a $0$-neighborhood we have, from $\FOU{\rho}(\epsilon\vec{p}\,)$, for $0 <\epsilon < 1$,
\begin{align}
\nonumber
     \iiint &| \FOU{J}(t,\vec{p}\,)|^2~d^3p
    \leq \frac{\Cst}{\epsilon^3}  \frac{\epsilon^{2q+2}}{((q+1)!)^2}
      \frac{1}{2m}~\\
\label{seri:34}
     \times &\Bigl| \int \cdots \int \sigma(\vec{z}\,)
    ~\FOU{\chi}(\epsilon\vec{y}\,) ~|\vec{z}\,|^{q+1}
    ~\bigl( \rmD^{q+1}\Xi\bigr) (\vec{y} - \theta\epsilon\vec{z}\,)
     ~d^3y ~d^3z
    \Bigr|^2
\end{align}
where $1/2m$ is the supremum of $1/2E_p$ over $\mathbb{R}^3$ in the $\vec{p}$ variable because $E_p = \sqrt{m^2 + \vec{p}\,|^2}$. Integration in $\vec{z}$ is ensured by $\sigma \in \mathcal{S}(\mathbb{R}^3)$.  Integration in $\vec{y}$ is ensured by $\FOU{\chi}$ and gives a bound $\Cst/\epsilon^3$ since $\FOU{\chi}$ has compact support. Finally, $| \bigl( \rmD^{q+1}\Xi\bigr) (\vec{y} - \theta\epsilon\vec{z}\,) |^2 \leq \Cst(q)$ since $\Xi$ is $\mathcal{C}^\infty$ with compact support.  Therefore,  
\begin{align}
\label{seri:35}
    \iiint &| \FOU{J}(t,\vec{p}\,)|^2~d^3p
    \leq \Cst(q)\epsilon^{2q+2-3-3},
\end{align}
i.e., $\| J \|_{\mathsf{L}^2(\mathbb{R}^3)} = \OOO(\epsilon^q), \forall q$, as requested. \END

  The next step is of course to interpret (\ref{seri:9}--\ref{seri:11}).  But, as was remarked in relation to \eqref{seri:12}, the calculations will be similar to those involved in the proof of Proposition~\ref{seri:prop:1}, which therefore leads to the corollary:
\begin{corollary}[Interpretation of equation (\ref{seri:9}--\ref{seri:11})]
%............................................................................
\label{seri:coro:1} 
~ Let $\bphi(\epsilon,t,\tau,\vec{x}\,)$ be defined by \eqref{seri:1} and $\bpi(\epsilon,t,\tau,\vec{x}\,)$ by \eqref{seri:2}.  Then $\forall \Xi(\vec{x}\,) \in \mathcal{C}^\infty_0(\mathbb{R}^3)$, and $\forall q \in \mathbb{N}$,
\begin{align}
\nonumber
     \iiint d^3x~ \Xi(\vec{x}\,) \Bigl\{
     \frac{\partial}{\partial t} \bpi(\epsilon,t,\tau,\vec{x}\,) 
 &-  \sum_{1\leq\mu\leq 3} \frac{\partial^2}{\partial{x_\mu}^2} \bphi(t,\vec{x}\,)\\
\label{seri:36}
  - m^2  \bphi(\epsilon,t,\tau,\vec{x}\,)
 &- g \bigl(\bphi(\epsilon,t,\tau,\vec{x}\,)\bigr)^N \Bigr\} = \OOO(\epsilon^q)~\mathbf{1},
\end{align}
where $\OOO(\epsilon^q)$ means that the left-hand side of the equation is contained in $\Cst(q)~\epsilon^q~\mathcal{B}$, ~where $\mathcal{B}$ is a bounded set independent of $q$ in our space of linear operators.
\end{corollary}

   In conclusion, by Proposition~\ref{seri:prop:1} and Corollary~\ref{seri:coro:1} the interpretation of (\ref{seri:8}--\ref{seri:11}) in $\mathcal{G}$ is given by the following theorem:
\begin{theorem}[Rigorous interacting-field equation]
%...................................................
\label{seri:theo:1} 
In the context of operator-valued generalized functions in $\mathcal{G}$, defined in Section~\eqref{embo:0}, one has, for all $\Xi(\vec{x}\,) \in \mathcal{C}^\infty_0$, i.e., $\forall \Xi \in \mathcal{C}^\infty$ with compact support in the variable $\vec{x}$,
\begin{align}
\label{seri:37}
     \iiint d^3x~ \Xi(\vec{x}\,) \Bigl\{
     \frac{\partial}{\partial t} \bphi  (t,\vec{x}\,) 
 &-  \bpi(t,\vec{x}\,) \Bigr\} = 0, \\
\nonumber 
     \iiint d^3x~ \Xi(\vec{x}\,) \Bigl\{
     \frac{\partial}{\partial t} \bpi(t,\vec{x}\,) 
 &-  \sum_{1\leq\mu\leq 3} \frac{\partial^2}{\partial{x_\mu}^2} \bphi(t,\vec{x}\,)\\
\label{seri:38}
 &- m^2  \bphi(t,\vec{x}\,) - g \bigl(\bphi(t,\vec{x}\,)\bigr)^N \Bigr\} = 0,
\end{align}
with the free-field operators as initial values at time $t=\tau$.
\end{theorem}
\begin{remark}[Differentiation of interacting-field equation]
%............................................................
\label{seri:rema:1}
 Differentiation in $\vec{x}$ inside the curly brackets of \emph{(\ref{seri:37}--\ref{seri:38})} is obvious since this is only differentiations by parts.  On the other hand, differentiation in $t$ is not obvious, but the proof of Proposition~\ref{seri:prop:1} shows that it can be done freely.  For instance,  \eqref{seri:37} implies
\begin{align}
\label{seri:139}
     \iiint d^3x~ \Xi(\vec{x}\,) \Bigl\{
     \frac{\partial^2}{\partial t^2} \bphi  (t,\vec{x}\,) 
 &-  \frac{\partial}{\partial t} \bpi(t,\vec{x}\,) \Bigr\} = 0, \\
\intertext{and therefore, using \eqref{seri:38},}
\nonumber 
     \iiint d^3x~ \Xi(\vec{x}\,) \Bigl\{
     \frac{\partial}{\partial t} \bpi(t,\vec{x}\,) 
 &-  \sum_{1\leq\mu\leq 3} \frac{\partial^2}{\partial{x_\mu}^2} \bphi(t,\vec{x}\,)\\
\label{seri:40}
 &- m^2  \bphi(t,\vec{x}\,) - g \bigl(\bphi(t,\vec{x}\,)\bigr)^N \Bigr\} = 0.
\end{align}
\end{remark}

   The concept of equality defined by Theorem~\ref{seri:theo:1} is far stronger than association but weaker than mere equality in $\mathcal{G}$.\footnote{This equality is reminiscent of a distributional concept, see \cite[p.\,200--201]{COLOM1984-}, which was introduced in the context of the convolution of scalar functions in $\mathcal{G}$, where the classic identity $\delta * \delta = \delta$ for the convolution of two $\delta$-functions does not hold, but is replaced by the somewhat weaker identity, $\forall \Xi \in \mathcal{C}^\infty_0$, $\int dx~ \Xi(x) (\delta * \delta)(x) = \int dx~ \Xi(x) \delta(x)$.}  Indeed, as specified by Definition~\ref{oper:defi:8}, association corresponds to the limit $\epsilon \rightarrow 0$ in which a $\mathcal{G}$-function can be identified with a Schwartz-distribution, which does not depend on $\epsilon$.  On the other hand, the proof of Theorem~\ref{seri:theo:1} is based on the notion of infinitesimal quantities such as $\Cst(q)~\epsilon^q~\mathcal{B}$ which vanish for all $\epsilon$ that are \emph{finite} but small enough, i.e., for \emph{all} $0 <\epsilon < 1$, because $q \in \mathbb{N}$ can be as large as we please.  Thus Theorem~\ref{seri:theo:1} gives a meaning to the interacting-field equation for $\bphi$ and $\bpi$ interpreted not as operator-valued distributions, as defined by Bogoliubov and Wightman \cite{BOGOL1990-,WIGHT1996-}, but as operator-valued generalized functions in $\mathcal{G}$, i.e., objects in which non-zero $\epsilon$-dependent contributions enable to continue nonlinear calculations which would be meaningless with distributions.  

  We do not know, however, if the rigorous interacting-field equation can be obtained with the mere equality in $\mathcal{G}$.  But it may also be that Theorem~\ref{seri:theo:1} is the most general statement with regards to the \emph{physical} interpretation of the interacting-field equation in the context of the Heisenberg-Pauli canonical formalism.  If this is so, the smooth function $\Xi(\vec{x}\,)$ is not a `test-function' in the sense of Schwartz-distribution theory, but an instance of the more fundamental `measurement-function' defining a finite space-time region (an `apparatus') over which a an averaging has to be made in order to get a physically meaningful measurement --- in accord with the universally accepted theory of measurement in quantum theory due to Bohr and Rosenfeld \cite{BOHR1950-}. 

%File: scat.26.tex     arXiv version 2              Date: 6 September 2008
%=====                            
%                                       
\chapter{The scattering operator}
%================================
\label{scat:0}

\def\Ie{{\text{\emph{I}}}}    % Interaction (no g) in "theorems"
\def\It{{\text{I}}}           % Interaction (no g) in text
\def\He{{\text{\emph{H}}}}    % Hilbert in "theorems"
\def\Ht{{\text{H}}}           % Hilbert in text

\section{Scattering operator: formal calculations}
%-------------------------------------------------
\label{scaf:0}
\setcounter{equation}{0}
\setcounter{definition}{0}
\setcounter{axiom}{0}
\setcounter{conjecture}{0}
\setcounter{lemma}{0}
\setcounter{theorem}{0}
\setcounter{corollary}{0}
\setcounter{proposition}{0}
\setcounter{example}{0}
\setcounter{remark}{0}
\setcounter{problem}{0}

In this section we present the formal calculations leading to the differential equation for the \emph{time evolution operator} $\mathbf{S}_\tau(t)$ introduced in Section \ref{summ:0}.4.  However, for the same reasons as those given in Section \ref{rigo:0} we will not write this operator in terms of the Hamiltonian $\mathbf{H}$ as in \eqref{summ:23}.
\begin{align}\label{scaf:1}
    \mathbf{S}_\tau(t) := \rme^{ i(t-\tau)\mathbf{H}_0}
                          \rme^{-i(t-\tau)\mathbf{H}^{(0)}}, 
\end{align}
where $\mathbf{H}_0=\mathbf{H}_0(\tau)$ is the Hamiltonian of a free field, and $\mathbf{H}^{(0)}=\mathbf{H}^{(0)}(\tau)$ is the Hamiltonian of the self-interacting field written in terms of the operators $\bphi_0(\tau,\vec{x}\,)$ and $\bpi_0(\tau,\vec{x}\,)$ of that free field, i.e., \eqref{rigo:19}.   As everything is defined with the same free field at a given initial reference time $\tau$, $\mathbf{H}^{(0)}$ reduces to $\mathbf{H}_0$ for $g=0$, i.e.,
\begin{align}
\label{scaf:2}
    \mathbf{H}^{(0)}(\tau) = \mathbf{H}_0(\tau)
                             + \frac{g}{N+1} \iiint_{\mathbb{R}^3} d^3x~
                               \bigl(\bphi_0(\tau,\vec{x}\,)\bigr)^{N+1}. 
\end{align}
\begin{remark}[Avoidance of interaction picture]\label{scaf:rema:1}
%................................................
Definition \ref{scaf:1} has the virtue that we avoid some of the pitfalls of the interaction picture concept (see Remark~\ref{summ:rema:2}) because we do not demand that a unitary transformation to that picture exists for any unspecified Hamiltonian $\mathbf{H}$.  In this respect, our method is comparable to that of Bogoliubov et al., see \emph{\cite[Sec.~9.4 and Chap.14]{BOGOL1990-}}.  To make this clear we shall not use the qualifier `interaction picture' in this chapter, even though some quantities will be formally very similar to those which arise in that picture.
\end{remark}

   We now consider Heisenberg's equation of motion \eqref{rigo:21}, first for the full Hamiltonian $\mathbf{H}^{(0)}$, i.e.,
\begin{align}
\label{scaf:3}
     \bphi(t,\tau,\vec{x}\,)   =
          \rme^{ i(t-\tau)\mathbf{H}^{(0)} }
                 ~ \bphi_0(\tau,\vec{x}\,) ~
          \rme^{-i(t-\tau)\mathbf{H}^{(0)} },
\end{align}
second for the free-field Hamiltonian $\mathbf{H}_0$, i.e.,
\begin{align}
\label{scaf:4}
     \bphi_I(t,\tau,\vec{x}\,)   =
          \rme^{ i(t-\tau)\mathbf{H}_0 }
                 ~ \bphi_0(\tau,\vec{x}\,) ~
          \rme^{-i(t-\tau)\mathbf{H}_0 }.
\end{align}
(Of course, \eqref{scaf:3} reduces to \eqref{scaf:4} when $g=0$.)
We have therefore
\begin{align}
\nonumber
     \bigl(\mathbf{S}_\tau(t)\bigr)^{-1}
                                  ~&\bphi_I(t,\tau,\vec{x}\,)~
                                        \mathbf{S}_\tau(t)\\
\nonumber
     =  \rme^{ i(t-\tau)\mathbf{H}^{(0)}}
        \rme^{-i(t-\tau)\mathbf{H}_0}
                                  ~&\bphi_I(t,\tau,\vec{x}\,)~
        \rme^{ i(t-\tau)\mathbf{H}_0}
        \rme^{-i(t-\tau)\mathbf{H}^{(0)}},\\
\label{scaf:5}
     = \rme^{ i(t-\tau)\mathbf{H}^{(0)}}
                                  ~&\bphi_0(\tau,\vec{x}\,)~
       \rme^{-i(t-\tau)\mathbf{H}^{(0)}},
\end{align}
which confirms \eqref{summ:24}, i.e.,
\begin{align}\label{scaf:6}
     \bphi(t,\tau,\vec{x}\,) = \bigl(\mathbf{S}_\tau(t)\bigr)^{-1}
                                  \bphi_I(t,\tau,\vec{x}\,)~
                                     \mathbf{S}_\tau(t),
\end{align}
provided we set $\bphi_{\text{ini}}=\bphi_0(\tau,\vec{x}\,)$ in \eqref{summ:25}.  Consequently, if an independent equation can be derived for $\mathbf{S}_\tau(t)$ we could calculate $\bphi(t,\tau,\vec{x}\,)$ from $\bphi_I(t,\tau,\vec{x}\,)$, which is known since it is given by \eqref{scaf:4} as a function of $\bphi_0(\tau,\vec{x}\,)$.  To find whether this is possible we differentiate \eqref{scaf:1}, i.e.,
\begin{align}
\nonumber
    \frac{\partial}{\partial t} \mathbf{S}_\tau(t)
    &= i\mathbf{H}_0\mathbf{S}_\tau(t)
     - i            \mathbf{S}_\tau(t)\mathbf{H}^{(0)},\\
\nonumber
    &= -i\Bigl(- \mathbf{H}_0
               + \mathbf{S}_\tau(t)
                 \mathbf{H}^{(0)} 
           \bigl(\mathbf{S}_\tau(t)\bigr)^{-1}
           \Bigr)\mathbf{S}_\tau(t),\\
\label{scaf:7}
    &= -i\Bigl(- \mathbf{H}_0
               + \rme^{ i(t-\tau)\mathbf{H}_0}
                 \mathbf{H}^{(0)} 
                  \rme^{-i(t-\tau)\mathbf{H}_0}
           \Bigr)\mathbf{S}_\tau(t),
\end{align}
where the simplification of the exponentials $\rme^{\pm i(t-\tau)\mathbf{H}^{(0)}}$ from the $\mathbf{S}_\tau^{\pm 1}(t)$ operators affecting $\mathbf{H}^{(0)}$ led to the last line.  Then, after substituting $\mathbf{H}^{(0)}$ given by \eqref{scaf:2}, the simplification of the remaining exponentials $\rme^{\pm i(t-\tau)\mathbf{H}_0}$ affecting the $\mathbf{H}_0$ term coming from $\mathbf{H}^{(0)}$ yields a $+\mathbf{H}_0$ that cancels the leading $-\mathbf{H}_0$ in \eqref{scaf:7}, so that we get the differential equation
\begin{align}\label{scaf:8}
    \frac{\partial}{\partial t} \mathbf{S}_\tau(t) 
        = -i \, \mathbf{H}_I(t,\tau) \, \mathbf{S}_\tau(t),
\end{align}
where, with the help of \eqref{scaf:4}, we have defined the \emph{interaction Hamiltonian}
\begin{align}
\label{scaf:9}
    \mathbf{H}_I(t,\tau) \DEF \frac{g}{N+1} \iiint_{\mathbb{R}^3} d^3x~
    \bigl(\bphi_I(t,\tau,\vec{x}\,)\bigr)^{N+1}, 
\end{align}
which, in contradistinction to $\mathbf{H}_0$ and $\mathbf{H}^{(0)}$, depends on $t$.  Together with the initial condition $\mathbf{S}_\tau(\tau) = \mathbf{1}$, Eq.~\eqref{scaf:8} is the sought after equation explicitly solving the interacting field equation, which was given without proof as Eqs.~(\ref{summ:26}--\ref{summ:27}).  In the limits  $\tau \rightarrow -\infty, t \rightarrow +\infty$, the operator $\mathbf{S}_\tau(t)$ becomes the  \emph{scattering operator}, i.e.,
\begin{align}\label{scaf:10}
    \mathbf{S} \DEF \lim_{t \rightarrow \infty}\mathbf{S}_{-t}(t).
\end{align}

\section{Scattering operator: rigorous calculations}
%---------------------------------------------------
\label{scar:0}
\setcounter{equation}{0}
\setcounter{definition}{0}
\setcounter{axiom}{0}
\setcounter{conjecture}{0}
\setcounter{lemma}{0}
\setcounter{theorem}{0}
\setcounter{corollary}{0}
\setcounter{proposition}{0}
\setcounter{example}{0}
\setcounter{remark}{0}
\setcounter{problem}{0}

We now proceed to derive the rigorous differential equation for the scattering operator by systematically reformulating the calculations made in the previous section in the $\mathcal{G}$-context.  Thus, instead of \eqref{scaf:1}, we define 
\begin{align}\label{scar:1}
    \mathbf{S}_\tau(\rho_\epsilon,\FOU{\chi}_\epsilon,t)
                       := \rme^{ i(t-\tau)\mathbf{H}_0^{<0>}}
                          \rme^{-i(t-\tau)\mathbf{H}^{<0>}}, 
\end{align}
where $\mathbf{H}^{<0>}(\rho_\epsilon,\FOU{\chi}_\epsilon,\tau)$ is the self-adjoint extension of the interacting-field Hamiltonian $\mathbf{H}^{(0)}$, and $\mathbf{H}_0^{<0>}$ that of the free field Hamiltonian $\mathbf{H}_0$ which is equal to $\mathbf{H}^{<0>}$ when $g=0$.  Using these notations, the $\mathcal{G}$-embedding of \eqref{scaf:2} is then
\begin{align}
\label{scar:2}
    \mathbf{H}^{<0>}(\tau) = \mathbf{H}_0^{<0>}(\tau)
              + \frac{g}{N+1}
                \iiint_{\mathbb{R}^3} d^3x~\FOU{\chi}(\epsilon \vec{x}\,)
                \bigl(\bphi_0(\rho_\epsilon,\tau,\vec{x}\,)\bigr)^{N+1}. 
\end{align}
Concerning the domains and ranges, writing for brevity $\mathbb{D}(t)$ for  $\mathbb{D}(\rho_\epsilon,\FOU{\chi}_\epsilon,t,\tau)$, we recall (\ref{anal:5}--\ref{anal:6})
\begin{align}
\label{scar:3}
   \rme^{-i(t-\tau)\mathbf{H}^{<0>} } ~ &: ~
   ~ \mathbb{D}(t) \rightarrow \mathbb{D},\\
\label{scar:4}
   \rme^{+i(t-\tau)\mathbf{H}^{<0>} } ~ &: ~
   ~ \mathbb{D}~~~~\, \rightarrow \mathbb{D}(t),
\end{align}
whereas
\begin{align}
\label{scar:5}
   \rme^{-i(t-\tau)\mathbf{H}_0^{<0>} } ~ &: ~
   ~ \mathbb{D}~~~~\, \rightarrow \mathbb{D},~~~\\
\label{scar:6}
   \rme^{+i(t-\tau)\mathbf{H}_0^{<0>} } ~ &: ~
   ~ \mathbb{D}~~~~\, \rightarrow \mathbb{D}.
\end{align}
Indeed, $\mathbf{H}_0^{<0>}$ extends the free-field Hamiltonian $\mathbf{H}_0$, given by \eqref{hamr:46}, which maps $\mathbb{D}$ into itself, what is also the case of $\rme^{+i(t-\tau)\mathbf{H}_0}$, as well of $\rme^{+i(t-\tau)\mathbf{H}_0^{<0>}}$. Consequently $\mathbf{S}_\tau(\rho_\epsilon,\FOU{\chi}_\epsilon,t)$ maps $\mathbb{D}(t)$ into $\mathbb{D}$.

Next we embed \eqref{scaf:3} as
\begin{align}
\label{scar:7}
     \bphi  (\rho_\epsilon,\FOU{\chi}_\epsilon,t,\tau,\vec{x}\,)   &=
          \rme^{ i(t-\tau)\mathbf{H}^{<0>} }
                 ~ \bphi_0(\rho_\epsilon,\tau,\vec{x}\,) ~
          \rme^{-i(t-\tau)\mathbf{H}^{<0>} },\\
\intertext{which maps $\mathbb{D}(t)$ into $\mathbb{D}(t)$, as well as \eqref{scaf:4} as}
\label{scar:8}
     \bphi_I(\rho_\epsilon,\FOU{\chi}_\epsilon,t,\tau,\vec{x}\,)   &=
          \rme^{ i(t-\tau)\mathbf{H}_0^{<0>} }
                 ~ \bphi_0(\rho_\epsilon,\tau,\vec{x}\,) ~
          \rme^{-i(t-\tau)\mathbf{H}_0^{<0>} },
\end{align}
which maps $\mathbb{D}$ into $\mathbb{D}$.

  The compositions involved in the calculations \eqref{scaf:5} therefore make sense, and consequently \eqref{scaf:6} in the form
\begin{align}\label{scar:9}
     \bphi(\rho_\epsilon,\FOU{\chi}_\epsilon,t,\tau,\vec{x}\,)
   = \bigl(  \mathbf{S}_\tau(\rho_\epsilon,\FOU{\chi}_\epsilon,t)\bigr)^{-1}
                \bphi_I(\rho_\epsilon,\FOU{\chi}_\epsilon,t,\tau,\vec{x}\,)~
             \mathbf{S}_\tau(\rho_\epsilon,\FOU{\chi}_\epsilon,t),
\end{align}
is rigorous and mapping $\mathbb{D}(t)$ into itself.

  The crucial step is of course the calculation of the $t$-derivative of $\mathbf{S}_\tau(\rho_\epsilon,\FOU{\chi}_\epsilon,t)$.  We do this according to Definition \ref{diff:defi:1} and therefore consider
\begin{align} \label{scar:10}
\nonumber
    \frac{\partial}{\partial t}
    \Bigl( \mathbf{S}_\tau(\rho_\epsilon,\FOU{\chi}_\epsilon,t) \Phi(t) \Bigr)
   &=\Bigl(  i\mathbf{H}_0^{<0>}
              \mathbf{S}_\tau(\rho_\epsilon,\FOU{\chi}_\epsilon,t)
            -i\mathbf{S}_\tau(\rho_\epsilon,\FOU{\chi}_\epsilon,t)
              \mathbf{H}^{<0>} \Bigr) \Phi(t)\\
   &+ \mathbf{S}_\tau(\rho_\epsilon,\FOU{\chi}_\epsilon,t) ~ \Phi'(t),
\end{align}
where the expression inside the big parentheses on the first line yields the $t$-derivative if the conditions of Definition \ref{diff:defi:1} are satisfied.  Thus, we take   $\Phi(t) \in \mathbb{D}(t)$ of the form $\Phi(t) = \rme^{i(t-\tau)\mathbf{H}^{<0>}}\Psi(t)$ with $\Psi \in \mathbb{D}$ as in \eqref{diff:2}. Then $\mathbf{H}^{<0>}\Phi = \mathbf{H}^{<0>}\rme^{i(t-\tau)\mathbf{H}^{<0>}}\Psi = \rme^{i(t-\tau)\mathbf{H}^{<0>}}\mathbf{H}^{<0>}\Psi$.
But $\mathbf{H}^{<0>}\Psi = \mathbf{H}^{(0)}\Psi \in \mathbb{D}$ and therefore $\mathbf{H}^{<0>}\Phi  \in \mathbb{D}(t)$.  Since $\mathbf{S}_\tau$ maps $\mathbb{D}(t)$ into $\mathbb{D}$, the first line of \eqref{scar:10} maps $\mathbb{D}(t)$ into $\mathbb{D}$ for all $\Phi \in \mathbb{D}(t)$, so that Definition \ref{diff:defi:1} is applicable.

%%%NOTE: $\mathbf{H}^{<0>}\Psi = \mathbf{H}^{(0)}\Psi \in \mathbb{D}$ is the same argument used to prove \eqref{seri:5}, i.e., (3.7.5) in self.25 

  Consequently, with the the $t$-derivative given by the expression inside the big parentheses on the first line of \eqref{scar:10}, the rigorous form of \eqref{scaf:7} is
\begin{align}
\nonumber
    \frac{\partial}{\partial t}
    \mathbf{S}_\tau(\rho_\epsilon,\FOU{\chi}_\epsilon,t)
    &= i\mathbf{H}_0^{<0>}\mathbf{S}_\tau(\rho_\epsilon,\FOU{\chi}_\epsilon,t)
      -i\mathbf{S}_\tau(\rho_\epsilon,\FOU{\chi}_\epsilon,t)\mathbf{H}^{<0>},\\
\nonumber
    &= -i\Bigl(- \mathbf{H}_0^{<0>}
               + \mathbf{S}_\tau(\rho_\epsilon,\FOU{\chi}_\epsilon,t)
                 \mathbf{H}^{<0>} 
           \bigl(\mathbf{S}_\tau(\rho_\epsilon,\FOU{\chi}_\epsilon,t)\bigr)^{-1}
           \Bigr)\mathbf{S}_\tau(\rho_\epsilon,\FOU{\chi}_\epsilon,t),\\
\label{scar:11}
    &= -i\Bigl(- \mathbf{H}_0^{<0>}
               + \rme^{ i(t-\tau)\mathbf{H}_0}
                 \mathbf{H}^{<0>} 
                  \rme^{-i(t-\tau)\mathbf{H}_0}
           \Bigr)\mathbf{S}_\tau(\rho_\epsilon,\FOU{\chi}_\epsilon,t).
\end{align}
With regards to domains these calculations are consistent when acting on $\mathbb{D}(t)$.  Moreover, we can simplify the exponentials as was done in the previous section, so that we finally get the rigorous differential equation
\begin{align}\label{scar:12}
    \frac{\partial}{\partial t}
    \mathbf{S}_\tau(\rho_\epsilon,\FOU{\chi}_\epsilon,t) 
        = -i \, \mathbf{H}_I(\rho_\epsilon,\FOU{\chi}_\epsilon,t,\tau) \, \mathbf{S}_\tau(\rho_\epsilon,\FOU{\chi}_\epsilon,t),
\end{align}
where the $t$-dependent interaction Hamiltonian is now
\begin{align}
\label{scar:13}
    \mathbf{H}_I(\rho_\epsilon,\FOU{\chi}_\epsilon,t,\tau)
   \DEF \frac{g}{N+1} \iiint_{\mathbb{R}^3} d^3x~\FOU{\chi}(\epsilon \vec{x}\,)
\bigl(\bphi_I(\rho_\epsilon,\FOU{\chi}_\epsilon,t,\tau,\vec{x}\,)\bigr)^{N+1}, 
\end{align}
in which the field operator $\bphi_I(\rho_\epsilon,\FOU{\chi}_\epsilon,t,\tau,\vec{x}\,)$ is defined by \eqref{scar:8}.

  Note that \eqref{scar:13} is consistent because both $\mathbf{H}_I$ and $\bphi_I$ map $\mathbb{D}$ into $\mathbb{D}$.  Therefore, in \eqref{scar:12}, both $\mathbf{S}_\tau$ and ${\partial_t} \mathbf{S}_\tau$ map $\mathbb{D}(t)$ into $\mathbb{D}$.

   In combination with equations \eqref{scar:9} and \eqref{scar:8} the differential equation \eqref{scar:12} with initial condition $\mathbf{S}_\tau(\rho_\epsilon,\FOU{\chi}_\epsilon,\tau) = \mathbf{1}$ solves the rigorous interacting field equation  (\ref{seri:37}--\ref{seri:38}) with the initial fields $\bphi_0(\rho_\epsilon,\tau,\vec{x}\,)$ and  $\bpi_0(\rho_\epsilon,\tau,\vec{x}\,) = \partial_t \bphi_0(\rho_\epsilon,\tau,\vec{x}\,)$ taken as those of a free field.  

   In the limits  $\tau \rightarrow -\infty, t \rightarrow +\infty$, which for now are formal limits, the operator $\mathbf{S}_\tau(\rho_\epsilon,\FOU{\chi}_\epsilon,t)$ becomes the  \emph{scattering operator}, i.e.,
\begin{align}\label{scar:14}
    \mathbf{S}(\rho_\epsilon,\FOU{\chi}_\epsilon) \DEF \lim_{t \rightarrow \infty}\mathbf{S}_{-t}(\rho_\epsilon,\FOU{\chi}_\epsilon,t),
\end{align}
which still depends on the mollifier $\rho_\epsilon$ and the damper $\FOU{\chi}_\epsilon$.

\section{Cancellation of zero-point energy}
%------------------------------------------
\label{canc:0}
\setcounter{equation}{0}
\setcounter{definition}{0}
\setcounter{axiom}{0}
\setcounter{conjecture}{0}
\setcounter{lemma}{0}
\setcounter{theorem}{0}
\setcounter{corollary}{0}
\setcounter{proposition}{0}
\setcounter{example}{0}
\setcounter{remark}{0}
\setcounter{problem}{0}

The free-field Hamiltonian $\mathbf{H}_{0}$, given by the explicit formula \eqref{hamr:46}, i.e.,
\begin{align}
\label{canc:1}
   \mathbf{H}_{0} = i\sum_{j=1}^n \frac{\partial }{\partial t_{\xi_j}}
                 + E_\mho(\rho_\epsilon,\FOU{\chi}_\epsilon),
\end{align}
depends on the zero-point energy $E_\mho$.  As explained in Sec.\ref{hamr:0}.5, that energy is bothersome since it is infinite in the limit $\epsilon \rightarrow 0$. 

   However, the zero-point energy $E_\mho$, which is also included in the extension $\mathbf{H}_{0}^{<0>}$ of $\mathbf{H}_{0}$, is a finite real constant as long as $\epsilon \neq 0$.  Thus, since $E_\mho$ is just a real number, it cancels out when calculating the interaction field-operator $\bphi_I$ given by \eqref{scar:8}.  For that reason the interaction Hamiltonian $\mathbf{H}_I$ defined by \eqref{scar:13} is also independent of $E_\mho$.

    Furthermore, on very general grounds, $\mathbf{H}_{0}$ does not depend on time, as was noted in Remark \ref{summ:rema:1}, and as can be seen in \eqref{canc:1}.  Thus, from the definition \eqref{scar:1} of $\mathbf{S}_\tau$, it is clear that a time-independent factor $\rme^{iE_\mho}$ can be simplified on both sides of the differential equation \eqref{scar:12} of the scattering operator.  Consequently, the zero-point energy can be `ignored' (i.e., equated to zero) when calculating the scattering operator, which implies that the physical results (i.e., the eigenvalues of the scattering operator) do not depend on $E_\mho$.

   In summary, while the zero-point energy $E_\mho$ is not `eliminated,' it cancels out in such a way that the physical observables are not affected by it.

%Alternate derivation  of differential equation for S

\section{Bypassing the `$t$-depending domains' difficulty}
%----------------------------------------------------------
\label{bypa:0}
\setcounter{equation}{0}
\setcounter{definition}{0}
\setcounter{axiom}{0}
\setcounter{conjecture}{0}
\setcounter{lemma}{0}
\setcounter{theorem}{0}
\setcounter{corollary}{0}
\setcounter{proposition}{0}
\setcounter{example}{0}
\setcounter{remark}{0}
\setcounter{problem}{0}

The scattering operator is mathematically simpler than the interacting field operator.  Also, using the explicit form \eqref{canc:1} of the free-field Hamiltonian, one can to a large extent circumvent the difficulties due to the time-dependence of the operators's domains, and therefore use a classical definition of the $t$-derivative.  This is the aim of this section which like most sections of this report should be considered as a basis for improvement.

Let $\mathbf{F} = \prod_{n=0}^\infty\mathsf{L}_S^2\bigl((\mathbb{R}^3)^n \bigr)$ be the infinite product of Hilbert spaces introduced in the proof of Th.~\ref{hami:theo:1}. An element of $\mathbf{F}$ is an arbitrary infinite sequence $(f_0,f_1,...,f_n,...), f_n \in \mathsf{L}_S^2\bigl((\mathbb{R}^3)^n$
without any restriction. We have
\begin{align}
\label{bypa:1}
   \mathbb{D} \subset \mathbb{F} \subset \mathbf{F}, 
\end{align}
(and $\mathbf{F}$ appears as the dual space of $\mathbb{D}$ in a natural sense developed classically in mathematics).  A bounded set in $\mathbf{F}$ is defined as a set of these infinite sequences such that all components (of order $n=0,1,2,...$) are bounded (in $\mathsf{L}_S^2\bigl((\mathbb{R}^3)^n\bigr)$ for the component of order $n$).  Therefore, it is contained in a product set $\prod_{n=0}^\infty B_n$, with $B_n$ bounded in $\mathsf{L}_S^2\bigl((\mathbb{R}^3)^n\bigr)$.  If $\mathfrak{B}$ is a bounded disk in $\mathbf{F}$, for instance $\mathfrak{B} = \prod_{n=0}^\infty B_n$ with $B_n$ bounded in $\mathsf{L}_S^2\bigl((\mathbb{R}^3)^n\bigr)$, we denote by $\mathbf{F}_\mathfrak{B}$ the vector space $\bigcup_n  n\mathfrak{B}$ normed by 
$\|x\| = \inf \{ \lambda > 0, \text{ such that } x \in \lambda \mathfrak{B} \}$.

Let $\mathbf{H}_{0} \DEF \mathbf{H}_{0}(\rho_\epsilon,\FOU{\chi}_\epsilon) \equiv \mathbf{H}_{0}(\rho_\epsilon,\FOU{\chi}_\epsilon,t)$ be the free-field Hamiltonian in the form \eqref{canc:1}, which explicitly shows that it does not depend on $t$ nor $\tau$, and only acts on state vectors, whose time-dependence is $t_\xi$.  Then, despite that it is an unbounded operator on $\mathbb{F}$, it has nice properties that will be basic in the sequel: $\mathbf{H}_{0}$ maps $\mathsf{L}_S^2\bigl((\mathbb{R}^3)^n\bigr)$ into itself, and further through its explicit formula \eqref{canc:1} it is bounded from $\mathsf{L}_S^2\bigl((\mathbb{R}^3)^n\bigr)$ into itself with a bound depending, of course, on $n$ and on $\epsilon >0$.\footnote{As well as on the mollifier used to embedded the states, as discussed in Sec.~\ref{embs:0}, but we shall keep this dependence implicit.}   Therefore, for fixed $\epsilon >0$ it maps $\mathbf{F}$ into itself, and any bounded set in $\mathbf{F}$ into another (larger) bounded set in $\mathbf{F}$.  From \eqref{canc:1} its exponential $\rme^{i(t-\tau)\mathbf{H}_{0}}$ is obviously well defined, and has the same properties as $\mathbf{H}_{0}$, with the bonus that it is a unitary operator on the Fock space $\mathbb{F}$.  Indeed, we do not need a proof of self-adjoint extension and to appeal to semigroup theory: The above direct properties suffice and are far better than those given by the abstract theory of Sec.~\ref{hami:0}.

   Let us consider, for fixed $\epsilon >0$, the operator
\begin{align}\label{bypa:2}
    \mathbf{S}_\tau(t) \DEF
    \mathbf{S}_\tau(\rho_\epsilon,\FOU{\chi}_\epsilon,t)
                        = \rme^{ i(t-\tau)\mathbf{H}_0^{<0>}}
                          \rme^{-i(t-\tau)\mathbf{H}^{<0>}}, 
\end{align}
where the first exponential is defined as above with $\mathbf{H}_0^{<0>}$ given by the explicit formula \eqref{canc:1} of $\mathbf{H}_0$, and the second stems from the abstract theory following Theorem~\ref{hami:theo:1}.
\begin{theorem}[Time-evolution operator equation]
%................................................
\label{bypa:theo:1}
For fixed $\epsilon > 0$ there exist an intermediate \emph{Hilbert} space $\mathbb{F}_\He$ is the position $ \mathbb{F} \subset \mathbb{F}_\He \subset \mathbf{F}$ with bounded inclusions such that $\forall \Phi \in \mathbb{D}$ the map
\begin{align}
\nonumber
  \mathbb{R} &\rightarrow \mathbb{F}_\He,\\
\label{bypa:3}
           t &\mapsto\mathbf{S}_\tau(t)~\Phi \in \mathbb{F} \subset \mathbb{F}_\He, 
\end{align}
is $\mathcal{C}^1$ with derivative
\begin{align}\label{bypa:4}
    \frac{\partial}{\partial t}
    \mathbf{S}_\tau(t) ~ \Phi
 = -i \, \mathbf{H}_I(t)
    \, \mathbf{S}_\tau(t) ~ \Phi,
\end{align}
where $\mathbf{H}_I(t) \DEF \mathbf{H}_I(\rho_\epsilon,\FOU{\chi}_\epsilon,t,\tau)$ is defined by \eqref{scar:13}.  Further, there is a bounded disk $\mathfrak{B}$ in $\mathbf{F}$ such that $\mathbb{F}_\He \subset \mathbf{F}_\mathfrak{B}$, and $\mathbf{H}_I(t)$ is linear continuous from $\mathbb{F}$ into the normed space $\mathbf{F}_\mathfrak{B}$.
\end{theorem}
Proof: Let us first consider the group of contractions $\rme^{-i(t-\tau)\mathbf{H}^{<0>}}$.  We refer to Ref.~\cite{TUCSN2004-} for references to classical results on semigroups of unbounded operators on a Hilbert space.

  Considering the restriction of $\mathbf{H}^{<0>}$ to its domain $\mathsf{D}\bigl(\mathbf{H}^{<0>}\bigr)$ equipped with the graph norm \cite{TUCSN2004-} as an unbounded operator on the Hilbert space $\mathsf{D}(\mathbf{H}^{<0>})$ with domain $\mathsf{D}\bigl((\mathbf{H}^{<0>})^2\bigr) \supset \mathbb{D}$, then from \cite[Prop.2.1.3 and corollaries]{TUCSN2004-} the map 
\begin{align}\label{bypa:5}
t \mapsto  \rme^{-i(t-\tau)\mathbf{H}^{<0>}}\Phi,
\end{align}
is $\mathcal{C}^1$ from $\mathbb{R}$ into $\mathbb{F}$ with derivative
\begin{align}\label{bypa:6}
    -i \mathbf{H}^{<0>} \rme^{-i(t-\tau)\mathbf{H}^{<0>}}\Phi 
=-i  \rme^{-i(t-\tau)\mathbf{H}^{<0>}}\mathbf{H}^{<0>}\Phi,
\end{align}
where $\rme^{-i(t-\tau)\mathbf{H}^{<0>}}$ maps $\mathsf{D}\bigl(\mathbf{H}^{<0>}\bigr)$ into $\mathsf{D}\bigl(\mathbf{H}^{<0>}\bigr)$ as well as  $\mathbb{F}$ into  $\mathbb{F}$.

   Next, concerning $\rme^{i(t-\tau)\mathbf{H}_0^{<0>}}$, it suffices to work componentwise:  Take as $B_n$'s suitable homothetics of the unit ball in $\mathsf{L}_S^2\bigl((\mathbb{R}^3)^n\bigr)$, ~$\mathfrak{B} = \prod_n B_n$, and for $\mathbb{F}_\Ht \supset  \mathbf{F}_\mathfrak{B}$ the completion of $\mathbf{F}_\mathfrak{B}$ for a suitable scalar product of the form
\begin{align}\label{bypa:7}
\langle(f_n),(g_n)\rangle = \sum \lambda_n  \langle f_n,g_n \rangle_{\mathsf{L}_S^2\bigl((\mathbb{R}^3)^n\bigr)},
\end{align}
with $\lambda_n$ real coefficients that tend to $0$ fast enough (depending on the $B_n$'s).

   Another way to deal with $\rme^{i(t-\tau)\mathbf{H}_0^{<0>}}$ consists in applying the general theory in \cite[Theorem\,2.2.2 and corollaries]{TUCSN2004-}:  $\mathbb{F}_\Ht$ is then the Hilbert space named $\mathbb{F}_{-1}$ in the terminology of Ref.~\cite{TUCSN2004-} with $\mathbf{A}$ being the self-adjoint version of $\mathbf{H}_0$ in our case.  Then $\mathbb{F}_\Ht$ is $\bigl(\mathsf{D}(\mathbf{H}_0^{<0>})\bigr)'$, i.e., the dual of $\mathsf{D}(\mathbf{H}_0^{<0>})$ equipped with the graph norm \cite{TUCSN2004-}.

   Now, we have,
\begin{align}
\label{bypa:8}
    \frac{\partial}{\partial t} \mathbf{S}_\tau(t) ~\Phi
    &= i\mathbf{H}_0\rme^{ i(t-\tau)\mathbf{H}_0}
                    \rme^{-i(t-\tau)\mathbf{H}^{<0>}}~\Phi\\
\nonumber
 &~~~~\mathbb{F}_\Ht \leftarrow
  \mathbb{F} \longleftarrow
  \mathbb{F} \longleftarrow
  \mathbb{F} \supset \mathbb{D}\\
\label{bypa:9}
    &- i            \rme^{ i(t-\tau)\mathbf{H}_0}
                    \rme^{-i(t-\tau)\mathbf{H}^{<0>}}\mathbf{H}^{<0>}~\Phi,\\
\nonumber
 &~~~~\mathbb{F}_\Ht \supset \mathbb{F} \leftarrow
  \mathsf{D}\bigl(\mathbf{H}_0\bigr)    \leftarrow
  \mathsf{D}\bigl(\mathbf{H}^{<0>}\bigr) \leftarrow
  \mathsf{D}\bigl((\mathbf{H}^{<0>})^2\bigr) \supset \mathbb{D}
\end{align}
where, below the two parts of the derivative, we show that the relations between the domains of the operators are correct.  Then, continuing,
\begin{align}
\label{bypa:10}
    \frac{\partial}{\partial t} \mathbf{S}_\tau(t) ~\Phi
    &= -i\Bigl(- \mathbf{H}_0
               + \rme^{ i(t-\tau)\mathbf{H}_0}
                 \mathbf{H}^{<0>} 
                 \rme^{-i(t-\tau)\mathbf{H}_0}
           \Bigr)\mathbf{S}_\tau(t) ~\Phi,\\
\label{bypa:11}
    &= -i\rme^{ i(t-\tau)\mathbf{H}_0}
          \Bigl( - \mathbf{H}_0 + \mathbf{H}^{<0>} \Bigr)
         \rme^{-i(t-\tau)\mathbf{H}_0} \mathbf{S}_\tau(t) ~\Phi,\\
\label{bypa:12}
    &= -i \, \mathbf{H}_I(t) \, \mathbf{S}_\tau(t) ~\Phi,\\
\nonumber
 &~~~~\mathbf{F} \leftarrow
  \mathbb{F} \leftarrow 
  \mathbb{F} \supset \mathbb{D}
\end{align}
so that \eqref{bypa:4} is proved.  Indeed, we know that $\mathbf{H}_I(t) \mathbf{S}_\tau(t)~\Phi \in \mathbb{F}_\Ht$.  But this is not obvious since it involves precise information on $\mathbf{S}_\tau(t)$ and $\mathbf{H}_I(t)$  that would amount to reconstructing Eqs.(\ref{bypa:8}--\ref{bypa:9}) by doing the above calculations in the reverse order.  But from \eqref{scar:8}, \eqref{scar:13} and the formula for $\bphi_0$ there is a bounded disk $\mathfrak{B}$ in $\mathbf{F}$ such that $\mathbf{H}_I(t)$ maps $\mathbb{F}$ into $\mathbf{F}_\mathfrak{B}$.  In fact, $\mathbf{H}_I(t)$ is even a linear bounded map from $\mathbf{F}$ into $\mathbf{F}$:  The exponential $\rme^{ i(t-\tau)\mathbf{H}_0}$ acts componentwise so that it maps $\prod_n B_n$ into $\prod_n \lambda_n B_n$ with $\lambda_n \rightarrow \infty$ in a suitable way.   $(\bphi_0)^{N+1}$ creates and annihilates only $N+1$ particles at most. Thus, from $\prod_{n=0}^\infty B_n$ one can easily construct (by induction on $n$) a sequence $(B_n')$ of bounded disks in $\mathsf{L}_S^2\bigl((\mathbb{R}^3)^n\bigr)$ such that $\int (\bphi_0)^{N+1} \FOU{\chi}(\epsilon x)\,dx$ maps $\prod_{n=0}^\infty B_n$ into $\prod_{n=0}^\infty B_n'$.  Set $\mathfrak{B} = \prod_{n=0}^\infty B_n'$, taking into account all factors in $\mathbf{H}_I(t)$, i.e., increasing the $n$'s and enlarging the $B_n$'s in $\prod_n B_n$ as much as necessary, and choose $\mathfrak{B}$ such that $\mathbf{H}_I(t)$ from $\mathbb{F}$ into $\mathbf{F}_\mathfrak{B}$ has operator-norm $\leq 1$ for convenience. ~~\END

\section{Attempt to perturbation expansion}
%------------------------------------------
\label{atte:0}
\setcounter{equation}{0}
\setcounter{definition}{0}
\setcounter{axiom}{0}
\setcounter{conjecture}{0}
\setcounter{lemma}{0}
\setcounter{theorem}{0}
\setcounter{corollary}{0}
\setcounter{proposition}{0}
\setcounter{example}{0}
\setcounter{remark}{0}
\setcounter{problem}{0}

In the standard formulation of quantum field theory the differential equation for the time-evolution operator, i.e., \eqref{scar:12} or \eqref{bypa:4}, is solved by iteration.  This gives a solution in the form of a power series in the coupling constant $g$, i.e., a `perturbation expansion' when $g$ is small. Then, to make contact with experiment, a number of assumptions are made to relate the Fock-space states to the asymptotic states (i.e., the states as $\tau \rightarrow -\infty, t \rightarrow +\infty$), as well as to relate the interaction Hamiltonian to the free-field Hamiltonian which is governing these states.  In the case of quantum electrodynamics, where $g \approx 10^{-2}$, the first terms in the series give then results which are in excellent agreement with the measurements, even though it has been proved that the full series is not convergent (i.e., it is only an asymptotic development).  

  Introducing these hypotheses and working out their consequences in the $\mathcal{G}$-formalism is beyond the scope of the present report. Nevertheless, it is of interest to derive the iterative solution to the differential equation for the time-evolution operator using the present stage of our formulation of quantum-field theory.  In particular, this gives an opportunity to calculate the remainder of the series, which will confirm that the present formulation is incomplete.  

   Therefore, considering \eqref{scar:12} or \eqref{bypa:4}, we put $g$ in evidence by defining $\mathbf{H}_\It \DEF g^{-1}\mathbf{H}_I$ so that
\begin{align}\label{atte:2}
    \frac{\partial}{\partial t}
    \mathbf{S}_\tau(t) ~ \Phi
 = -i g \, \mathbf{H}_\It(t)
    \, \mathbf{S}_\tau(t) ~ \Phi,
   \qquad \text{with} \qquad \mathbf{S}_\tau(\tau) = \mathbf{1}.
\end{align}
Then, writing $\mathbf{S}_\tau(t)$ for $\mathbf{S}_\tau(t,g)$, we study the map 
\begin{align}
\nonumber
  \mathbb{R} &\rightarrow \mathbb{F}_\He,\\
\label{atte:1}
           g &\mapsto \mathbf{S}_\tau(t)
            \equiv \mathbf{S}_\tau(\rho_\epsilon,\FOU{\chi}_\epsilon,t,g)~\Phi
            \in \mathbb{F} \subset \mathbb{F}_\He, 
\end{align}
in the perspective of determining whether a Taylor development of $\mathbf{S}_\tau(t,g)$ makes sense in the limit $g \rightarrow 0$. 
\begin{theorem}[Iterative solution of time-evolution operator equation]
%......................................................................
\label{atte:theo:1}
~ Let $n \in \mathbb{N}$, then $\exists \mathfrak{B}_n$ bounded disk in $\mathbf{F}$ such that, $\forall \Phi \in \mathbb{F}$,
\begin{align}
\nonumber
 \mathbf{S}_\tau(t,g)~\Phi
     &=   \mathbf{S}_{[0]}~\Phi
      + g \mathbf{S}_{[1]}(t)~\Phi
      + g^2 \mathbf{S}_{[2]}(t)~\Phi
      + ...\\
\label{atte:3}
     &+ g^n \mathbf{S}_{[n]}(t)~\Phi
      + \mathbf{R}_{[n]}(t,\tau,g)~\Phi,
\end{align}
where $\mathbf{S}_{[0]} = \mathbf{1}$, and
\begin{align}
\label{atte:4}
   \mathbf{S}_{[n]}(t) = -i\int_\tau^{t} du~
                                \mathbf{H}_\It(u) \, \mathbf{S}_{[n-1]}(u),
\end{align}
and where the remainder is such that
\begin{align}
\label{atte:5}
    \| \mathbf{R}_{[n]}(t,\tau,g)~\Phi \|_{ \mathbb{F}_{\mathfrak{B}_n} }
    \leq  g^n \| \Phi \|_\mathbf{F} \frac{(t - \tau)^n}{n!}.
\end{align}
\end{theorem}
Remark: The theorem as it is gives no hope of convergence of the series because $\mathfrak{B}_n$ changes with $n$.  As expected, it makes sense only for fixed $n$.

\noindent Proof: The proof is by induction on $n$.  We give only the first few steps since the following ones are clearly similar. Let $\Phi \in \mathbb{F}$:

\noindent {\bf Step 0:} We start from
%......................
%
\begin{align}\label{atte:6}
           \mathbf{S}_g'(t) ~ \Phi
 = -i g \, \mathbf{H}_\It(t)
        \, \mathbf{S}_g(t) ~ \Phi,
   \qquad \text{with} \qquad \mathbf{S}_g(\tau)~ \Phi = \Phi \in \mathbb{F}.
\end{align}
Let
\begin{align}\label{atte:7}
           \mathbf{Y}_g(t) = \Bigl( \mathbf{S}_g(t) - \mathbf{1} \Bigr) \Phi,
\end{align}
Then
\begin{align}\label{atte:8}
           \mathbf{Y}_g'(t)
 = -i g \, \mathbf{H}_\It(t)
        \, \mathbf{S}_g(t) ~ \Phi,
   \qquad \text{and} \qquad \mathbf{Y}_g(\tau) = 0.
\end{align}
Thus we set
\begin{align}\label{atte:9}
      \mathbf{Y}_g(t) \DEF \int_\tau^t du~ \mathbf{Y}_g'(u) ~\Phi
      = -i g \int_\tau^t du~ \mathbf{H}_\It(u) \, \mathbf{S}_g(u) ~\Phi.
\end{align}
Recall from the end of the proof of Theorem~\ref{bypa:theo:1} that $\|\mathbf{S}_g(t)~\Phi\|_\mathbb{F} = \|\Phi\|_\mathbb{F}$, and  $\mathbf{H}_\It(t): \mathbb{F} \rightarrow \mathbf{F}_{\mathfrak{B}_0}$ with operators norm $\leq 1$.  Then, 
\begin{align}\label{atte:10}
\nonumber
       \|\Bigl(\mathbf{S}_g(t) - \mathbf{1}\Bigr)\Phi\|_{\mathfrak{B}_0}
    &= \|\mathbf{Y}_g(t)~\Phi \|_{\mathfrak{B}_0}
     = g \| \int_\tau^t du~ \mathbf{H}_\It(u) \,
                            \mathbf{S}_g(u) ~\Phi \|_{\mathfrak{B}_0}\\
    &\leq g~(t-\tau)~\|\Phi\|_\mathbb{F}.
\end{align}

\noindent {\bf Step 1:}  Let 
%......................
%
\begin{align}
\label{atte:11}
   \mathbf{S}_{[1]}'(t) = -i\int_\tau^{t} du~
                                \mathbf{H}_\It(u).
\end{align}
Now set
\begin{align}\label{atte:12}
   \mathbf{Y}_g(t) = \Bigl( \mathbf{S}_g(t) - \mathbf{1}
                        - g \mathbf{S}_{[1]}(t)
                     \Bigr)\Phi.
\end{align}
Then
\begin{align}\label{atte:13}
           \mathbf{Y}_g'(t)
  = -ig \, \mathbf{H}_\It(t) \, \mathbf{S}_g(t) ~\Phi
    +ig \, \mathbf{H}_\It(t) ~ \Phi
  = -ig \, \mathbf{H}_\It(t) \, \Bigl( \mathbf{S}_g(t) - \mathbf{1} \Bigr) \Phi,
\end{align}
and $\mathbf{Y}_g(\tau) = 0$. Thus
\begin{align}\label{atte:14}
      \mathbf{Y}_g(t) \DEF \int_\tau^t du~ \mathbf{Y}_g'(u) ~\Phi
      = -i g \int_\tau^t du~ \mathbf{H}_\It(u) \, 
      \Bigl( \mathbf{S}_g(u) - \mathbf{1} \Bigr) \Phi.
\end{align}
Let $\mathfrak{B}_1$ be a bounded disk in $\mathbf{F}$ such that $\mathbf{H}_\It(u) \mathfrak{B}_0 \subset \mathfrak{B}_1, \forall t$ (easy construction by induction on the components $n$ from the formula of $\mathbf{H}_\It$ for fixed $\epsilon$).  Then
\begin{align}\label{atte:15}
   \| \mathbf{Y}_g(t) \|_{\mathfrak{B}_1}
 = \| \Bigl( \mathbf{S}_g(t) - \mathbf{1}  
         - g \mathbf{S}_{[1]}(t)
      \Bigr)\Phi \|_{\mathfrak{B}_1}
 \leq g^2 \frac{(t-\tau)^2}{2} \| \Phi \|_\mathbb{F},
\end{align}
where we used \eqref{atte:10} and $\int_\tau^t (u-\tau)\,du = (t-\tau)^2/2$.

\noindent {\bf Step 2:}  Let 
%......................
\begin{align}
\label{atte:16}
   \mathbf{S}_{[2]}'(t) = -i\int_\tau^{t} du~
                                \mathbf{H}_\It(u)\mathbf{S}_{[1]}(u).
\end{align}
Now set
\begin{align}\label{atte:17}
   \mathbf{Y}_g(t) = \Bigl( \mathbf{S}_g(t) - \mathbf{1}
                   - g   \mathbf{S}_{[1]}(t) 
                   - g^2 \mathbf{S}_{[2]}(t)
                     \Bigr) \Phi.
\end{align}
Then
\begin{align}
\nonumber
           \mathbf{Y}_g'(t)
 &= -ig \, \mathbf{H}_\It(t) \, \mathbf{S}_g(t) ~\Phi
    +ig \, \mathbf{H}_\It(t) ~ \Phi
    +ig^2\,\mathbf{H}_\It(t) \, \mathbf{S}_{[1]}(t)~ \Phi\\
\label{atte:18}
 &= -ig \, \mathbf{H}_\It(t) \, \bigl( \mathbf{S}_g(t) - \mathbf{1} 
       -g \mathbf{S}_{[1]} \bigr) \Phi.
\end{align}
Let $\mathfrak{B}_2$ be a bounded disk in $\mathbf{F}$ such that $\mathbf{H}_\It(u) \mathfrak{B}_1 \subset \mathfrak{B}_2, \forall t$ (induction on the components).  Then,
\begin{align}
\nonumber
   \| \mathbf{Y}_g(t) \|_{\mathfrak{B}_2}
 &=\| \Bigl( \mathbf{S}_g(t) - \mathbf{1} 
         - g \mathbf{S}_{[1]}(t)
       - g^2 \mathbf{S}_{[2]}(t)
      \Bigr) \Phi \|_{\mathfrak{B}_1}\\
\label{atte:19}
 &\leq g^3 \frac{(t-\tau)^2}{3!} \| \Phi \|_\mathbb{F},
\end{align}
where we used \eqref{atte:15} and $\int_\tau^t (u-\tau)^2/2\,du = (t-\tau)^3/3!$.

\noindent {\bf Step 3:}  Let 
%......................
\begin{align}
\label{atte:20}
   \mathbf{S}_{[3]}'(t) = -i\int_\tau^{t} du~
                                \mathbf{H}_\It(u)\mathbf{S}_{[2]}(u).
\end{align}
Now set
\begin{align}\label{atte:21}
   \mathbf{Y}_g(t) = \Bigl( \mathbf{S}_g(t) - \mathbf{1} 
                   - g   \mathbf{S}_{[1]}(t)
                   - g^2 \mathbf{S}_{[2]}(t)
                   - g^3 \mathbf{S}_{[3]}(t)
                     \Bigr) \Phi.
\end{align}
We obtain in the same way, using \eqref{atte:19},
\begin{align}\label{atte:22}
   \| \mathbf{Y}_g(t) \|_{\mathfrak{B}_4}
               \leq g^4 \frac{(t-\tau)^4}{4!} \| \Phi \|_\mathbb{F},
\end{align}
And so on.  \END

   As an obvious corollary, we can calculate the transition probabilities of the time-evolution operator.  Thus, from definition \eqref{summ:29}:
\begin{corollary}[Transition probabilities of time-evolution operator]
%.....................................................................
\label{atte:corol:1}
~ Let $n \in \mathbb{N}$, then $\forall \Phi \in \mathbb{F}$, and  $\forall \Psi \in \mathbb{D}$
\begin{align}
\nonumber
 \BRA \Psi \| \mathbf{S}_\tau(t,g)~\Phi \KET
     &= \BRA \Psi \|\Phi \KET
      + g \BRA \Psi \|\mathbf{S}_{[1]}(t)~\Phi \KET
      + g^2 \BRA \Psi \|\mathbf{S}_{[2]}(t)~\Phi \KET
      + ...\\
\label{atte:23}
     &+ g^n \BRA \Psi \|\mathbf{S}_{[n]}(t)~\Phi \KET
      + R_{[n]}(t,\tau,g),
\end{align}
with remainder
\begin{align}
\label{atte:24}
    | R_{[n]}(t,\tau,g) |
    \leq  \Cst \, g^n \, \| \Psi \|_\mathbf{F}
                      \, \| \Phi \|_\mathbf{F} \, \frac{(t - \tau)^n}{n!},
\end{align}
where $\Cst$ depends on $n$, and a priori also on $\rho, \FOU{\chi},$ and $\epsilon$.
\end{corollary}

Of course, as for the theorem, nothing can be said about convergence.  In particular, taking two normalized states $\Phi$ and $\Psi$, we do not get a well defined result for the transition probability \eqref{summ:29}, i.e,
\begin{equation}
\label{atte:25}
\mathcal{P}(\Phi \rightarrow \Psi) =
  \lim_{\substack{
       \tau \rightarrow -\infty\\
          t \rightarrow +\infty}} ~
       |\BRA \Psi|\mathbf{S}_\tau(t)\Phi\KET_\mathbb{F}|^2 = ~ ?
\end{equation}
But this could be expected from what we said in the introduction to this section.

The calculations of these probabilities will be considered in Part II of this report.

%File:  nota.8           arXiv version 2       Date: 6 September 2008
%=====                                 

\chapter{Notations}
%==================
\label{nota}

\noindent{\bf Some general notations}
%....................................

\begin{itemize}

\item In general, boldface types are operators: $\mathbf{A}, \mathbf{a}, \bphi, \bpi, ...$

\item In general, `phi's and `psi's are quantum fields or states: $\Phi, \phi, \varphi, \psi, \bphi, ...$

\item The symbol $\ASS$ is used for `association' (rather than $\thickapprox$).

\item The symbol $\odot$ is used to emphasize that a product is calculated in $\mathcal{G}$.

\end{itemize}

\noindent{\bf Some specific notations}
%.....................................

\begin{itemize}

\item[$\mathbf{A}$] unspecified operator

\item[$\mathbf{a}^\pm$] creation/annihilation operators

\item[$\mathcal{A}$] action

\item[$\mathcal{A}_\infty$] set of suitable mollifiers

\item[$\mathcal{B}_\infty$] set of suitable dampers

\item[$\mathbf{B}$] unspecified operator

\item[$B$]  bounded set in $\mathbb{D}$

\item[$\mathfrak{B}$]  bounded set in $\mathsf{L}(\mathbb{D})$

\item[$\mathcal{C}$] space of continuous functions

\item[$\mathcal{C}^\infty$] algebra of smooth functions

\item[$\rmD$]   differentation

\item[$\mathcal{D}$] space of test functions

\item[$\mathcal{D}'$] space of distributions

\item[$\mathbb{D}$] dense space of states with finite number of particles

\item[$\mathbf{D}^{<0>}$] domain of $\mathbf{H}^{<0>}$

\item[$E_\mho$] zero-point energy

\item[$\mathcal{E}_{\text{M}}$] algebra of moderate functions

\item[$\mathbb{F}$] Fock space

\item[$\mathsf{F}$] Fourier transformation

\item[$\bphi$] interacting-field operator

\item[$\bphi_0$] free-field operator

\item[$g$] coupling constant

\item[$\mathcal{G}$] algebra of nonlinear generalized functions

\item[$\mathcal{H}$] Hilbert space

\item[$\pmb{\mathcal{H}}$] Hamiltonian density

\item[$\mathbf{H}$] Hamiltonian (general, e.g., of a self-interacting field)

\item[$\mathbf{H}_0$] Hamiltonian of a free field

\item[$\mathbf{H}^{(0)}$] Hamiltonian of a self-interacting field written in terms of the operators of a free field.

\item[$\mathbf{H}^{<0>}$] self-adjoint extension of $\mathbf{H}^{(0)}$.

\item[$\imath$] infinitesimal function

\item[$\iota(~)_\epsilon$] embedding of $(~)$

\item[$K$] compact set

\item[$\FOU{\chi}$] suitable damper

\item[$L$] Lagrange function

\item[$\mathsf{L}$]  linear space

\item[$\mathsf{L}^2$] Hilbert space of square-integrable functions

\item[$\mathcal{N}$] algebra of negligible functions

\item[$N$] some (possibly fixed) integer, e.g., $\exists N \in \mathbb{N}$

\item[$\OOO$] on the order of, e.g., $\OOO(\epsilon^q)$

\item[$\mathcal{O}_{\text{M}}$] space of functions of growth less than $|x|^N$ at infinity

\item[$\Omega$] open set in $\mathbb{R}^n$

\item[$\bpi$] conjugate interacting-field operator

\item[$\bpi_0$] conjugate free-field operator

\item[$q$] any integer, e.g., $\forall q \in \mathbb{N}$

\item[$\rho$] suitable mollifier

\item[$\mathbf{S}$] scattering operator

\item[$\mathbf{S}_\tau(t)$] time evolution operator

\item[$\mathcal{S}$] space of smooth functions of steep descent

\item[$t$] time

\item[$T$] test function in $\mathcal{D}$

\item[$\tau$] initial time

\item[$x$] space-time dependence of operators,
           i.e., $x = \{t, \vec{x}\,\}$

\item[$\xi$] space-time dependence of states,
             i.e., $\xi = \{t_\xi, \vec{\xi}\,\}$

\item[$\mho$] vacuum state

\end{itemize}

%%%%%%%%%%%%%%%% Beginning of conclusion %%%%%%%%%%%%%%%%%%%

%%\chapter {Conclusion}
%%=====================
%%\label{conc:0}

\chapter{References}
%==================
\label{bibl}

\end{document}